\documentclass[fleqn,usenatbib]{mnras}

\usepackage{newtxtext,newtxmath}
\usepackage{subfig}

\usepackage[T1]{fontenc}

\DeclareRobustCommand{\VAN}[3]{#2}
\let\VANthebibliography\thebibliography
\def\thebibliography{\DeclareRobustCommand{\VAN}[3]{##3}\VANthebibliography}

\usepackage{graphicx}	
\usepackage{amsmath}	
\usepackage{pdflscape}
\usepackage{tablefootnote}

\title[Probing the formation of QSO hosts]{Probing the rapid formation of black holes and their galaxy hosts in QSOs}

\author[Cutiva-Alvarez et al.]{
Karla Alejandra Cutiva-Alvarez,$^{1}$\thanks{E-mail: ka.cutivaalvarez@ugto.mx}
Roger Coziol,$^{1}$
Juan Pablo Torres-Papaqui$^{1}$
\\
$^{1}$Departamento de Astronom\'ia, Universidad de Guanajuato, Apartado Postal 144, 36000, Guanajuato, Mexico
}

\date{Accepted XXX. Received YYY; in original form ZZZ}

\pubyear{2015}

\begin{document}
\label{firstpage}
\pagerange{\pageref{firstpage}--\pageref{lastpage}}
\maketitle

\begin{abstract}
Using the modelling code X-CIGALE, we reproduced the SEDs of 1,359 SDSS QSOs within the redshift range $0 < z < 4$, for which we have NIR/MIR fluxes with the highest quality and spectral data characterizing their SMBHs. Consistent with a rapid formation of the host galaxies, the star formation histories (SFHs) have small e-folding, at most 750 Myrs using an SFH function for Spiral or 1000 Myrs using one for Elliptical. Above $z \sim 1.6$, the two solutions are degenerate, the SEDs being dominated by the AGN continuum and high star formation rates (SFRs), typical of starburst galaxies, while at lower redshifts the starburst nature of the host, independent from its morphology, is better reproduced by an Spiral SFH. In general, the SFR increases with the redshift, the mass of the bulge, the AGN luminosity and Eddington ratio, suggesting there is no evidence of AGN quenching of star formation. Comparing the specific BHAR with specific SFR, all the QSOs at any redshift trace a linear sequence below the Eddington luminosity, in parallel and above the one-to-one relation, implying that QSOs are in a special phase of evolution during which the growth in mass of their SMBH is more rapid than the growth in mass of their galaxy hosts. This particular phase is consistent with a scenario where the galaxy hosts of QSOs in the past grew in mass more rapidly than their SMBHs, suggesting that a high star formation efficiency during their formation was responsible in limiting their masses.

\end{abstract}




\section{Introduction}

Since their discovery in the 1960s, theories about quasars have experienced a dramatic paradigm shift. Because they were discovered as strong radio sources, at first it was thought that quasars are examples of a rare and short phase in the evolution of radio galaxies (RGs), with a typical duration time of the order of a few $10^6$ years \citep{1965Sandage,1967Gold}. However, based on a photometric survey of blue quasi-stellar objects (QSOs), \citet{1969Sandage} made clear that radio loud quasars only constitute a few percent of a much larger population of QSOs that are intrinsically radio quiet \citep[see][and references therein]{2017Coziol}, suggesting that this phenomenon must represent a more common phase of galaxy evolution. This last conclusion was readily accepted once the cosmological nature of the redshift was recognized and different surveys revealed QSOs were more common in the past, their density and luminosity significantly increasing at high redshifts \citep[e.g.,][]{2009Croom}.

The other breakthrough in QSO studies consisted in identifying the physical source of their activity: a super-massive black hole (SMBH) accreting matter at the centre of their galaxy hosts \citep{1978Rees,1982Soltan}.
This implies that QSOs represent a primordial phase in the formation of galaxies during which SMBHs are actively forming at the centre of their nuclei \citep{1986Cavaliere,2006Lapi,2010Letawe}. Adopting this new paradigm transformed the study of QSOs into an investigation about galaxy formation and the role the formation of SMBHs plays during this process \citep[e.g.,][]{1998Silk}. 

There are three key observations to consider in search for such a connection. The first one is that SMBHs with masses of the order of $10^9$~M$_\odot$ are found to exist well above a redshift $z \sim 6$ \citep{2003Fan,2011Mortlock,2015Wu,2018Banados,2020YangJ,2021Wang}. This observational fact emphasizes how fast must be the formation process of SMBHs \citep{2019Woods,2020Inayoshi,2022Pacucci}. For example, consider J0313$ - $1806, which is the farthest QSOs discovered so far at $z = 7.642$ \citep{2021Wang}. Based on its emission lines, and adopting a standard $\Lambda$CDM cosmology (with $H_0 = 70$ km s$^{-1}$ Mpc$^{-1}$, $\Omega_{DM} = 0.30$, and $\Omega_{\Lambda} = 0.70$), this QSO, appearing at a relatively young age of the universe, that is, a cosmological time (CT) of only 0.667 Gyrs, would already have reached a mass of $(1.6\ \pm\ 0.4) \times 10^9$ M$_\odot$. Now, assuming the formation of the SMBH started just after recombination at $z = 1,100$, this implies a formation time $\Delta t \sim 7 \times 10^8$ years. How can that be possible? For instance, assuming the black hole (BH) mass increases solely by accretion, at a rate $\dot{M}_{BH} = (1- \eta) \dot{m}_{acc}$, adopting a typical radiation efficiency $\eta = 0.1$ would imply a constant rate of the order $\dot{m}_{acc} \sim 3$ M$_\odot /{\rm yr}$, which is slightly lower than necessary to produce its observed bolometric luminosity, $L_{bol} = 3.6 \times 10^{13}$ L$_\odot$, or $\sim 0.7$\ $L_{\rm Edd}$. However, assuming a constant accretion rate just after recombination is highly improbable, considering that $\dot{m}$ varies as the mass of the BH grows, and its growth itself is entangled with convoluted initial conditions related to the formation of its host galaxy, which, within the context of hierarchical structures formation (massive galaxies forming by the mergers of massive, gas-rich protogalaxies), is a highly complex, non-linear physical phenomenon; this explains why some researchers have considered super massive BHs (SMBHs) form by accreting matter at super Eddington rates, that is, well above the Eddington limit. 

The second clue about the formation of SMBHs is that their masses are correlated with the masses of the bulges of their host galaxies \citep[the so-called BH mass-stellar velocity dispersion relation, or $M_{\rm BH}-\sigma$;][]{1998Magorrian,2000Ferrarese,2004Haring,2009Gultekin,2011Graham}. 
For J0313$ - $1806, assuming the relation deduced based on local AGNs applies at high redshifts \citep[since the formation of SMBH is possibly fixed early on;][]{2006Fan,2019Shen}, would yield a velocity dispersion of stars in the bulge equal to $\sigma \sim 323$ km s$^{-1}$ \citep[using the relation in][]{2011Graham}, which is typical of elliptical galaxies with minimal mass $\sim 1.5 \times 10^{11}$ M$_\odot$ \citep[according to][]{2015Reines}. For QSOs with SMBHs, this not only suggests their host galaxies formed at the same time as their SMBHs, but also that they form the bulk of their stellar populations extremely rapidly, a formation process that is typical of massive, bulge-dominated, early-type galaxies \citep[][]{1986Sandage,2003Warner,2021Bischetti}. 

A rapid formation of galaxy host (or galaxy bulge) is also implied by the third observational clue, which is that in any QSO at any redshift, the gas in the broad line region surrounding the accretion disk has a solar or higher than solar metallicity \citep{1993Hamann,2007Jiang,2009Juarez,2021Sniegowska}. Since metals are formed by stars, the high metallicity of the gas accreting on a SMBH could only mean higher star formation rates in the past accompanied the rapid formation of the SMBHs (the process possibly favouring massive stars; a starburst with flat initial mass function, IMF). A fast formation process for galaxies might also explain the various mass-metallicity relations observed \citep{2003DietrichA,2011NeriLarios,2018Matsuoka}.

But probably the most constraining facts about the abundances is that the line ratio \ion{Fe}{ii}/\ion{Mg}{ii} observed in any QSO at any redshift is relatively high \citep[between 2 and 4;][]{2014DeRosa,2022ChuWang}, which, taken as a proxy for the ratio \ion{Fe}/$\alpha$-elements, suggests a rapid enrichment in \ion{Fe}\ at an epoch much earlier than the CT \citep{2003DietrichB,2003Barth,2007Kurk,2019Shin}. That characteristic could represent a problem for QSOs at high redshifts, because, according to the standard interpretation, while $\alpha$-elements are produced by SNe \rm{II} and \rm{I}b, and thus their abundances in the interstellar medium are expected to increase rapidly during a massive burst of star formation, \ion{Fe}\ is produced by SNe \rm{I}a and thus its abundance in the interstellar medium should be delayed compared to the $\alpha$-elements by a at least 1 Gyr, which is much longer than the typical time-scale for the growth of SMBH by accretion \citep[a few $\times 10^8$ years][]{2020Inayoshi}. However, this apparent difficulty might also translates into an intriguing alternative, which is that the protogalaxies that merged to form the bulges of galaxies \citep{1979Tinsley,1981Silk} were populated by very massive and metal-poor stars, consistent with the elusive Pop \rm{III} \citep{2014DeRosa}: since these stars are metal-poor and massive they would eject in the ISM very few \ion{Mg}{ii}\ compared to \ion{Fe}\ as they evolve, explaining the typical high ratio observed in QSOs \citep{1999Thomas,2002Heger}. 

The hypothesis of protogalaxies formed by Pop \rm{III} stars would also readily fit within a model for the formation of SMBH seeds based on the very popular hierarchical galaxy formation model \citep{2009Devecchi,2016Yajima,2017Sakurai,2018Reinoso}: 1- a huge quantity of pristine gas fall into dark matter (DM) mini-haloes, with masses of the order $10^{5-6}$ M$_\odot$, to form protogalaxies where the IMF is top-heavy, 2- the most massive stars ($m_* \ge 10^2$ M$_\odot$) at the centre of these protogalaxies rapidly evolve and merge to form SMBH seeds with masses of the order $M_{BH} \sim  10^{3-4}$~ M$_\odot$, 3- the mass of these seeds rise rapidly up to $M_{BH} \sim  10^{5-6}$ M$_\odot$ through haloes/protogalaxies/BHs mergers, forming galaxies with massive bulges where a central SMBH rapidly increases its mass by gas accretion up to $M_{BH} \sim  10^{8-9}$ M$_\odot$. 

However, this is the naive scenario, and details how it really happened are still open to discussion \citep[see][]{2020Inayoshi}. Some important questions that need to be answered are the following. Does the formation process of SMBH in high density regions (at high z) differ from what happens in lower density regions (intermediate and low z)? One possibility is that SMBHs observed at high redshifts are rare objects forming only in the most massive DM haloes ($> 10^{8-9}$ M$_\odot$), suggesting that the majority of SMBHs have much lower masses and thus their formation process (and that of their host galaxies) might have been different. Some researcher also claimed that the most massive SMBHs at high redshifts reside in slightly less massive galaxies than observed at lower redshifts, which would imply their formation pre-dated the formation of their hosts \citep{2019Shimasaku,2021Vayner}. However, this would be difficult to explain, considering the spectral similarities and metallicities of QSOs at any redshift.
Related topics consist in searching for correlations between the BH accretion rate (BHAR) and star formation rate (SFR) in AGNs at different redshifts \citep[e.g.,][]{2021McDonald} and determining what role \citep[if any; e.g.,][]{2007Peng,2011Jahnke} AGN feedback (outflows or winds) play in establishing the $M_{\rm BH}-\sigma$ relation
\citep[e.g.,][]{2018Harrison,2020Torres-Papaqui}.

In this article, we present the results of a study that aims to better constrain the formation galaxies hosting SMBHs by tracing the change of SFRs with redshift in a sample of QSOs with different BHAR. Although the UV-Opt part of the spectrum of a QSO is dominated by the AGN continuum and intense broad line emission, information about star formation in their host galaxies still appears in their SEDs through their dominant stellar populations and dust emission in MIR and FIR. In principle, therefore, the host galaxy SFR in an AGN can be estimated by reproducing its SED synthetically using sophisticated tools like \textsl{X-CIGALE} \citep{2019Boquien,2022Yang},\footnote{https://cigale.lam.fr}  which allows distinguishing in a SED the stellar population components (young and old) from the SBMH component. The organization of this paper is the following: in Section~\ref{SD}, our sample of QSOs is described, and their characteristics in terms of BH mass and BHAR are discussed in Subsection~\ref{QSOprop}. Our method of analysis using \textsl{X-CIGALE} is explained in Section~\ref{Method} and the results are presented in Section~\ref{Res}, followed by a brief discussion in Section~\ref{Disc}. Our conclusions can be found in Section~\ref{Con}.

\section{Sample and data}
\label{SD}

To better constrain the SEDs of QSOs using \textsl{X-CIGALE}, we need the largest sample possible with the largest range in high quality fluxes measured. Starting with the QSOs compilation based on SDSS DR12 built by \citet{2017Paris} and the AllWISE catalog \citep{2014Cutri}, both available through CDS VizieR,\footnote{https://vizier.cds.unistra.fr/viz-bin/VizieR} we used the X-Match application tool to retrieve a preliminary list of detected QSOs in the MIR by cross-correlating the position of each entry, obtaining 190,415 candidates \citep[64\% of the QSOs in][]{2017Paris}. However, because the fluxes in WISE do not have all the same quality, we restricted our selection to those QSOs having the highest level: Flag A on all four magnitudes, W1 (3.368\ $\mu$m), W2 (4.618\ $\mu$m), W3 (12.082\ $\mu$m), and W4 (22.194\ $\mu$m). This reduced our final list of high-quality WISE QSOs (hereafter HQWISE QSOs) to 1,359 QSOs covering a range in redshift from $0 < z < 4$. Initially we contemplated adding more fluxes in NIR or even FIR. However, due to the small number of HQWISE QSOs and scarcity of information in other bands, cross-correlation with other data banks (in particular, IRAS for data in the FIR) turned out to be fruitless. 

\begin{figure}
	\includegraphics[width=\columnwidth]{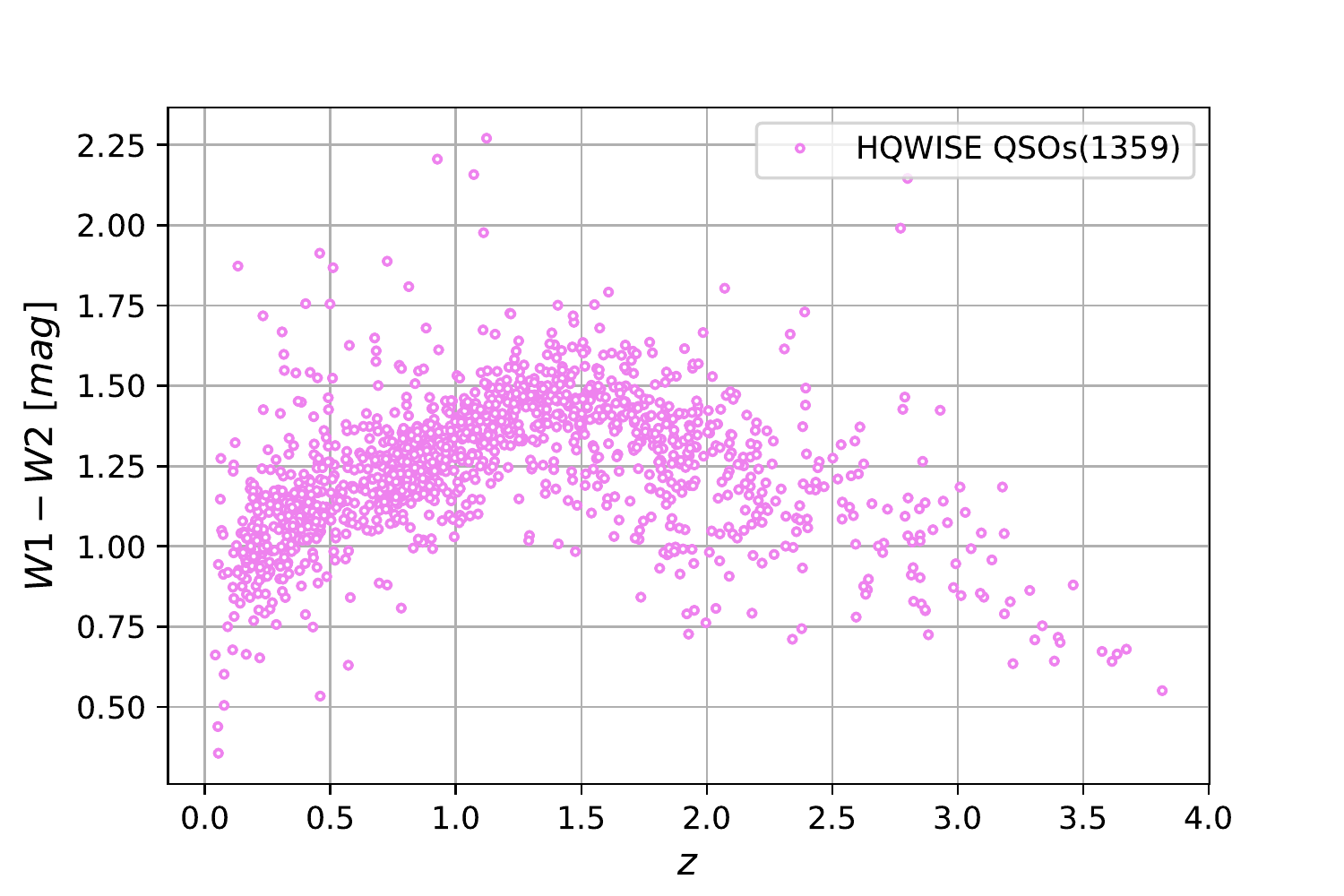}
	\includegraphics[width=\columnwidth]{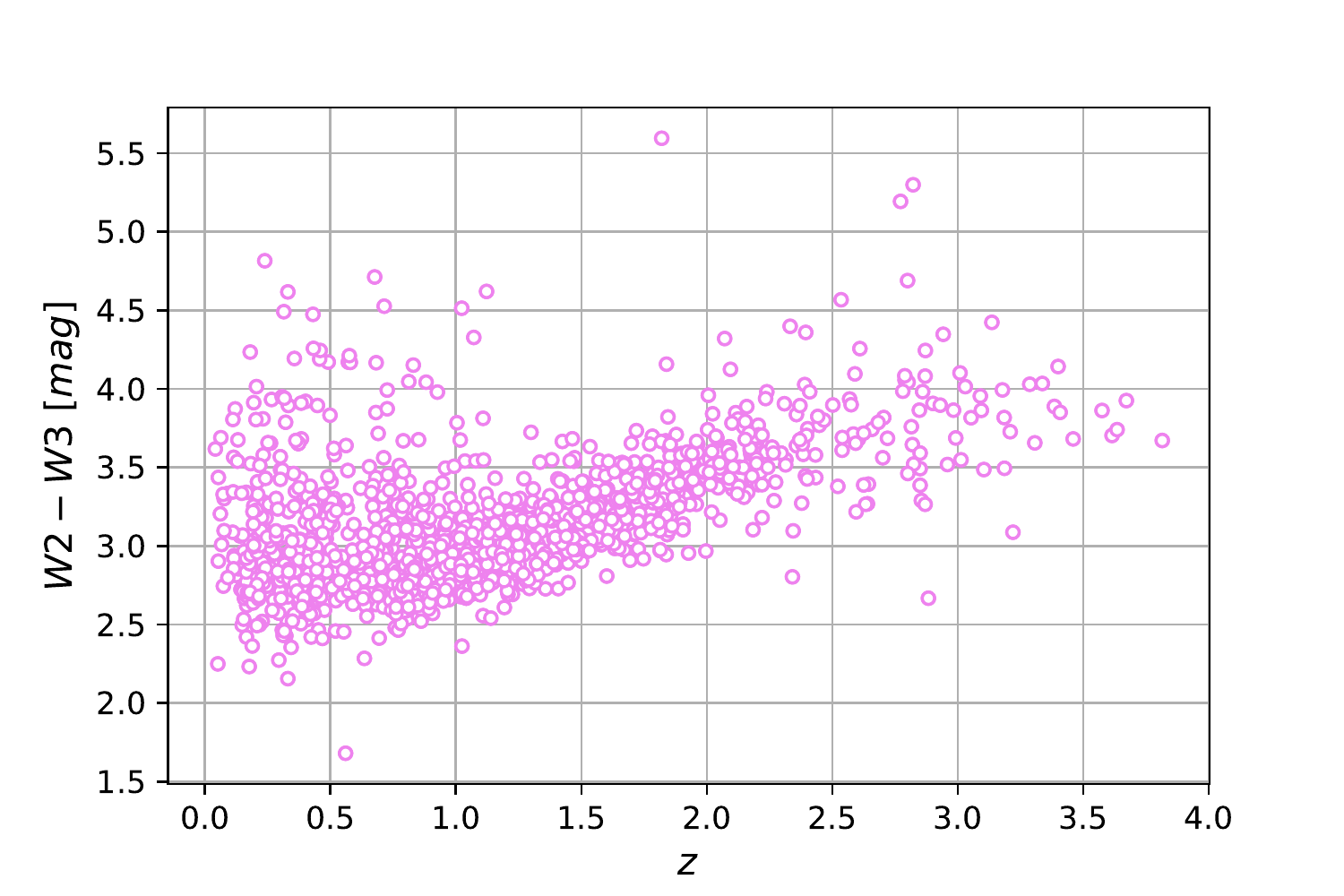}
	\includegraphics[width=\columnwidth]{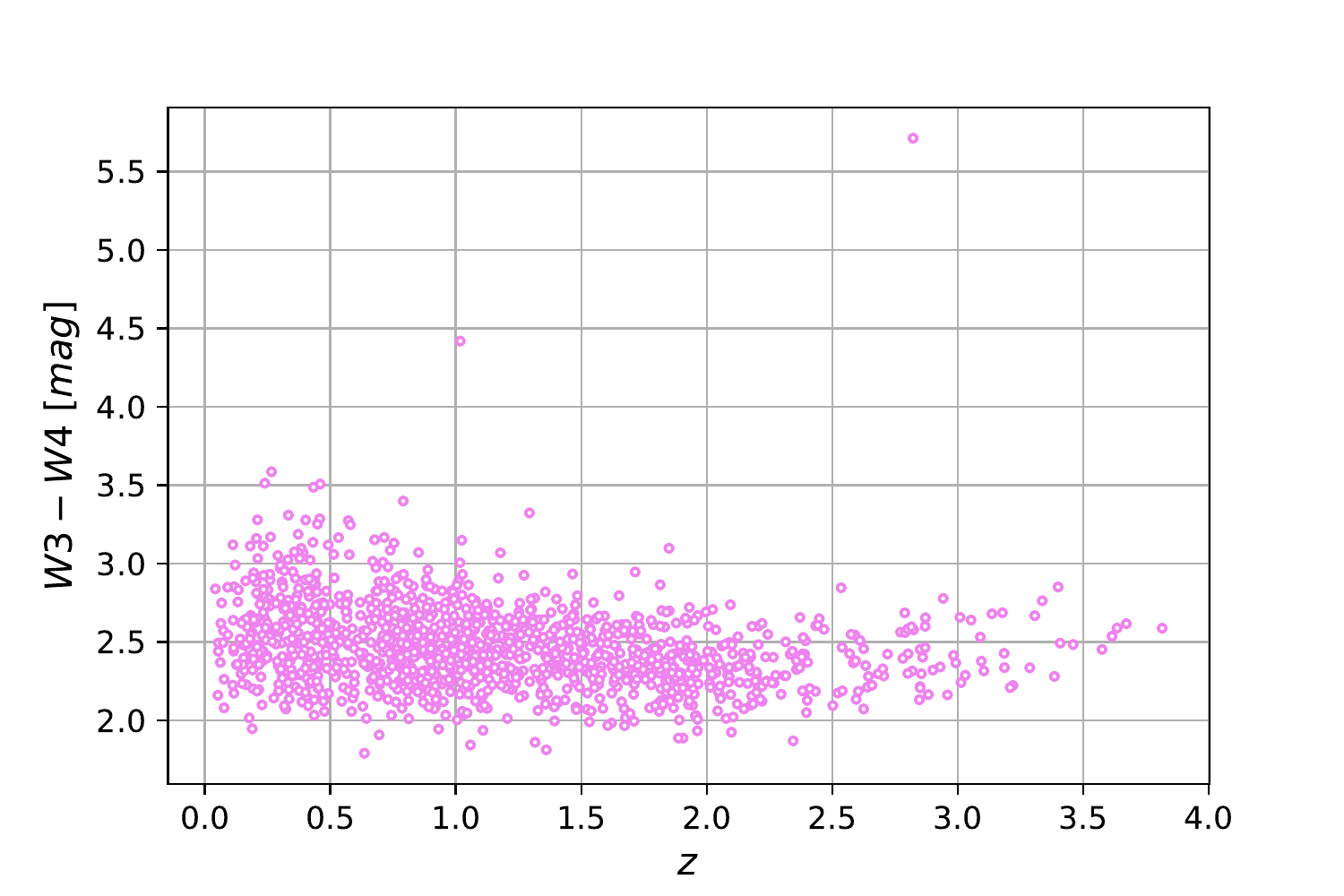}
    \caption{Variations of WISE colours with the redshift.}
    \label{fig1}
\end{figure}

Note that we favor the highest quality fluxes rather than high number of objects in order to fully understand the physical consequences related with the variations of the SED of QSOs at different redshifts. To explain better this last point, we trace in Figure~\ref{fig1} the variations with redshift of the IR colours of the HQWISE QSOs. At any redshift, we observe a relatively high variance, which, considering our strict selection criterion for the fluxes, suggests its physical cause must be intrinsic. For the color W2-W3 (middle panel) an increase in variance is most obvious at low redshift (below $z \sim 1$), where many sources have colours typical of high redshift QSOs, suggesting they might share similar characteristics. One possibility is higher than normal star formation (SF) enshrouded in dust, which could be evidence at low redshifts of ultra luminous IR galaxies (ULIRGs), that is, starburst galaxies where SMBHs hide behind thick veils of dust. In 1988, \citet{1988Sanders} proposed that the ULIRG state is a normal phase in the formation/evolution of QSOs and confirming their presence in our sample would thus be a decisive step. This justifies using high quality IR fluxes, since using lower quality fluxes would increase the variance (creating false results) and impede us to have a clear view about this phenomenon.

One important consequences of the ULIRGs model is the connection implied between Starbursts and AGNs. According to this model, an important part of the dust in QSOs must be heated by star bursts, and as SF in galaxies increases at high redshifts \citep[][]{2014Madau}, an increase of SFR in QSOs would explain why the W2-W3 get redder. At low redshifts ($z <0.25)$, such a reddening sequence, Liners$\rightarrow$Seyfert 2 (Sy2)$\rightarrow$Star Forming Galaxies (SFGs), was found by our group to be empirically connected with a genuine increase of SF \citep[][]{2013Torres-Papaqui,2014Coziol,2015Coziol}. What happens in QSOs however is more complicated. Comparing the colors produced by different SED models for QSOs at low redshifts \citep[Figure 27 in][]{2015Coziol} suggested mild star formation rates, between 0.3 M$_\odot$ yr$^{-1}$ and 10 M$_\odot$ yr$^{-1}$ \citep[consistent with SFRs as measured by][]{2012Diamond-Stanic,2015Xu}. However, this would imply as W2-W3 gets redder in QSOs at higher redshifts, an increase in SFR by a factor 10 or even 100 \citep[e.g.,][]{2014Drouart,2016Dong}. Note that the variation of the W3-W4 colour in the lower panel of Figure~\ref{fig1} seems consistent with such scenario, the colours at low redshifts being comparable to those of Sy2s (local AGNs with high SFRs), becoming bluer at high redshifts; the stellar component of the SED becoming flatter as the SFR increase with the redshift. 

However, the behaviour of the W1-W2 colour with the redshift in the upper panel of Figure~\ref{fig1} is more complicated to explain. At first, W1-W2 becomes redder as the redshift increases, but after $z \sim 1.4$ it becomes bluer. Note that to locate this inflection precisely, we need high quality fluxes. Adopting the standard interpretation of WISE colours \citep[Figure~12 in][also based on high quality fluxes]{2010Wright}, a change in AGN type could be suggested, since Seyfert~1 (Sy1s) are bluer than QSOs in W1-W2. However, the shift in colour happening at high redshift contradicts this explanation, since Sy1s (mostly spiral galaxies) appear only at low redshifts \citep[cf.][]{2020Torres-Papaqui}. Also, the fact that the W2 fluxes show no evidence of a similar shift in W2-W3 at this redshift suggests this particularity of the W1-W2 colour must be related to a change in the SED of the QSOs. But why is that change happening at $z = 1.4$ is perplexing. This cannot be due to a spectral line entering or getting out of one of the WISE filter pass bands, since at this redshift the most intense emission line in the red in AGNs, H$\alpha$, is still in the blue, at 1.6 $\mu$m, while PAH emission pass bands \citep[e.g.][]{1996Cesarsky} fall way farther to the red than W1 and W2; besides, PAH emission is usually weak or absent due to the strong AGN continuum component in QSOs \citep{2007Draine}. 

Interestingly, \citet{2009Richards} observed the same trend for the colours  3.6-4.5 $\mu$m (consistent with W1-W2) for a sample 5,546 quasars observed with Spitzer (IRAC instrument). To explain this feature, the authors alluded to the inflection at 1 $\mu$m, separating the blue and red part of the SED, as was detected previously by \citet{1994Elvis} using a sample of 47 quasars at low redshifts. However, the covering range in redshifts does not match. This is better seen in the study made by \citet[][see their Figure~3]{2008Labita}: the inflection at low redshift is at a frequency of $10^{14.5}$ Hz, and the blue bump at $10^{15}$ Hz, while at $z = 1.4$, the part of the SED affecting W1 is only at $10^{14.3}$ Hz, the 1 $\mu$m only reaching the W1 bandpass at $z = 2.5$, far from the inflection of W1-W2 colours. The 1 $\mu$m inflection consequently cannot explain the W1-W2 change in colours towards the blue. What is needed is a component of the SED that increases the flux in W1 relative to W2, making this part of the SED flatter. Since the UV bump is still far to the blue at $z = 1.4$, this might suggest an extra component, like a star burst in the circumnuclear region \citep[e.g.,][]{2021Xie}. This once again legitimizes using high quality fluxes to distinguish what component of the SED can have this property. 

Researching ADS for the literature and using NED and CDS VizieR tool to compare our data with available catalogues, we found no evidence of detection in radio or X-ray for any of the HQWISE QSO, which makes this sample a purely UV-Opt/MIR selected sample. For the present study, our data are thus restricted to the five SDSS fluxes in u, g, r, i and z filters and the four high quality fluxes in WISE. In SDSS the magnitudes are expressed as inverse hyperbolic sine function (or asinh), as described by \citep{1999Lupton}.\footnote{https://www.sdss.org/dr12/algorithms/magnitudes/} The transformation from linear flux measurements to asinh magnitudes was designed to be identical to the standard astronomical magnitude at high signal-to-noise ratio (S/N) and to behave reasonably at low S/N. The magnitudes are characterized by a softening parameter $b$, which is the typical 1-sigma noise of the sky in a PSF aperture for a 1~arcsecond seeing. The detected flux, $f$, is then obtained by inverting the following relation for the asinh magnitude: 

\begin{equation}
m = \frac{-2.5}{\ln 10} \cdot [{\rm asinh}(\frac{f/f_0}{2b})+\ln b]
\end{equation}

\noindent Where $f_0$ is the flux of an object with conventional magnitude of zero. The quantities $b$ as measured relative to $f_0$ are given for each filter in Table~21 of \citet{1999Lupton}, along with the asinh magnitude associated with a zero flux calibrator. The table also lists the flux corresponding to $10 \times f_0$, above which the asinh magnitude and the traditional logarithmic magnitude differ by less than 1\% in flux. 

The instruction how to transform the WISE magnitudes in fluxes can be found in \citet{2010Wright}.\footnote{https://wise2.ipac.caltech.edu/docs/release/allsky/expsup/sec4{\_}4h.html} According to the colour distributions shown in Figure~\ref{fig1}, we opted to use the standard calibration for the power law $F_\nu \propto \nu^{-2}$.

\subsection{Properties of HQWISE QSOs.}
\label{QSOprop} 

\begin{table}
\begin{tabular}{cccccc}
\hline
\hline
(1) & (2) & (3) & (4) & (5) & (6)  \\
\#  & z bin  & N    &  Log M$_{BH}$     &   Log L$_{AGN}$     & $nEdd$  \\
    &        &      & (M$_{\odot}$) & (erg s$^{-1}$)  &           \\
\hline\noalign{\smallskip}
0  & $(0.00 - 0.25]$ & 0/103   &  -- 	  & --      & -- \\
1  & $(0.25 - 0.50]$ & 103/188 & 7.53 & 44.12 & -1.17 \\
2  & $(0.50 - 0.75]$ & 120/135 & 8.36 & 45.02 & -0.75 \\
3  & $(0.75 - 1.00]$ & 211/226 & 8.97 & 45.71 & -0.70 \\
4  & $(1.00 - 1.25]$ & 171/177 & 9.25 & 45.98 & -0.72 \\
5  & $(1.25 - 1.50]$ & 121/124 & 9.39 & 46.22 & -0.58 \\
6  & $(1.50 - 1.75]$ & 109/112 & 9.48 & 46.38 & -0.52 \\
7  & $(1.75 - 2.00]$ & 113/118 & 9.60 & 46.50 & -0.58  \\
8  & $(2.00 - 2.25]$ &  69/70  & 9.58 & 46.59 & -0.44  \\
9  & $(2.25 - 2.50]$ &  21/31  & 9.43 & 46.66 & -0.56 \\
10 & $(2.50 - 2.75]$ &  19/23  & 9.89 & 47.00 & -0.47 \\
11 & $(2.75 - 3.00]$ &  21/27  & 9.51 & 46.83 & -0.53 \\
12 & $(3.00 - 3.25]$ &  11/13  & 9.06 & 46.94 & -0.61  \\
13 & $(3.25 - 4.00]$ &   9/12  & 9.90 & 47.07 & -0.67  \\
\hline
\end{tabular}
\caption{\small Median properties of HQWISE QSOs in 14 redshift bins. The numbers in col.~3 give the number of QSOs with BH masses measured using the \ion{Mg}{II} or \ion{C}{IV} emission lines, compared with the total number of HQWISE QSOs in each bin. Note that when both lines are present in a spectrum, we only used the mass determined using the \ion{Mg}{II} line.}
\label{tab:bins}
\end{table}

Information gleaned from two spectral analysis of SDSS spectra allows us to characterize these QSOs in terms of AGN luminosity, BH mass and Eddington ratio. The first analysis is from \citet{2017Kozlowski}. Using QSOs from the \citet{2017Paris}'s sample, within redshift range $0.1 \leq z \leq 5.5$, Koz{\l}owski measured in the extinction-corrected SDSS spectra the fluxes and FWHM of the broad \ion{Mg}{II} and \ion{C}{IV} emission lines, and their adjacent fluxes in the continuum, respectively at $\lambda = 3000$ and 1350 \AA, using these measurements to estimate the bolometric luminosities, L$_{bol}$, AGN monochromatic luminosities, L$_{AGN} =\lambda {\rm L}_{\lambda}$, virial masses, M$_{BH}$ and Eddington ratios, $nEdd = {\rm Log}({\rm L}_{bol}/{\rm L}_{Edd})$ \citep[see details about measurements and calculations in][]{2017Kozlowski}. From his compilation, we retrieved the characteristics of the BHs for 1,098 of the HQWISE QSOs (only those with \ion{Mg}{II} and \ion{C}{IV} emission lines; Koz{\l}owski only used these two lines). For our analysis, we separated the HQWISE QSOs in 14 redshift bins with width 0.25, except for the last bin which contains only 12 QSOs within the range $3.25 < z \leq 4$. We report the medians of the parameters characterizing the QSOs in each bin in Table~\ref{tab:bins}. Note that there is no data for the QSOs in bin \#0 because at this redshift the \ion{Mg}{ii} line does not appear in the SDSS spectra. The number of QSOs with measurements in each bin is indicated in col.~3, compared to the number of HQWISE QSOs in the bin considered. For the BH mass and luminosity we utilized the values based on the \ion{Mg}{ii} line, only using the \ion{C}{iv} line at high redshifts where the \ion{Mg}{ii} line is not visible anymore. 

\begin{figure}
	\includegraphics[width=\columnwidth]{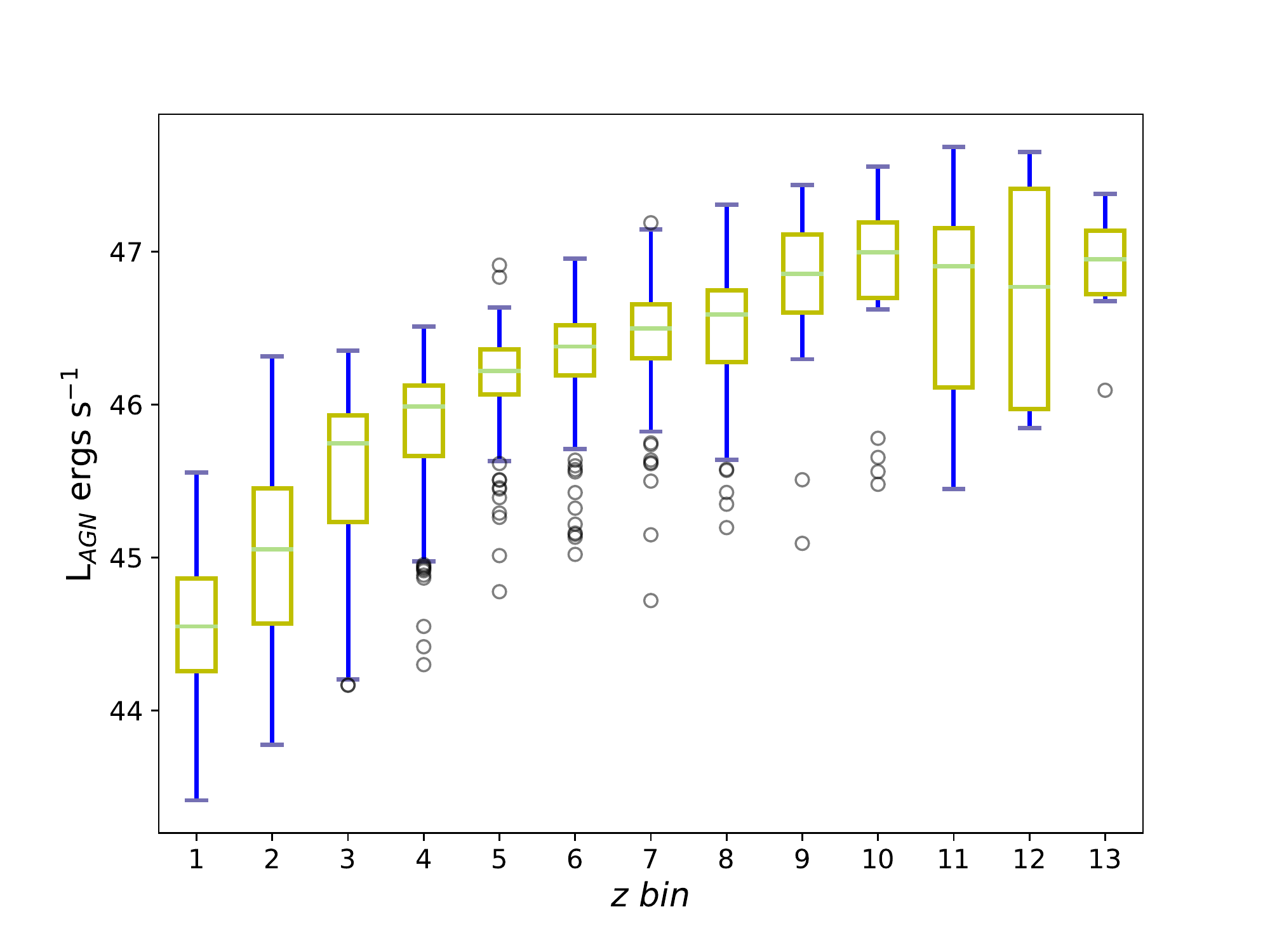}
	\includegraphics[width=\columnwidth]{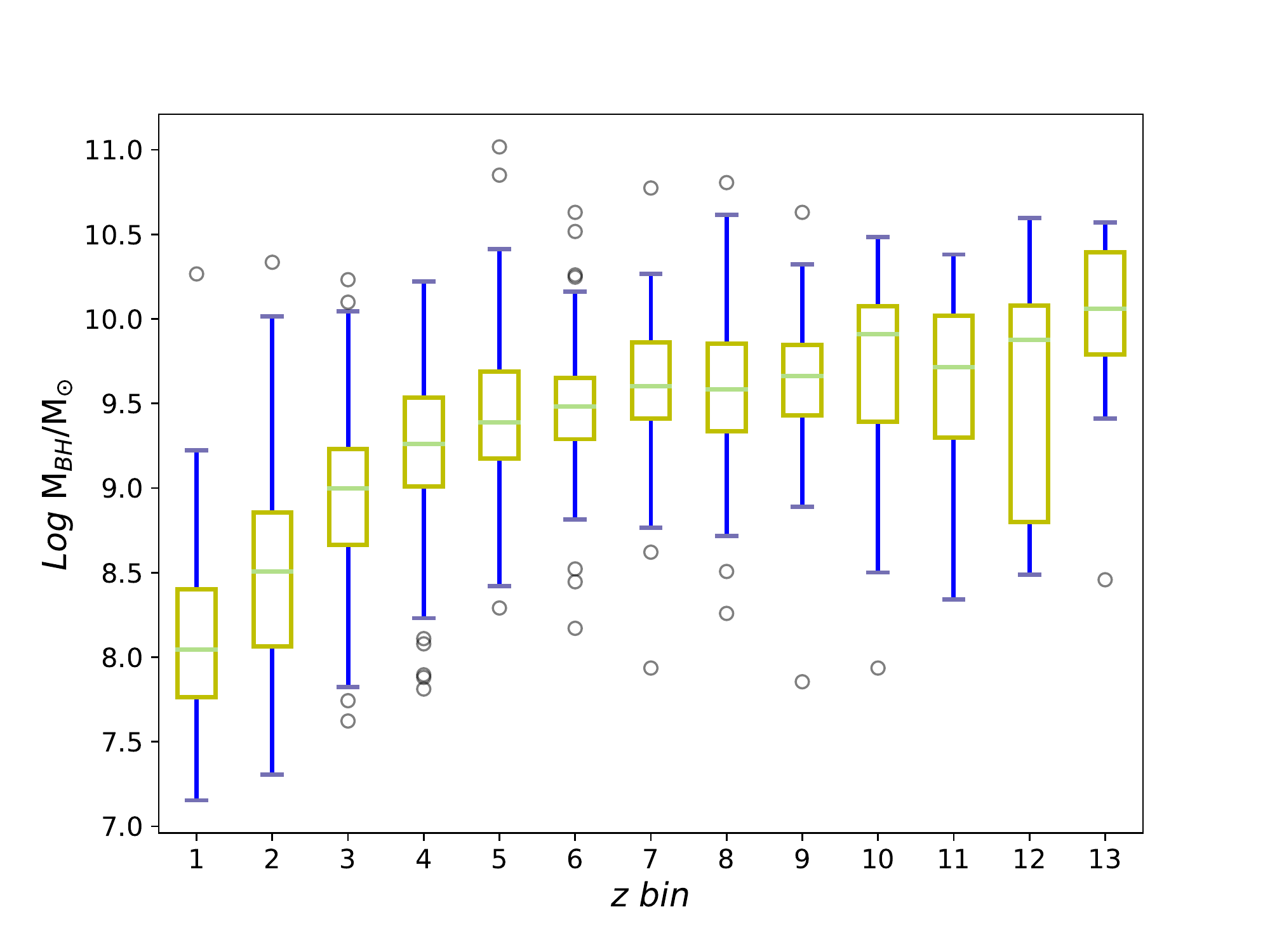}
	\includegraphics[width=\columnwidth]{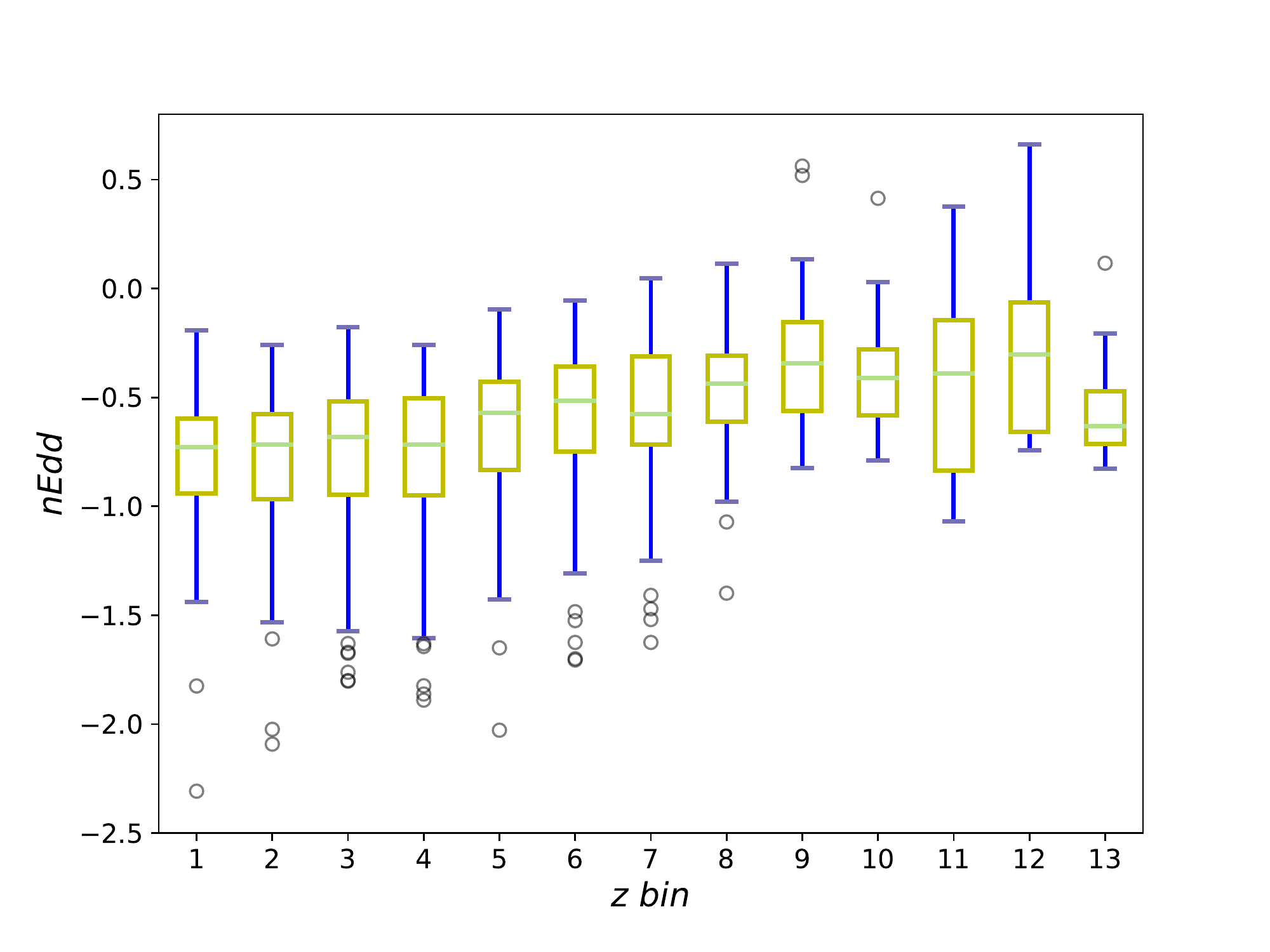}
    \caption{Box-whisker plots showing the QSOs characteristics in each redshift bin. A box is formed by the medians ($Q2$; green line), the lower ($Q1$) and upper ($Q3$) quartiles, yielding for the whiskers, upper bound $= Q3 + 1.5 \times$ IRQ and lower bound $= Q1 - 1.5 \times$ IQR, where IQR $= Q3-Q1$ is the interquartile; the open circle are data considered outsiders following Tuckey's definition.}
    \label{fig2}
\end{figure}

In figure~\ref{fig2}, we present the box-whisker plots for L$_{AGN}$, M$_{BH}$ and $nEdd$ for our sample of HQWISE QSOs. As expected, L$_{AGN}$ (upper panel) continuously rises with the redshift, by almost three orders, increasing rapidly up to $z \sim 1.62$ (bin \#6) and more slowly after that, until $z \sim 2.38$ (bin \#9), where the statistics becomes less secured as the number of QSOs in the bins decrease. Because the BH mass estimation depends on both the luminosity and FWHM \citep[cf. equation~1 in][]{2017Kozlowski}, its uptrend with the redshift (middle panel) naturally follows the increase in luminosity. However, this is only until bin \#6 where M$_{BH}$ is almost constant. Due to the lower number of QSOs above $z = 2.38$ in our sample, we cannot be sure the trend towards higher BH masses is physical or due to a bias in luminosity. Note that the redshift range $1.63 \leq z \leq 2.12$ (bin \#6 to bin \#8), which is well covered in our sample, includes the epoch in the evolution of the universe when the SFR in galaxies peaks and start to decrease at low redshifts \citep[cf.][]{2014Madau}. The trend for the BH mass in QSOs to peak at almost the same epoch is usually interpreted as evidence for a connection between the formation of SMBH and their host galaxies \citep{1998Boyle,2014Madau}. 

\begin{figure}
	\includegraphics[width=\columnwidth]{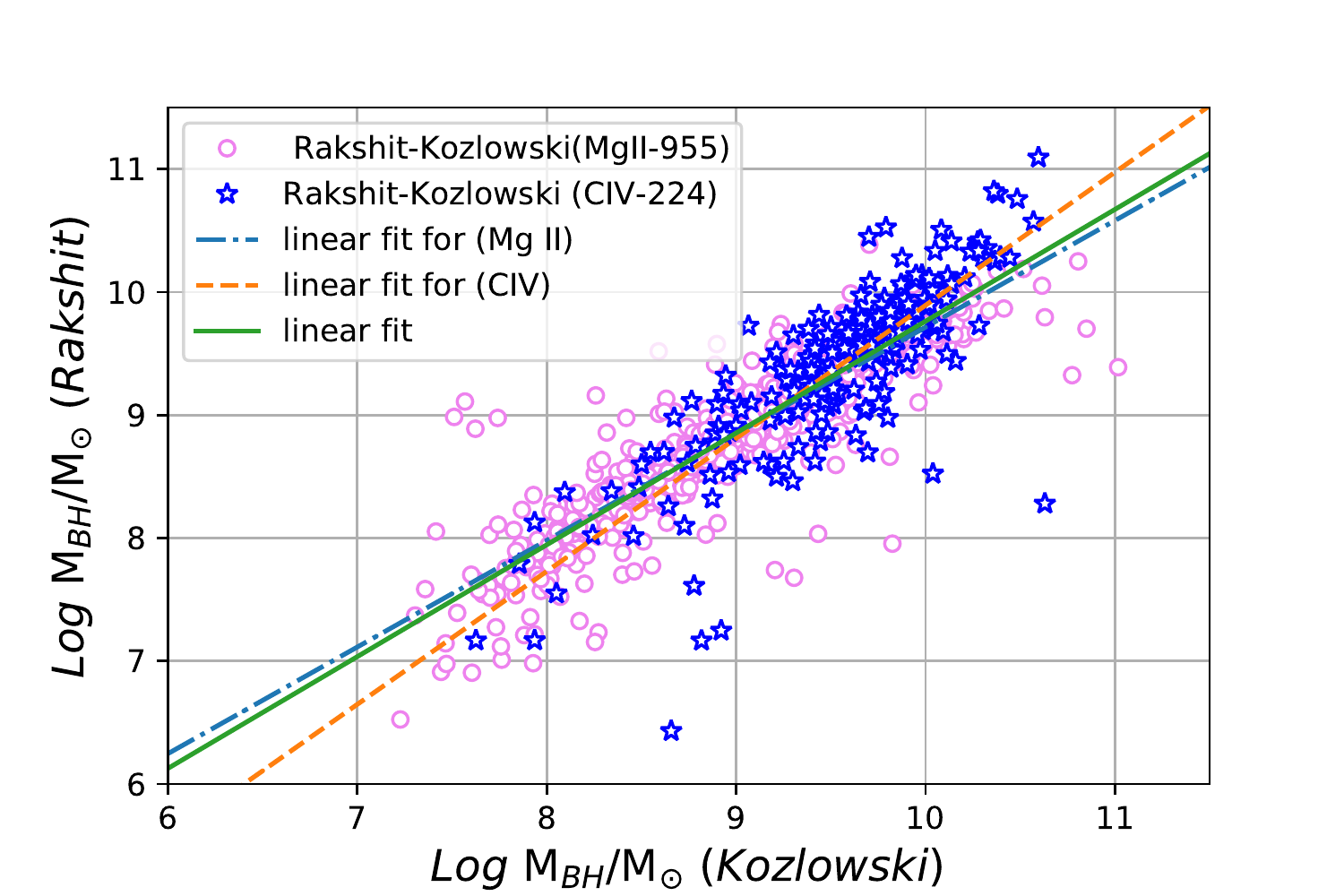}
    \caption{Comparing in $X$ the BH masses in \citet{2017Kozlowski} with in $Y$ the BH masses in \citet{2020Rakshit}, the three linear fits are: using only \ion{Mg}{II}, $Y = 0.86 X + 1.04$, with Pearson's coefficient $r=0.91$, or only \ion{C}{IV}, $Y = 1.08 X - 0.92$, $r=0.82$ and fitting both lines (green line), $Y = 0.91 X + 0.67$, $r=0.89$.} 
    \label{fig3}
\end{figure}

\begin{figure}
	\includegraphics[width=\columnwidth]{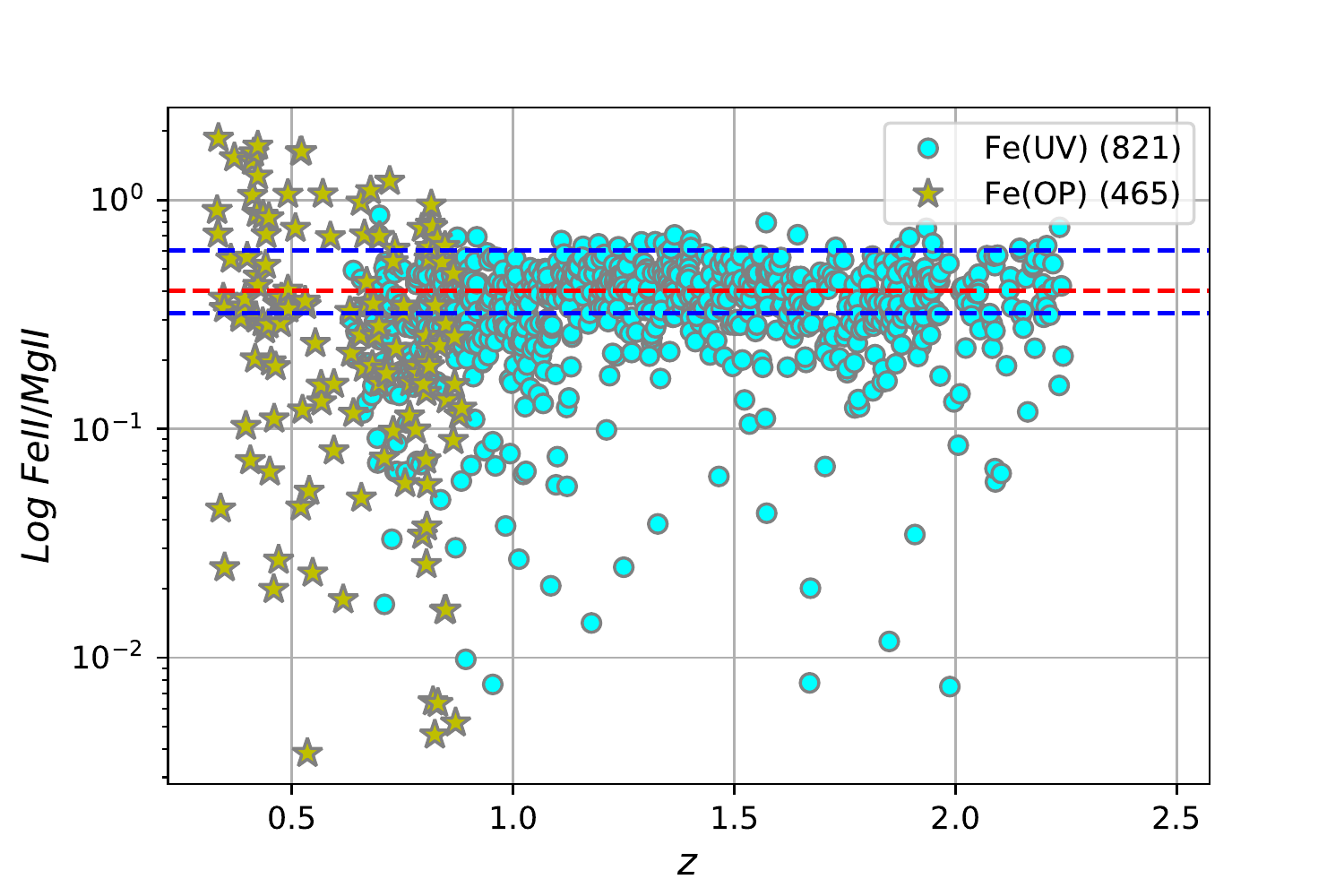}
    \caption{Ratio of abundances \ion{Fe}{ii}/\ion{Mg}{ii} as calculated by \citet{2020Rakshit} for the HQWISE QSOs sample. The identification Fe(UV) and Fe(OP) refer to different templates used by the authors, one in the UV and other in the optical. The median (red line) with bootstrap confidence interval (blue lines) estimated by \citet[][]{2022ChuWang} is included: \ion{Fe}{ii}/\ion{Mg}{ii} $ = 2.54_{-0.43}^{+1.12}$.}
    \label{fig4}
\end{figure}

Because the Eddington ratio (lower panel) corresponds to the ratio of accretion rate (BHAR) to BH mass ($nEdd \propto {\rm BHAR}/{\rm M}_{BH}$) and the BH mass grows with the luminosity, the fact that we observe a lower amplitude variation of this parameter with the redshift is not unexpected. However, there seems to be a weak trend for $nEdd$ to increase above $z = 1.37$ (bin \#5), suggesting higher BHAR at high redshifts than at low redshifts in more massive BH and luminous QSOs. 

\begin {table*}
\caption{Summary of inputs used in \textsl{X-Cigale} for ultimate run with polar emission.} 
\label{MtA}
\begin{tabular}{p{0.15\textwidth}p{0.25\textwidth}p{0.45\textwidth}}
\hline
\hline
Parameters & Values & Descriptions \\
\hline
\multicolumn{3}{c}{Star formation history (SFH) functions: \texttt{sfhdelayed} and \texttt{sfh2exp}}\\
$\tau_{\textnormal{main}}$ & 50, 100, 500, 1000, 2000, 4000, 7000 & e-folding time in Myr.\\
Age & CT & Age of the oldest stars (Myr); we used the cosmic time (CT) at the redshift of the QSOs\\
\hline
\multicolumn{3}{c}{fritz06}\\
R$_{\textnormal{max}}$/R$_{\textnormal{min}}$ & 10.0, 30.0, 60.0, 100.0, 150.0 & Ratio of the maximum to minimum radii of the dust torus. \\
$\tau$ & 0.1, 0.6, 1.0, 6.0, 10.0 & Optical depth at 9.7 $\micron$. \\
$\beta$ & $-$1.00, $-$0.75, $-$0.50, $-$0.25, 0.00 & Beta describing the dust density distribution with r the radius.\\
$\gamma$ & 0.0, 2.0 & Gamma from the power-law density distribution for the polar component of the dust torus.\\
Opening Angle ($\theta$) & 60.0, 100.0, 140.0 & Full opening angle of the dust torus. \\
$\psi$ & (20.000, 25.000, 30.000), 89.990 & Angle between equatorial axis and line of sight; values within parenthesis were tried in different run omitting value out of parenthesis. \\
$\delta$ & -1, -0.5, -0.7, 0 & Power-law index $\delta$ modifying optical slope of AGN continuum. Negative values make slope steeper/positive values make slope shallower.\\
$f_{AGN}$ & 0.1, 0.5, 0.9 & Fraction of AGN contribution to the IR luminosity.\\
Extinction Law & 1 & Extinction law of the polar dust: 0 (SMC), 1 (Calzetti 2000), or 2 (Gaskell et al. 2004).\\
$E(B-V)$ & 0, 0.05, 0.1, 0.15, 0.2, 0.3, 0.4 & E(B-V) for the extinction in the polar direction in magnitudes.\\
Temperature & 100.0, 500.0, 1000.0 & Temperature of the polar dust in K.\\
\hline
\multicolumn{3}{c}{Dust emission: dl2014}\\
$q_{\rm PHA}$ & 0.45 (0.47, 2.50, 4.58, 5.95, 7.32) & Mass fraction of PAH (in percentage); values within parenthesis were tried in different run omitting value out of parenthesis. \\
umin & 50 &  Minimum radiation field. \\
$\alpha$ & 1.0 &  Powerlaw slope  $dU/dM \propto U^{\alpha}$\\
$\gamma$ & 0.1 & Fraction illuminated from Umin to Umax. \\
\hline
\multicolumn{3}{c}{Single-age stellar population (SSP): bc03}\\
$Z$ & (0.0004, 0.004, 0.008), 0.02, 0.05 &  Metallicity; values within parenthesis were tried in different runs omitting values out of parenthesis.\\
Separation Age & 10 & Age difference between the youngest and the oldest stellar populations (Myr).\\
\hline
\multicolumn{3}{c}{Dust attenuation: dustatt\_modified\_starburst }\\
E$(B-V)_{\rm line}$ & 0.005, 0.05, 0.5, 0.10, 0.70 & The colour excess of the nebular lines light for both the young and old population.\\
E$(B-V)_{\rm factor}$ & 0.25, 0.5, 0.75 & Reduction factor E$(B-V)_{line}$ of  stellar continuum attenuation reduction.\\
Power-law slope & 0.0, 0.25, 0.5& Slope of the power law modifying the attenuation curve.\\
\hline
\end{tabular}
\end{table*}

The second spectral analysis useful for our study was done by \citet{2020Rakshit}, who contrary to \citet{2017Kozlowski} used spectra in SDSS DR14 instead of DR12. 
Comparing M$_{BH}$ as calculated in these two different sources in Figure~\ref{fig3}, we found the results to be almost the same. The linear fit shows that in general the BH masses in \citet{2020Rakshit} tend to be slightly lower than in \citet{2017Kozlowski}, the difference most probably being due to the subtraction by the latter of \ion{Fe}{}templates. Indeed, to determine the contribution of \ion{Fe}{}in each spectrum, \citet{2020Rakshit} did a robust fitting analysis of different templates, which allowed them not only to correct the spectra but also estimate the \ion{Fe}{} and \ion{Mg}{ii} abundances. In Figure~\ref{fig4}, we show the ratios \ion{Fe}{}/\ion{Mg}{ii} for these two elements in our HQWISE QSO sample based on the abundances calculated by \citet{2020Rakshit}. This result shows the typical high ratios found in QSOs independently of the redshift \citep[compare with Figure 11 in][]{2014DeRosa}. Note that to preserve the homogeneity of our analysis, we will not use the results of \citet{2020Rakshit} for M$_{BH}$ based on measurements of the broad component of H$\beta$ at low redshifts.

\section{Method of analysis}
\label{Method}

Our analysis consists in reproducing the SEDs of the 1,359 HQWISE QSOs using \texttt{X-CIGALE} \citep[][]{2019Boquien,2022Yang}.\footnote{\texttt{X-CIGALE} manual can be found here: https://cigale.lam.fr/documentation/} The key parameter to estimate how fast the galaxy hosts of quasars formed the bulk of their stellar populations is the star formation history (SFH) function of the host galaxies. In our universe, galaxies form their stars following only two different paths, which in great part also explains their morphology \citep{1986Sandage,1991Charlot}: they either form most of their stars early on, like elliptical galaxies, or form their stars more gradually like in spiral galaxies, the time-scale difference being due to the high angular momentum of gas collapsing in spirals to form large-scale disks. In \texttt{X-CIGALE} the SFH functions that reproduce these two patterns are called \texttt{sfhdelayed}, for Elliptical-like or early-type, massive bugle galaxies, and  \texttt{sfh2exp}, an exponential SFH typical of smaller bulge, spiral galaxies \citep[see Fig.~1 in][]{2019Boquien}. Because we cannot assume a priori the morphology of the QSO hosts in our sample both SFH functions were tried using \texttt{X-CIGALE}. 

How fast galaxies with different SFH function form their stars is determined by the e-folding time, $\tau_{\rm main}$, which corresponds to the time it takes a galaxy to form 69\% of its stars. The smaller $\tau_{\rm main}$ and the faster the formation of the galaxy. Theoretically \citep{1986Sandage} elliptical galaxies have smaller e-folding time than spiral galaxies and very large e-folding times (a few Gyrs) are rare, since they imply SF in these galaxies have been delayed over a long period of time (the cases of small-mass irregular galaxies). However, assuming the host galaxies of QSOs are massive and the $M_{\rm BH}-\sigma$ relation is universal, star formation cannot be delayed over arbitrarily long periods of time compared to the formation of their SMBHs, and we would thus expect the $\tau_{\rm main}$ produced by \texttt{X-CIGALE} to be relatively small. How small can it be, this is the condition that was tested (as shown in Table~\ref{MtA}) by trying different values for $\tau_{\rm main}$, varying between 50 Myrs to 7000 Myrs, that is, all the values allowed in \texttt{X-CIGALE} (see Table~\ref{MtA}).  

One important module in \texttt{X-CIGALE} tackles the AGN contribution. Because QSOs are Type-1 AGNs, we expect the AGN continuum to dominate the SED in the UV-Opt and NIR (most specifically the W1 and W2 passbands). Having experimented with the two main options, \texttt{skirtor2016} \citep{2016Stalevski} and \texttt{fritz2006} \citep{2006Fritz}, and having obtained slightly better fits (lower $\chi^2$) with the latter, we decided to use only this module for our final testing. The main difference between the two models lies in the isotropy of re-emission by dust, the torus in \texttt{fritz2006} model being less clumpy than in the \texttt{fritz2006} model, producing more isotropic re-emission. Consistent with Type-1 AGN in the Fritz module, we fixed the line of sight to $89.99^\circ$ (implying we see the torus of dust face on) but left the program free to try different inputs: most specifically, varying the fraction of AGN, f$_{AGN}$, and opening angle, OA, of the dust torus \citep[see manual of \texttt{X-CIGALE} and original article by][for detailed descriptions]{2006Fritz}. 

One crucial improvement to \texttt{X-CIGALE} made in 2022 by \citet{2022Yang} is the addition of a $\delta$ parameter, which fixes the slope of the AGN continuum between $0.125\ \micron <\lambda \leq 10\ \micron$ ($\lambda {\rm L}_{\lambda} =\lambda^{-0.5+\delta}$). Without this parameter, the contribution of the AGN to the SED would be underestimated, increasing artificially the stellar contribution in the UV-Opt, which then translates into higher SFRs. Having experimented with the previous model of \texttt{X-CIGALE} without the slope correction, we did observed such SFR overestimation in the host galaxies of our QSOs, although not as high as we could have expected. In our modelling, different values of $\delta$ were tried, varying between -1 and 0 (see Table~\ref{MtA}).

Another module that is important for the MIR range (W3 and W4 in WISE) is the module related to re-emission in IR by dust of UV light due to SF. Experimenting with the different options offered by the program \citep[see Figure~5 in ][]{2019Boquien}, we have adopted the \texttt{dl2014} module \citep{2014Draine}. After experimenting with different values for the PAH mass fraction, we used in our final modelling the minimal value $q_{\rm PAH} \sim 0.45$\% \citep[][]{2007Draine}. Indeed, even with high SFR in the QSO hosts, the PAH features should not play a great role due to the dominant AGN component \citep[e.g.,][]{2018Marshall}; the fine dust producing PAH emission can also be destroyed by the intense AGN luminosity \citep{2023Almeida}. 

For the single stellar population (SSP), we used the module \texttt{bc03} based on \citet{2003Bruzual}, adopting a Salpeter's IMF. After experimenting with different values for the metallicity, from $Z = 0.004$ to $Z = 0.05$, we determined that our best fits were those where the metallicity was solar or higher at any redshift (consistent with observations; cf. Figure~\ref{fig4}). For the attenuation of UV-Opt light due to dust extinction, we used the module \texttt{dustatt\_modified\_starburst} \citep[based on][]{2000Calzetti,2002Leitherer}, which includes a bump at 2200~\AA\ with varying amplitudes. 


Finally, after experimenting with different SED models without polar emission, we realized that we can greatly improve the fits by including such component in all the HQWISE QSOs, at all redshift. This was done using the special module introduced in \texttt{X-CIGALE} by \citet{2020Yang}. In the literature, the existence of polar dust components \citep[excess of dust emission emitted in a direction perpendicular to the obscuring torus;][]{2014Tristram} was already noted and thoroughly studied by various authors in different AGNs \citep [e.g.,][]{2021Toba,2022Isbell,2022Lyu}, suggesting this characteristic is more common than previously thought. The inclusion of polar dust in our models systematically produced best fits with $\chi^2 <3$ for most of the QSOs in our sample (87\% using \texttt{sfhdelayed} and 83\% using \texttt{sfh2exp}). 

In Table~\ref{MtA}, we summarize the inputs used in \texttt{X-CIGALE} for our final solutions, indicating the range of values for the parameters that we allowed to varied to get the best SED models (i.e., the models that minimize the $\chi^2$). Note that for some parameters, we tried more values (indicated in parenthesis) in separated runs, to compare with our best models. The mock option was also used to verify the robustness of our final solutions \citep[see full decription in][]{2021Mountrichas}.  

\section{Results}
\label{Res}

\begin{figure*}
\includegraphics[width=0.92\columnwidth]{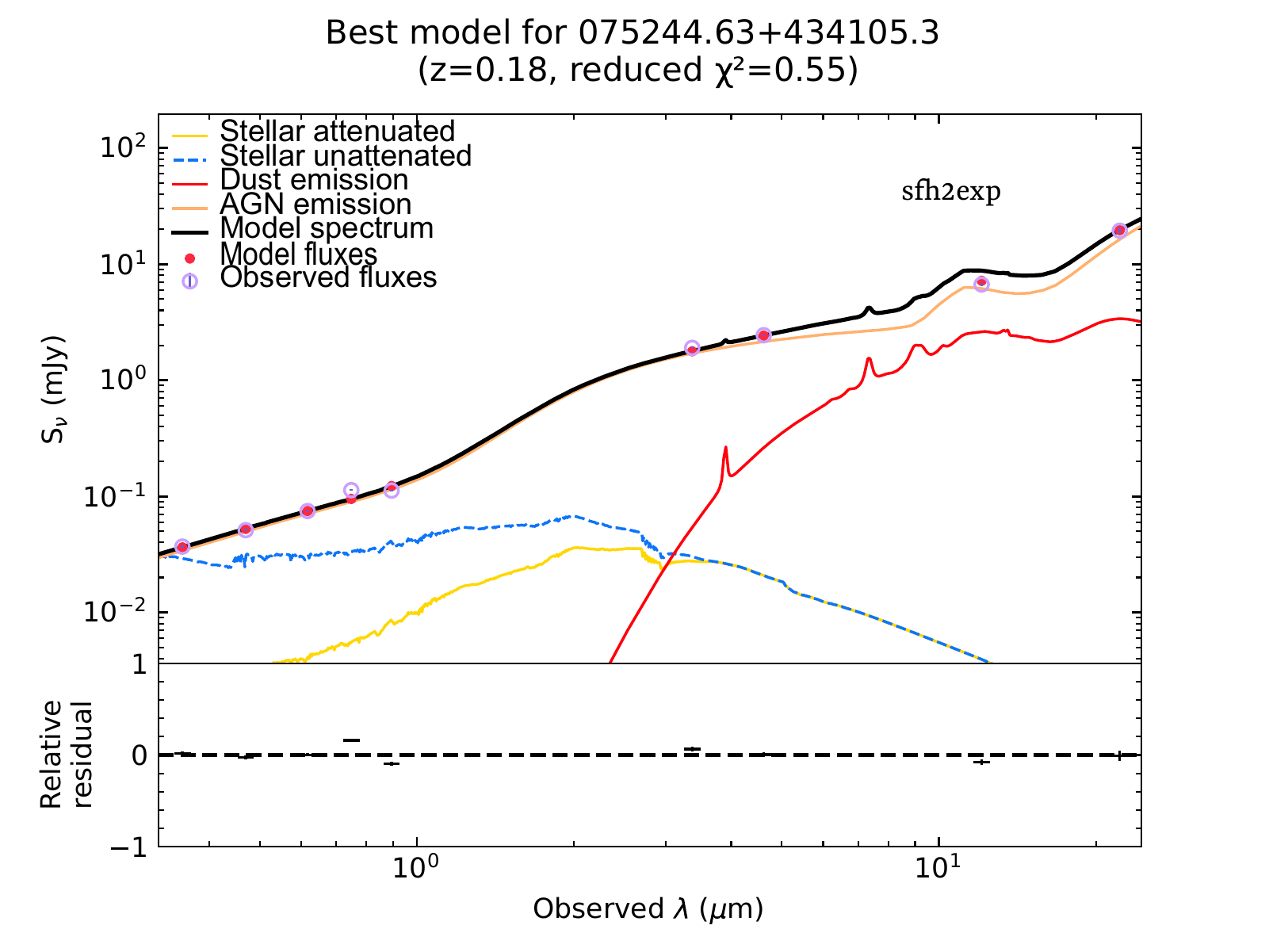}
\includegraphics[width=0.92\columnwidth]{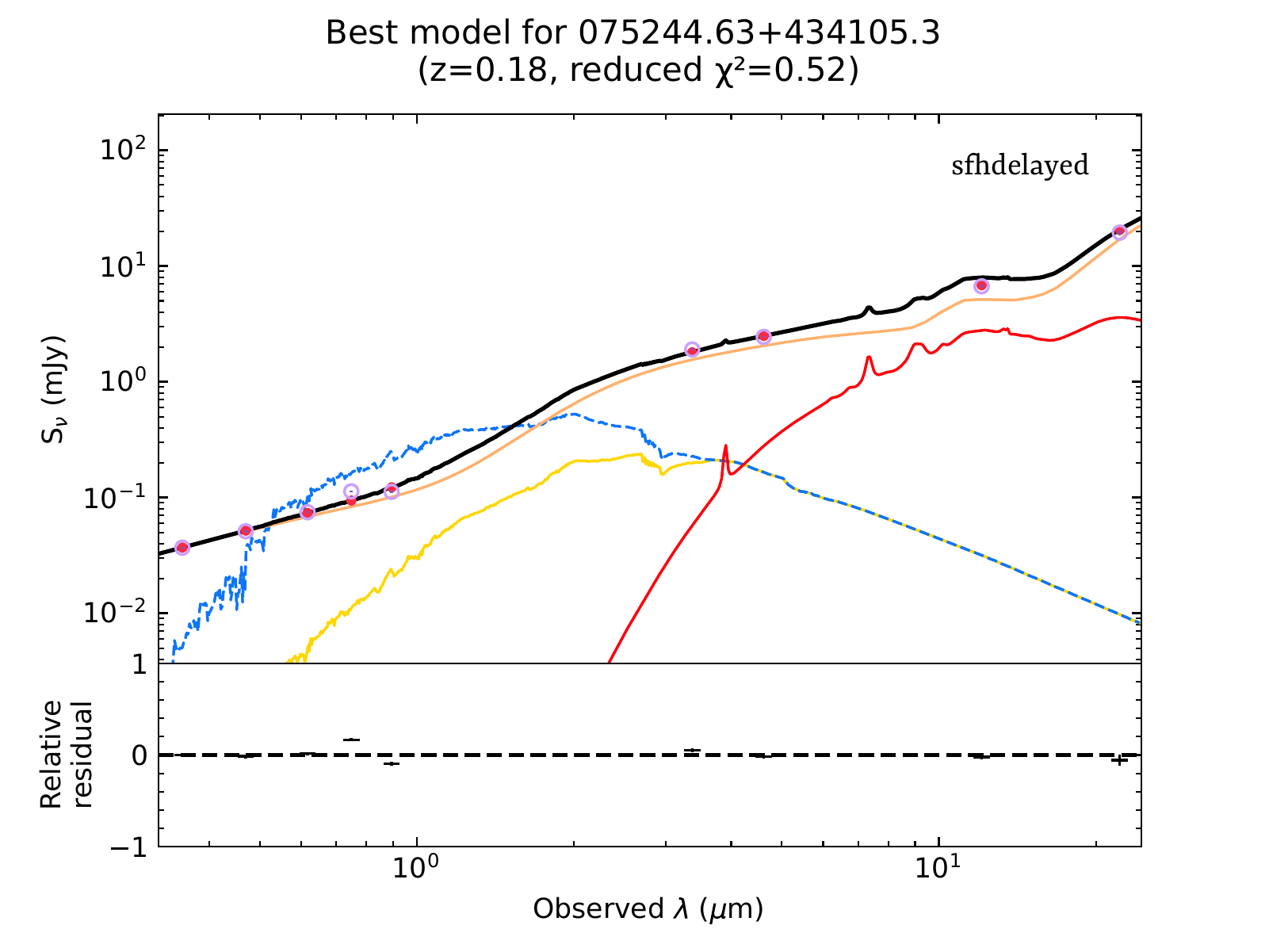}
\includegraphics[width=0.92\columnwidth]{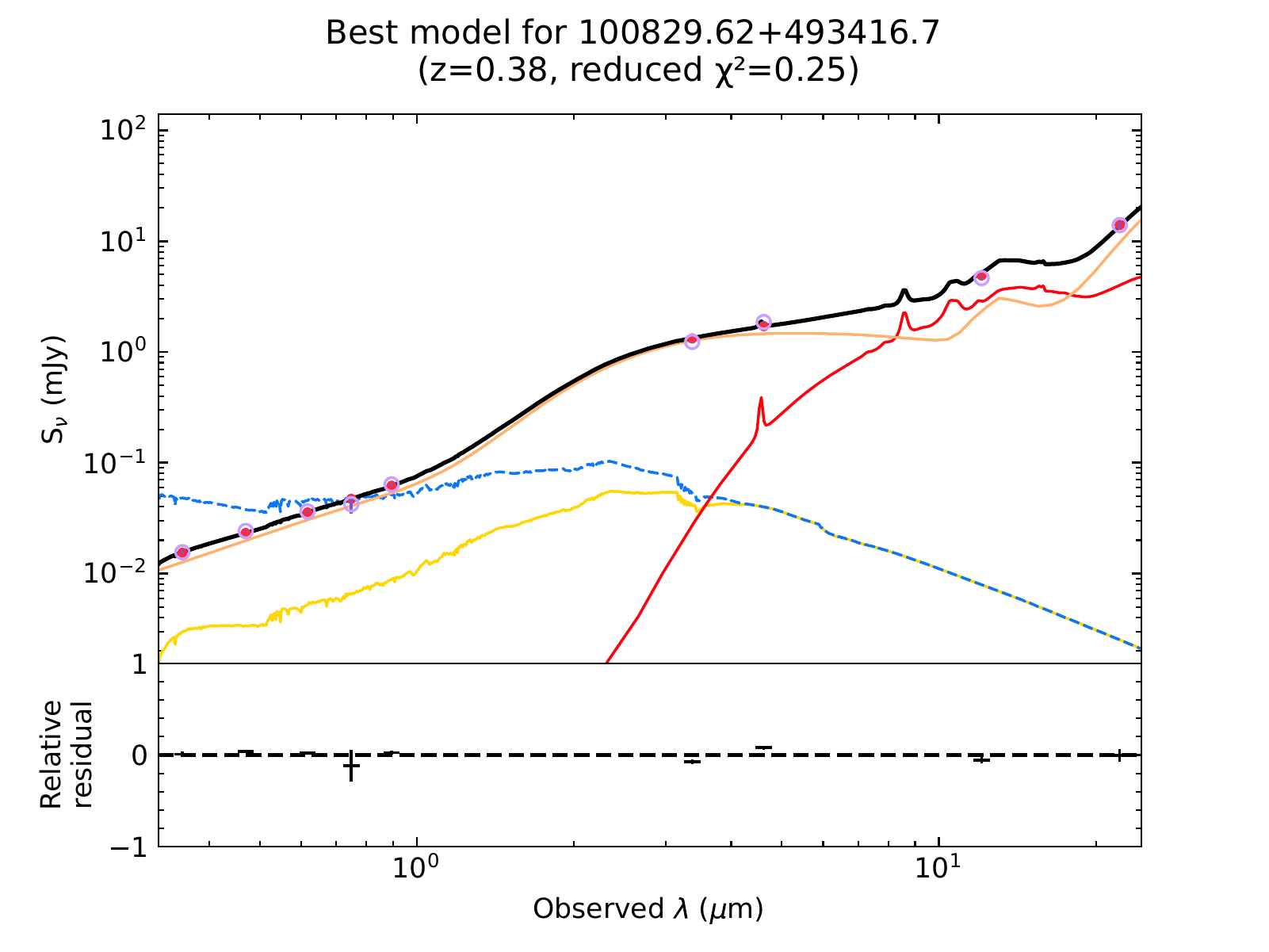}
\includegraphics[width=0.92\columnwidth]{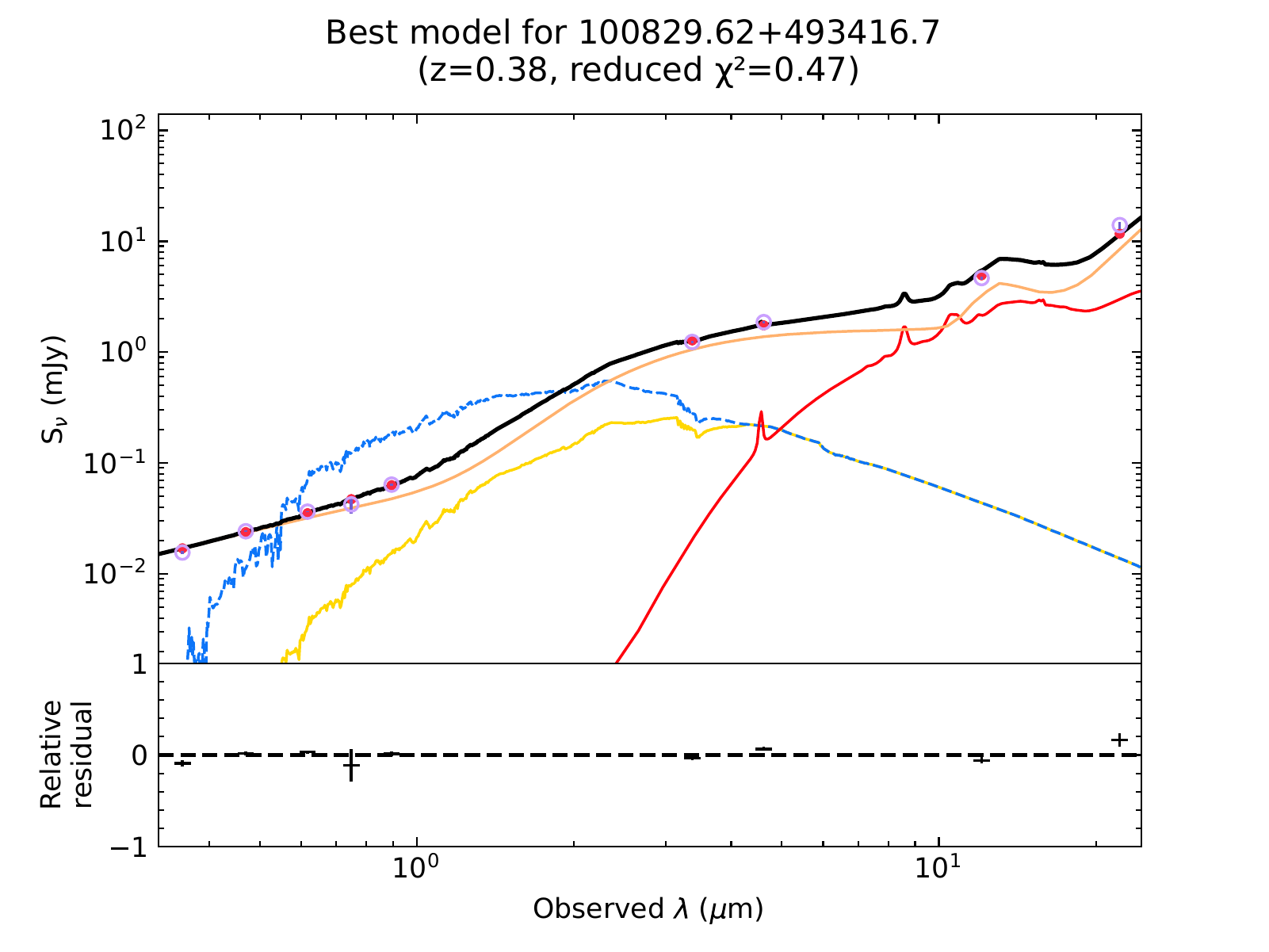}
\includegraphics[width=0.92\columnwidth]{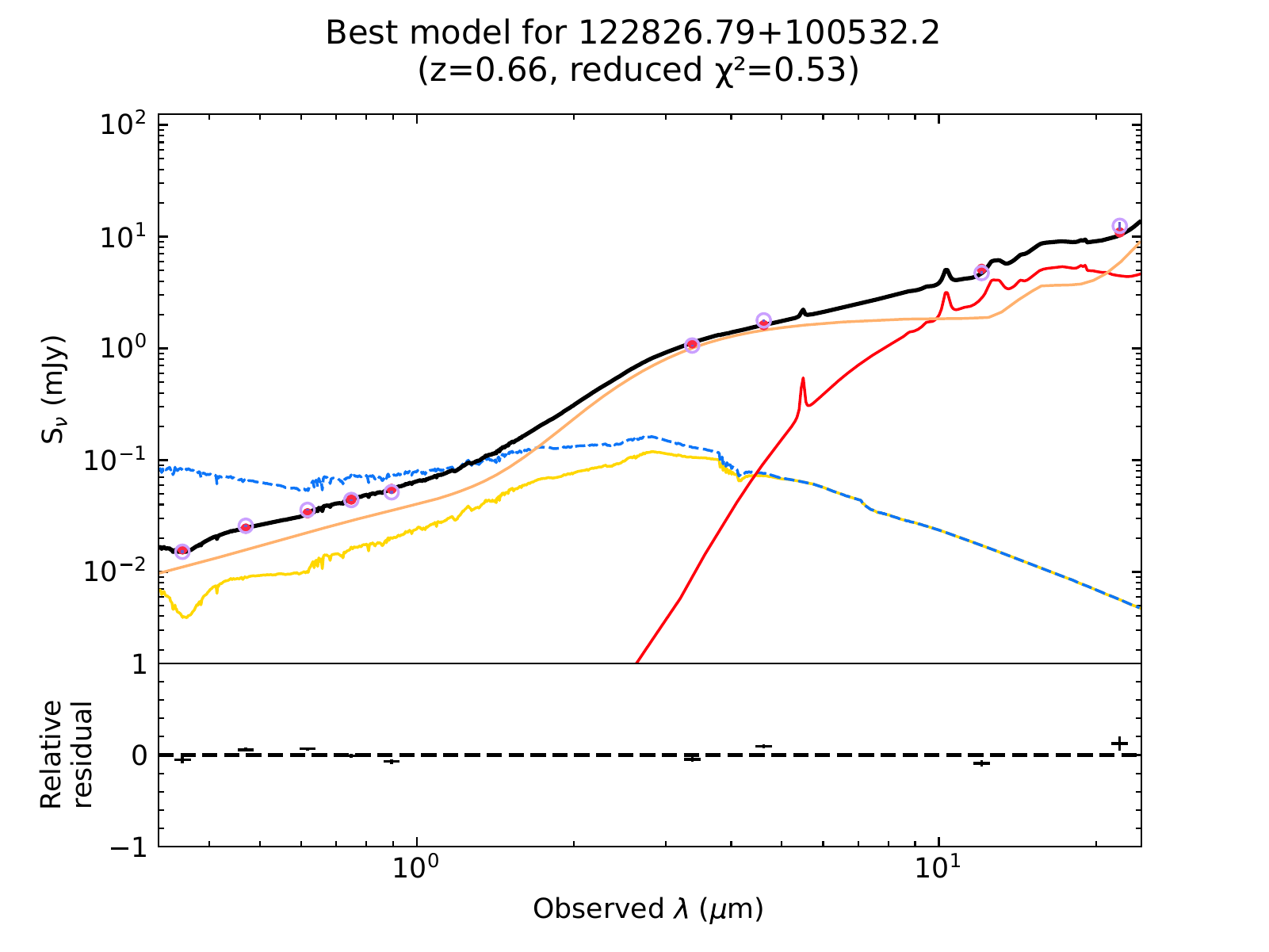}
\includegraphics[width=0.92\columnwidth]{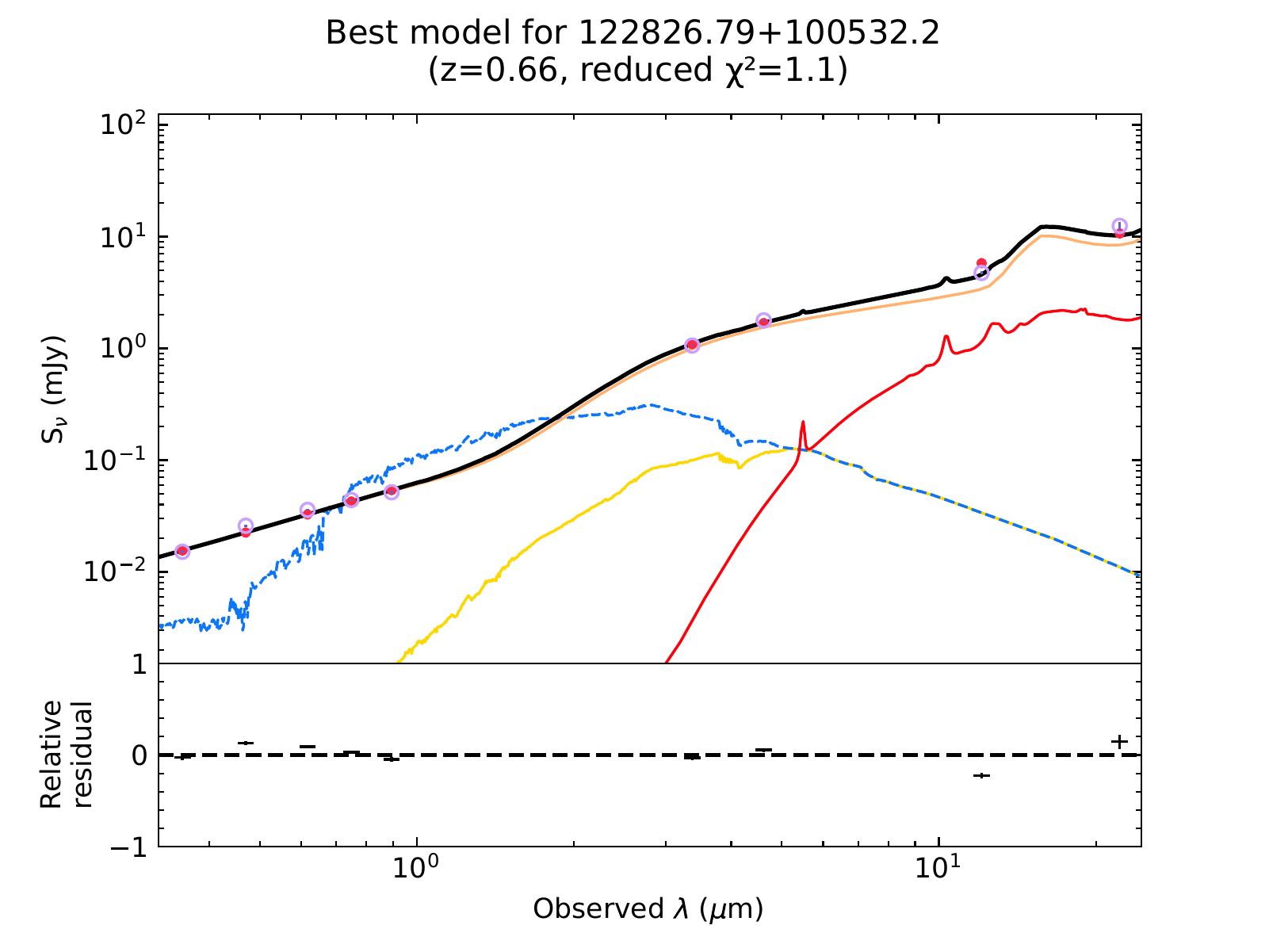}
\includegraphics[width=0.92\columnwidth]{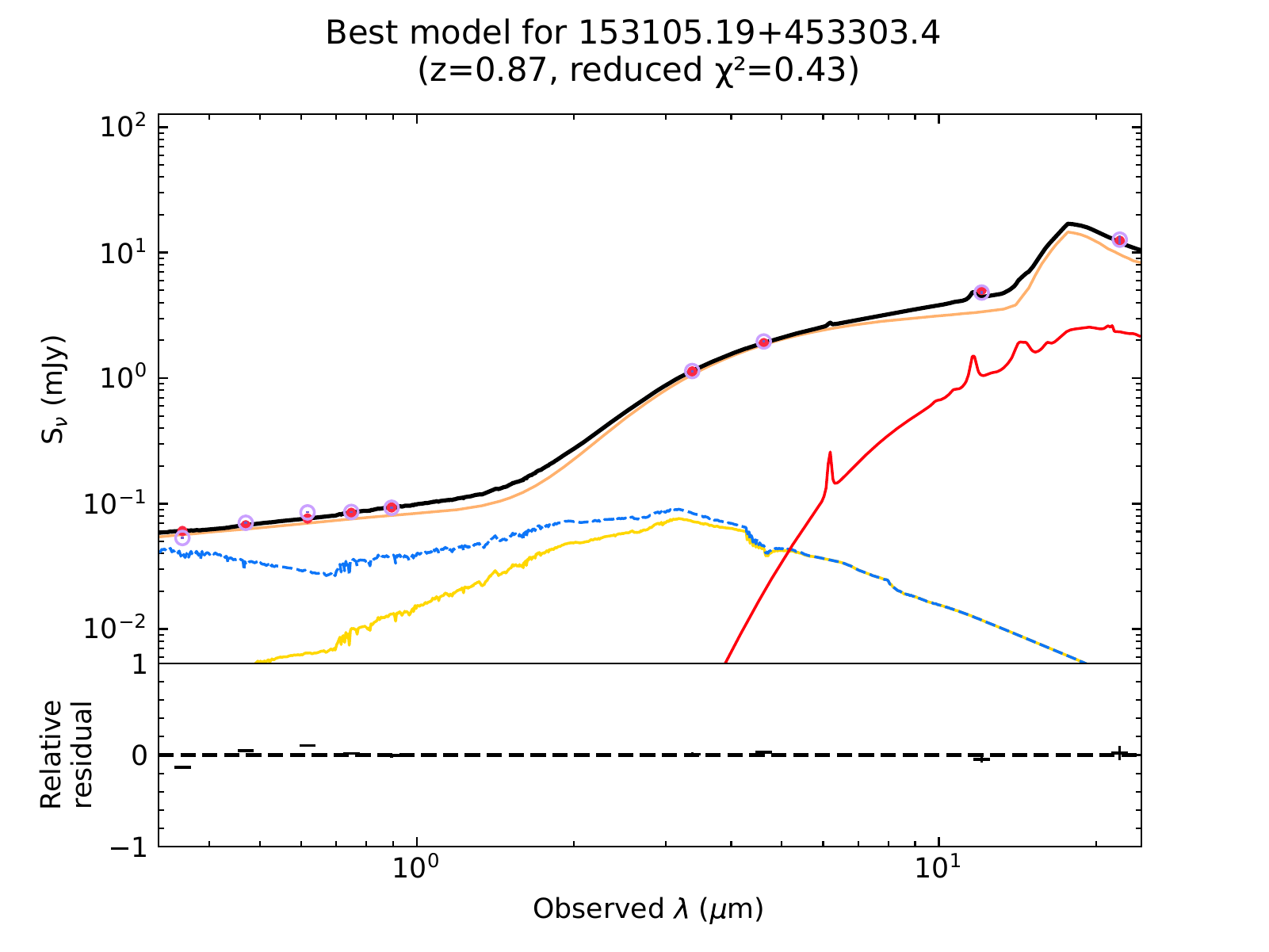}
\includegraphics[width=0.92\columnwidth]{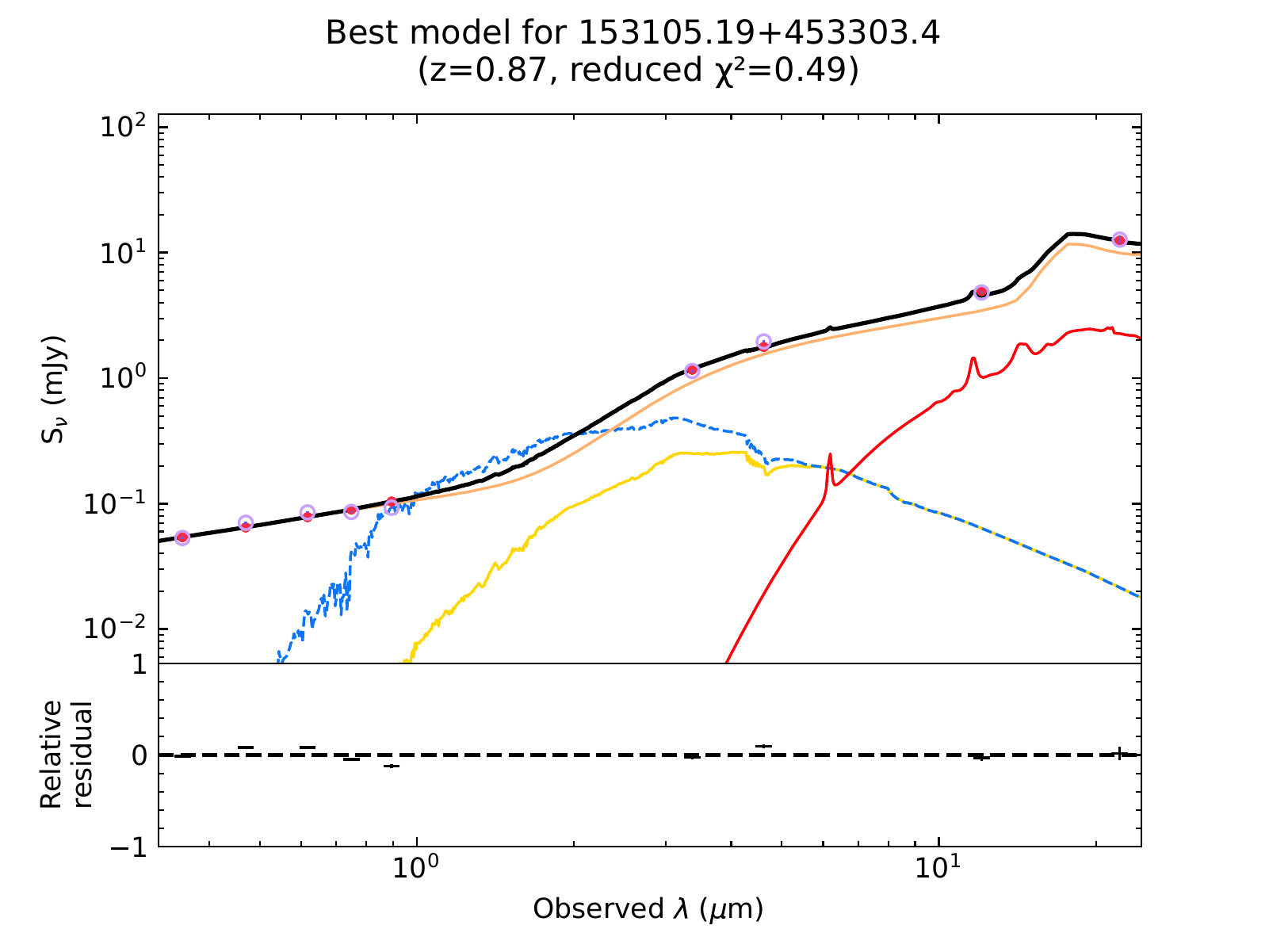}
\caption{Comparing results for SFH2exp (left panels) and SFHdelayed in bins 0 to 3.}
    \label{figA1}
\end{figure*}

\begin{figure*}
\ContinuedFloat
\includegraphics[width=0.92\columnwidth]{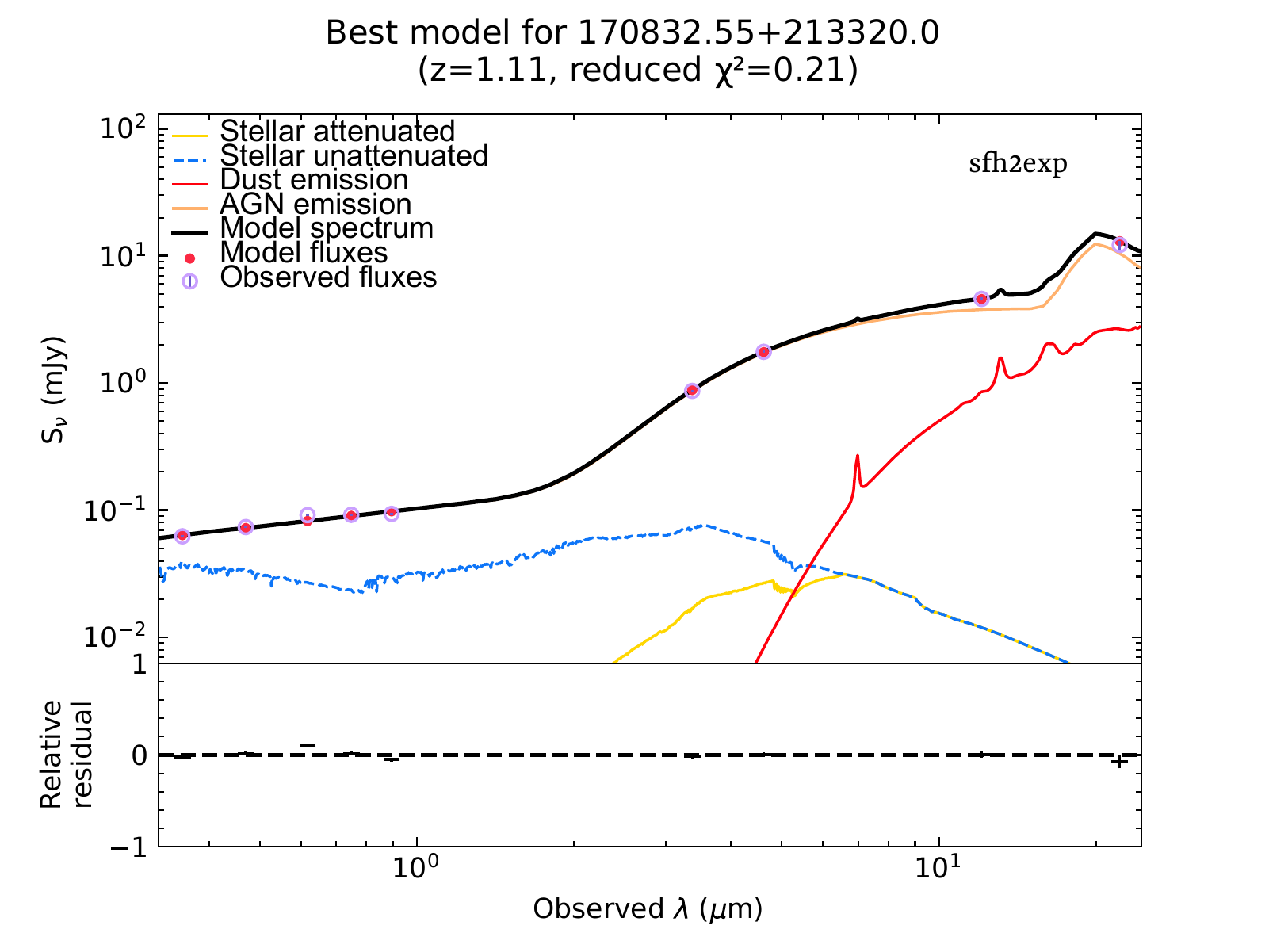}
\includegraphics[width=0.92\columnwidth]{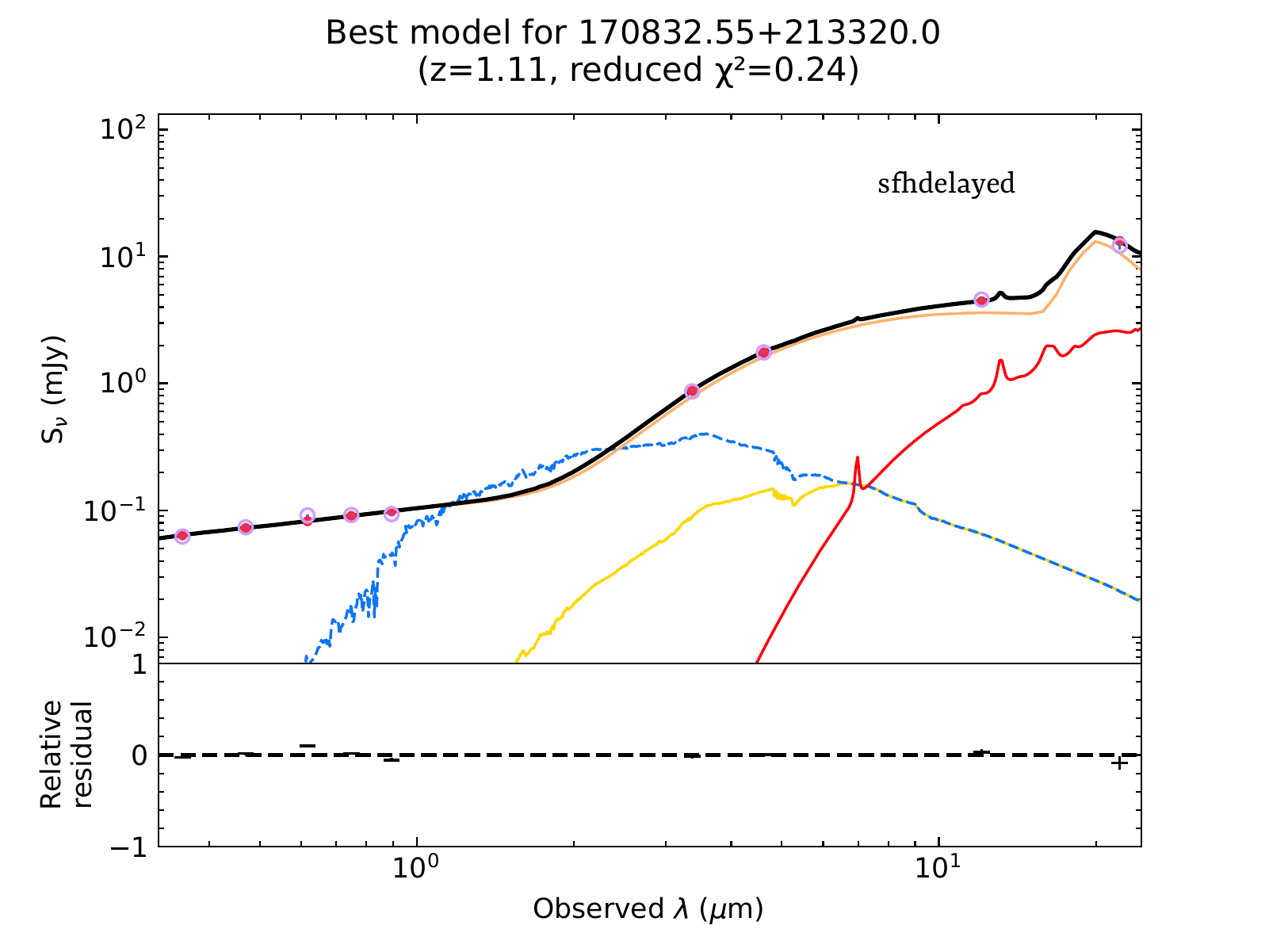}
\includegraphics[width=0.92\columnwidth]{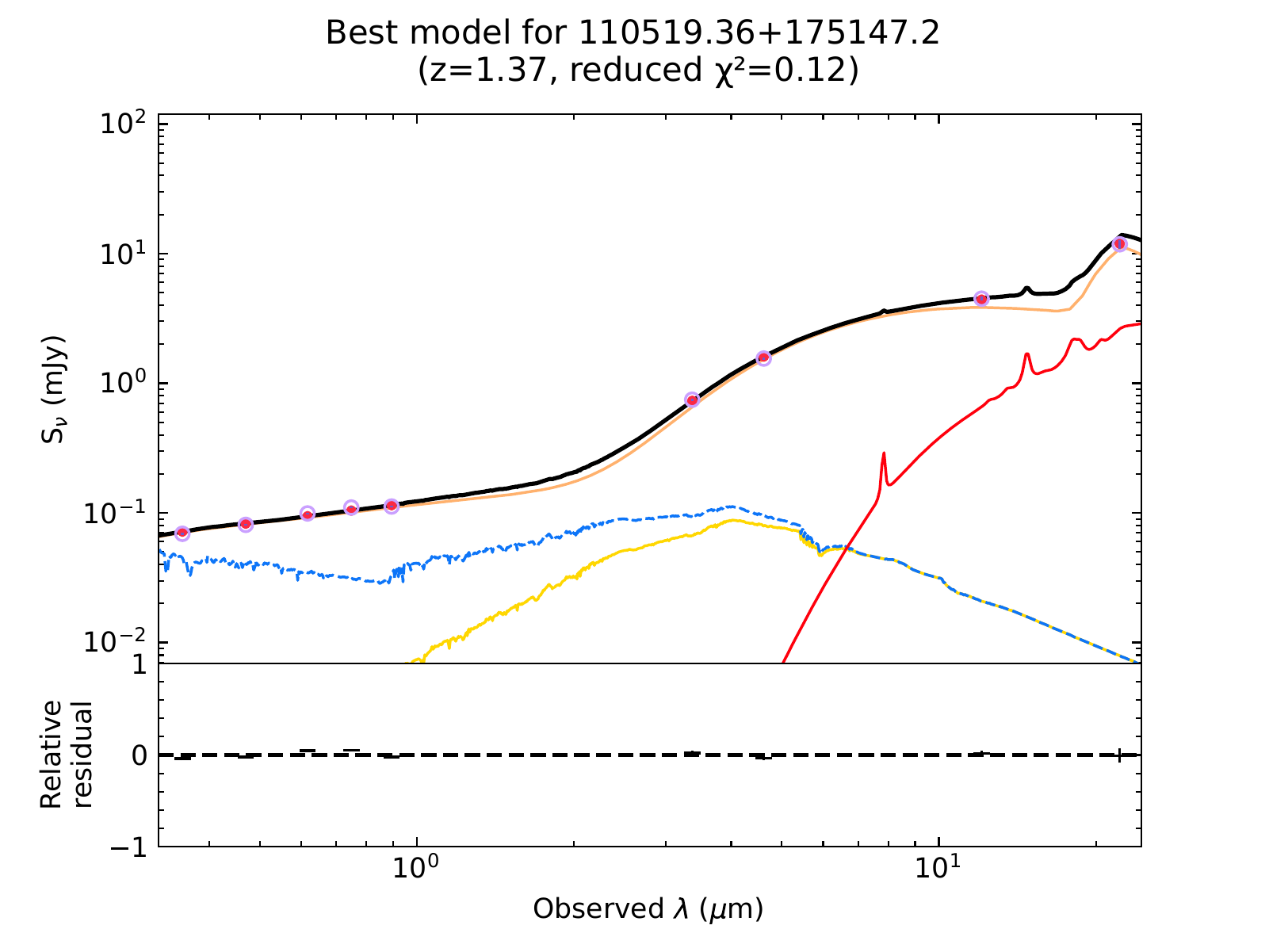}
\includegraphics[width=0.92\columnwidth]{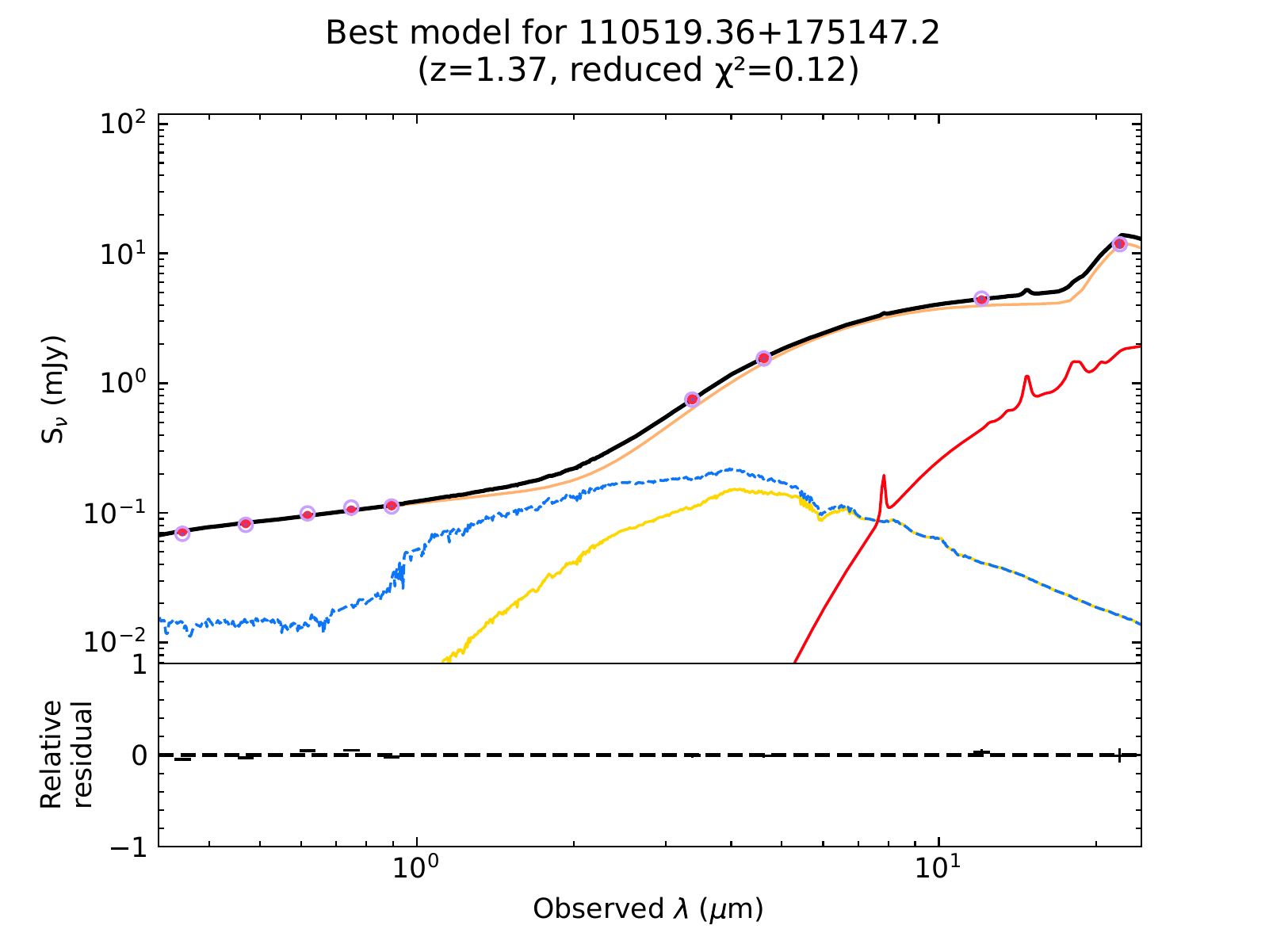}
\includegraphics[width=0.92\columnwidth]{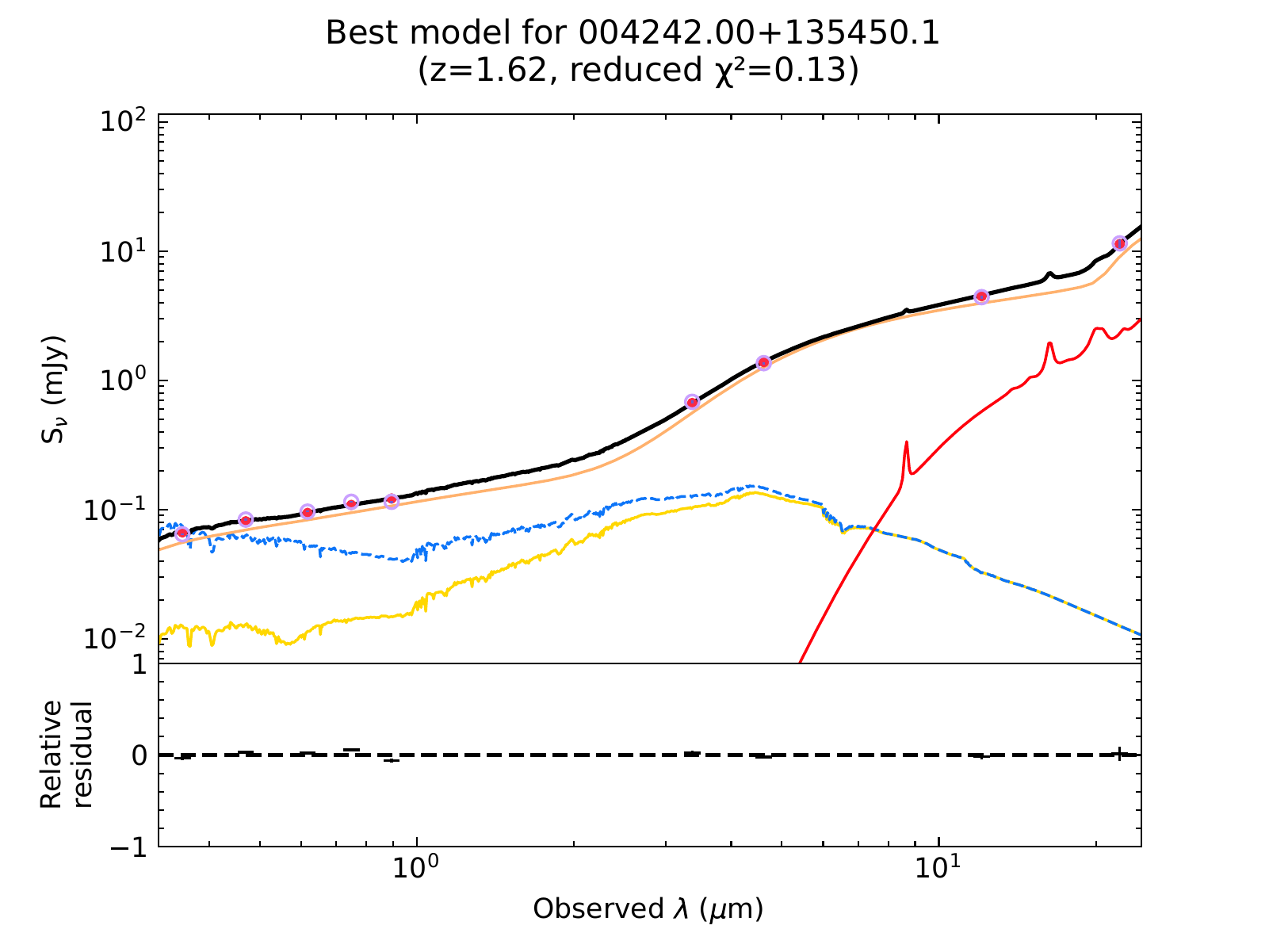}
\includegraphics[width=0.92\columnwidth]{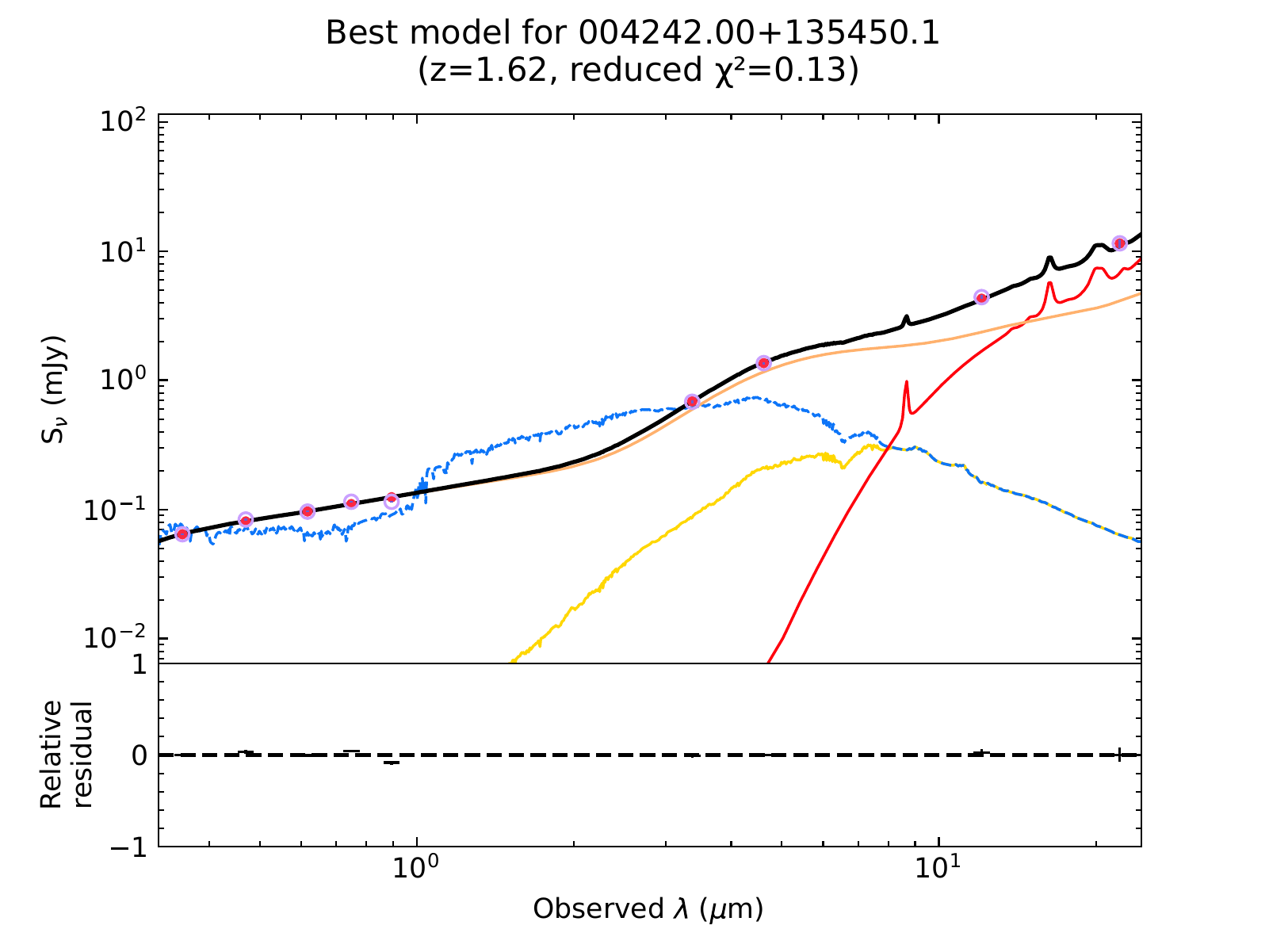}
\includegraphics[width=0.92\columnwidth]{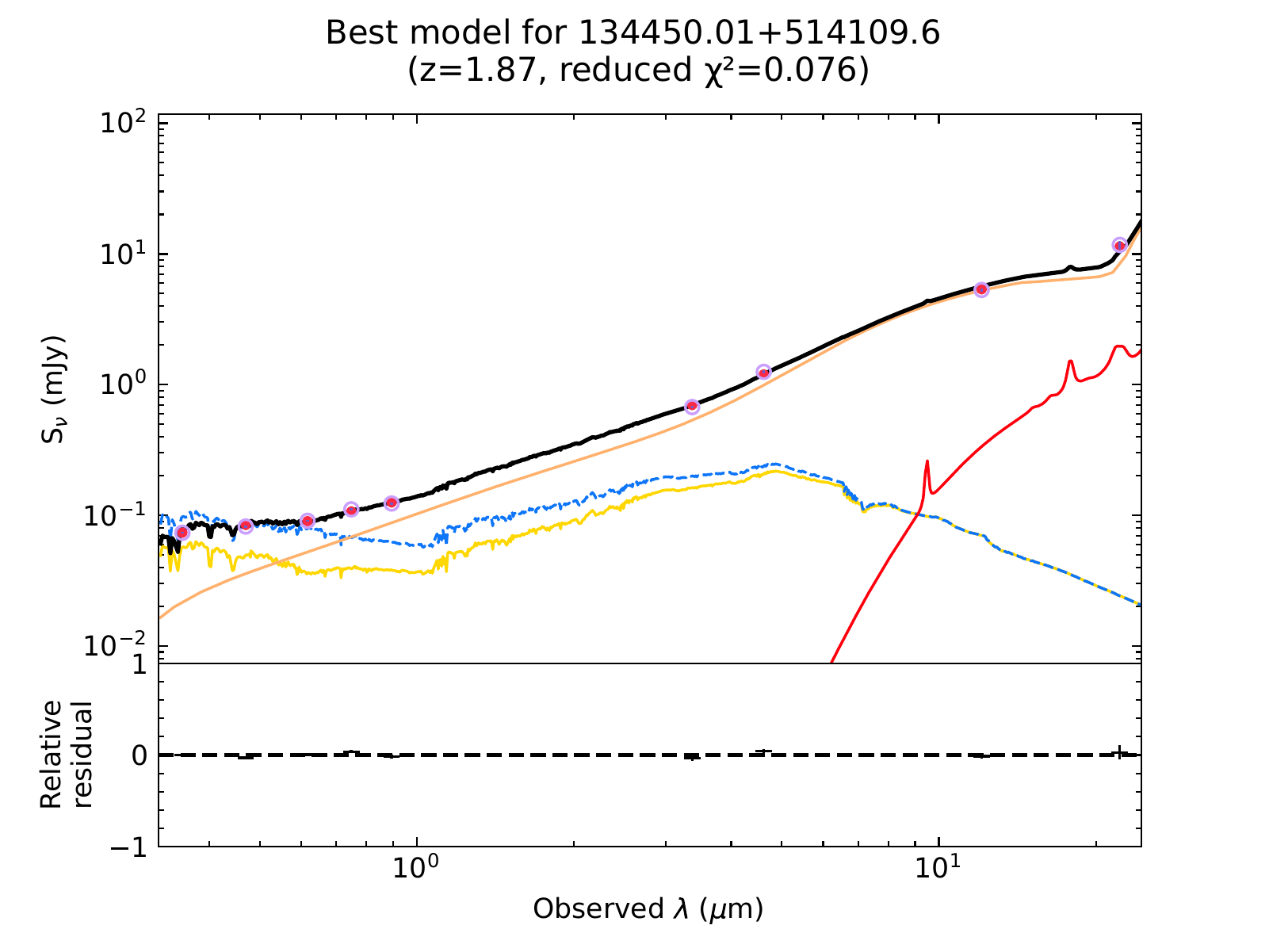}
\includegraphics[width=0.92\columnwidth]{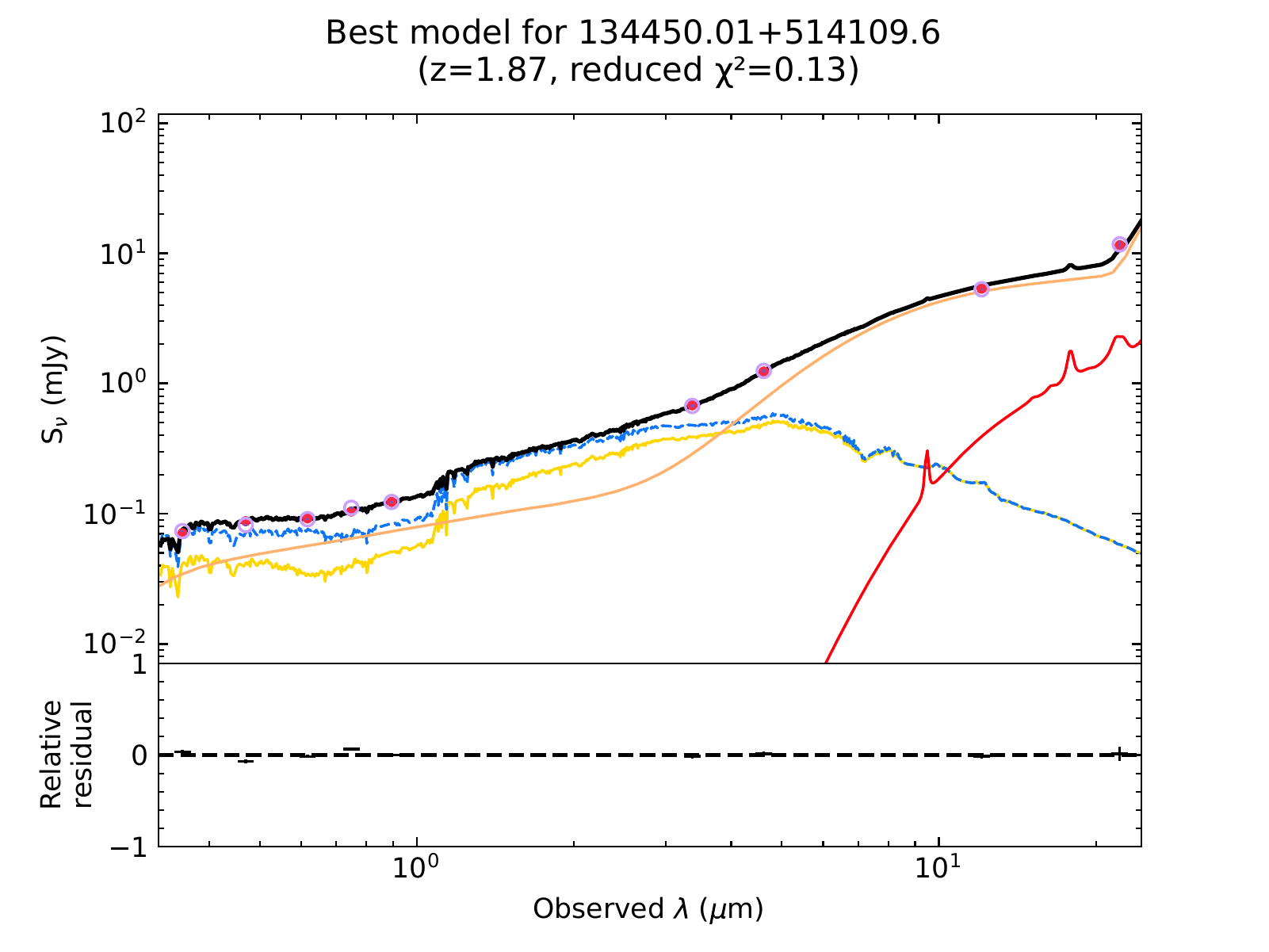}
\caption{Continued: bins 4 to 7.}
    \label{figA2}
\end{figure*}

\begin{figure*}
\ContinuedFloat
\includegraphics[width=0.92\columnwidth]{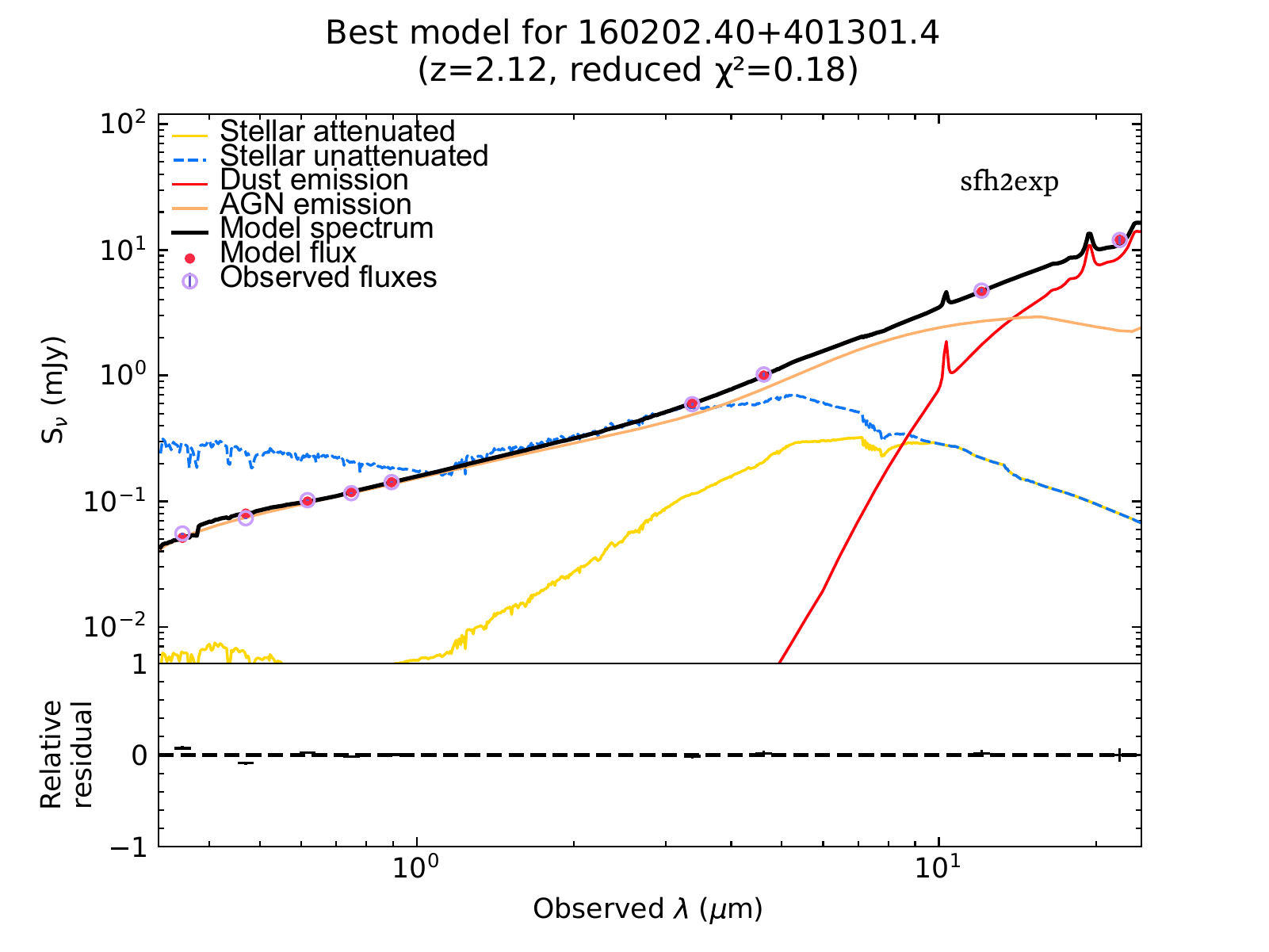}
\includegraphics[width=0.92\columnwidth]{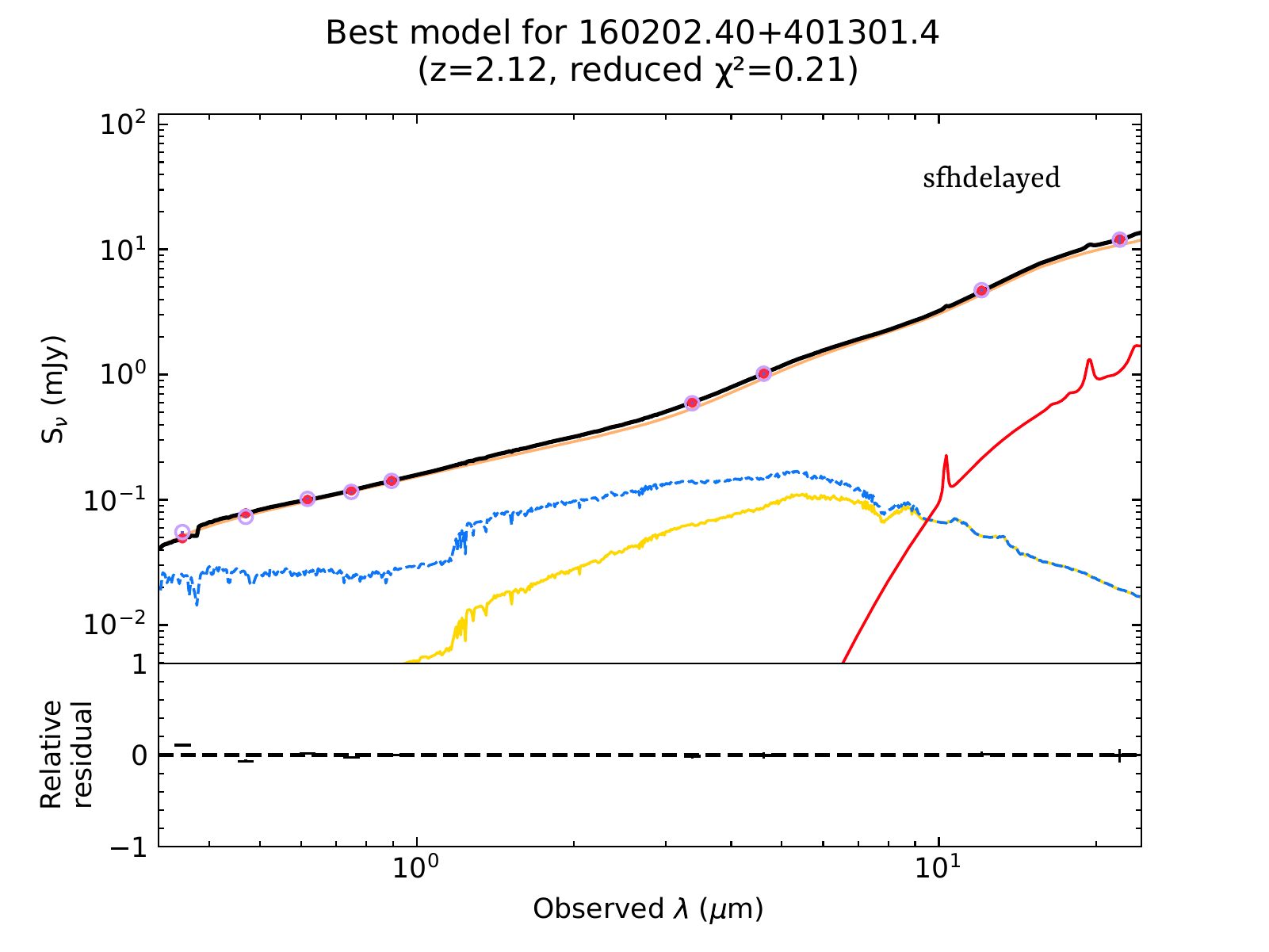}
\includegraphics[width=0.92\columnwidth]{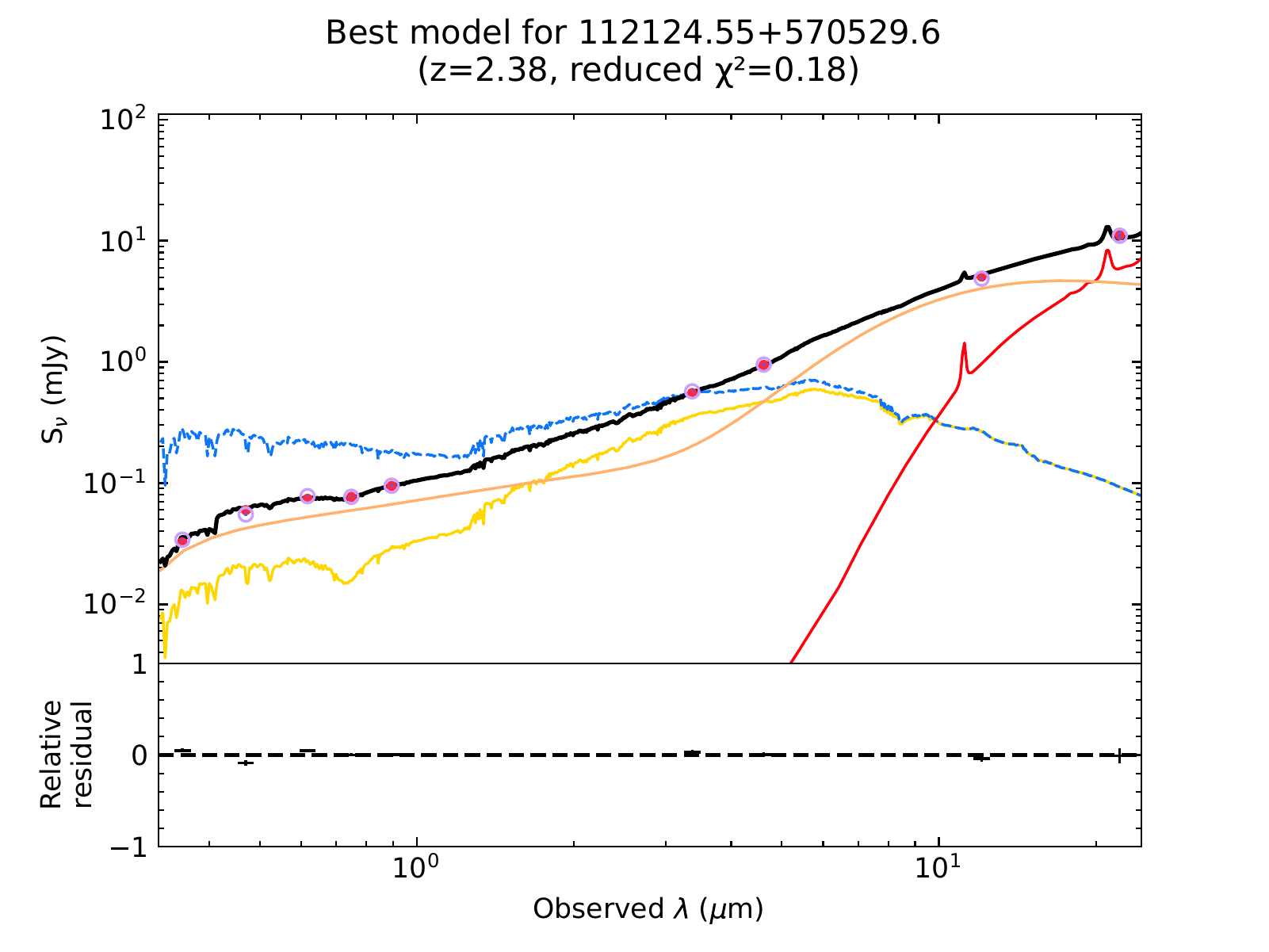}
\includegraphics[width=0.92\columnwidth]{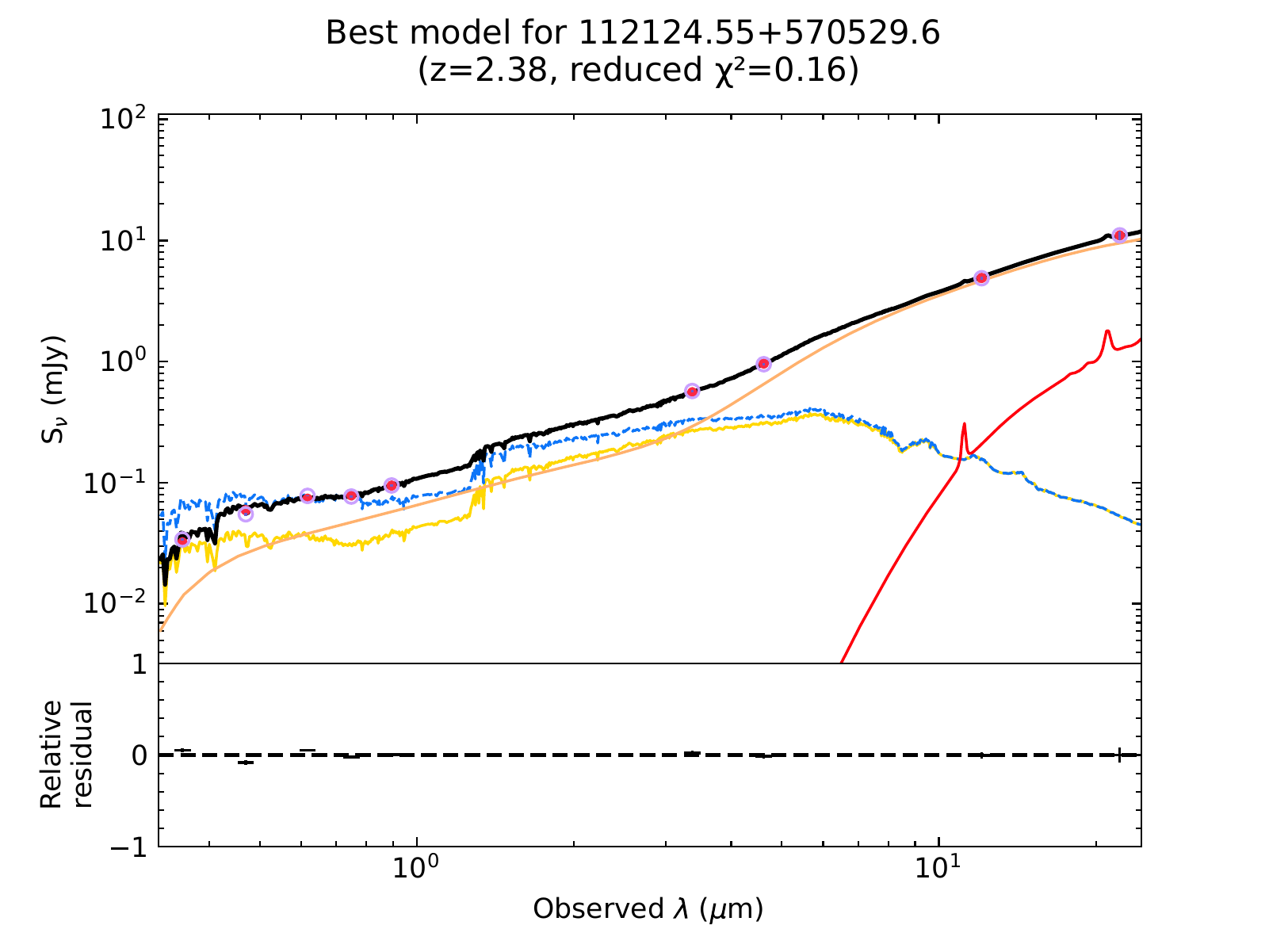}
\includegraphics[width=0.92\columnwidth]{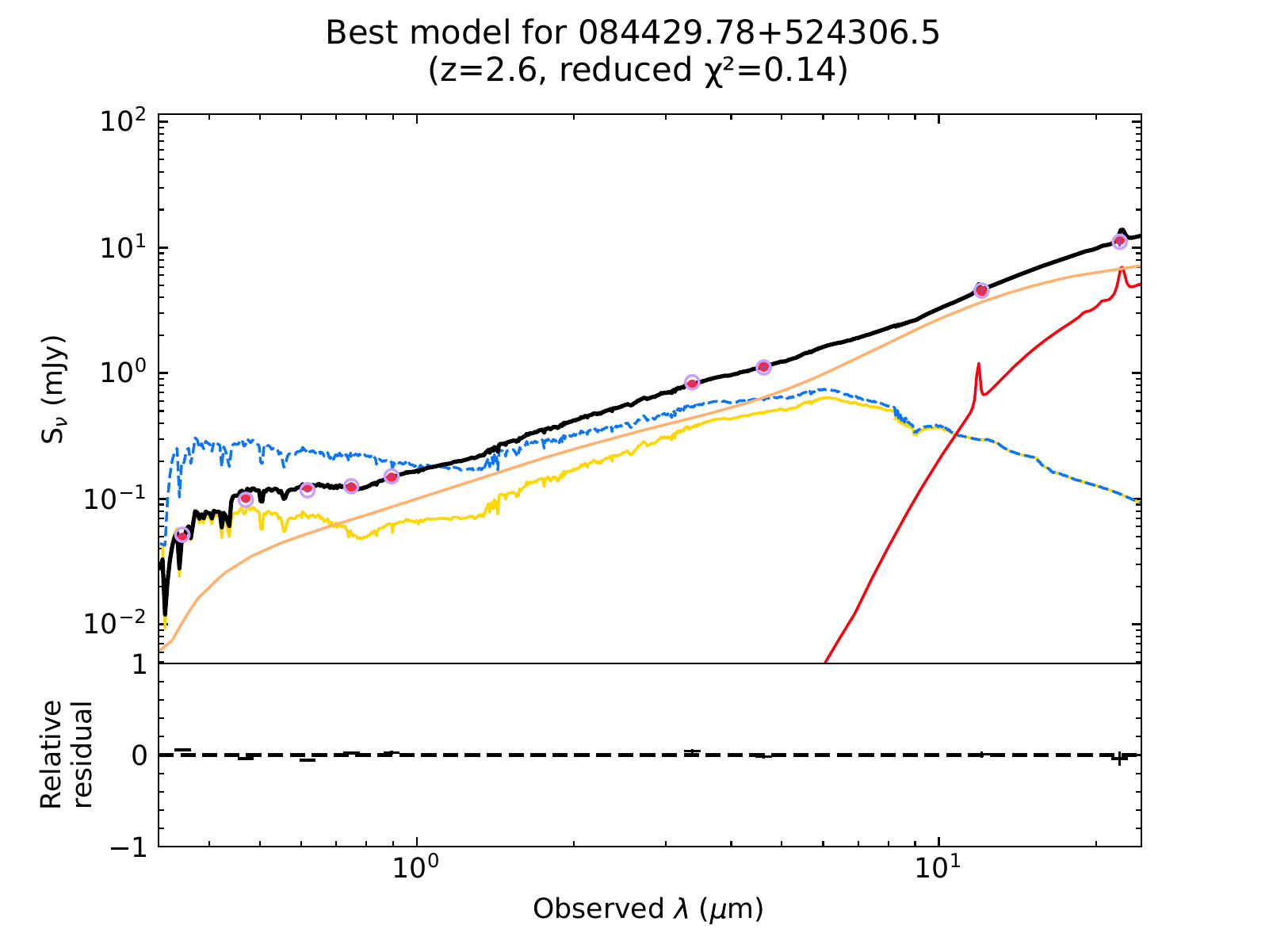}
\includegraphics[width=0.92\columnwidth]{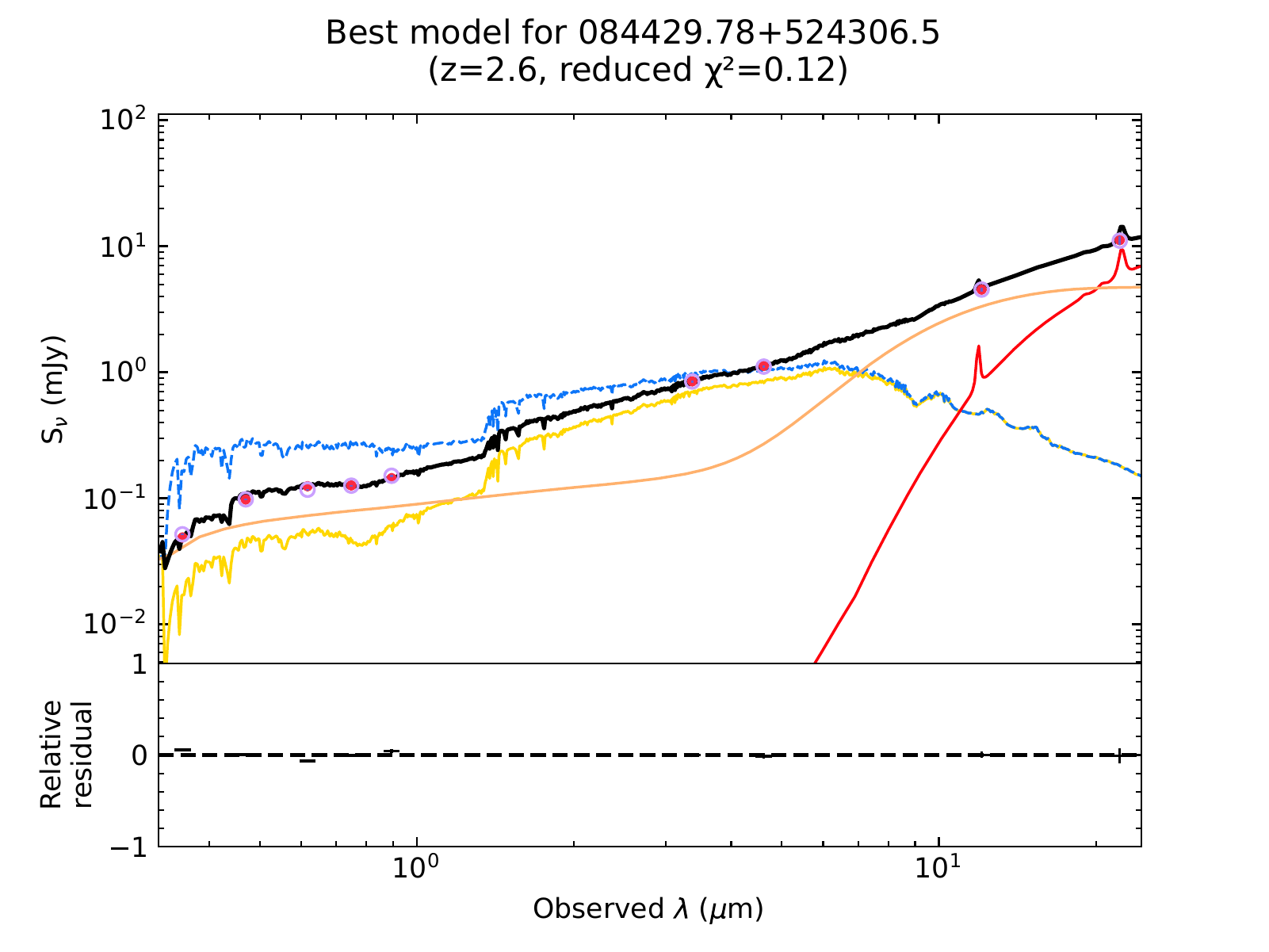}
\includegraphics[width=0.92\columnwidth]{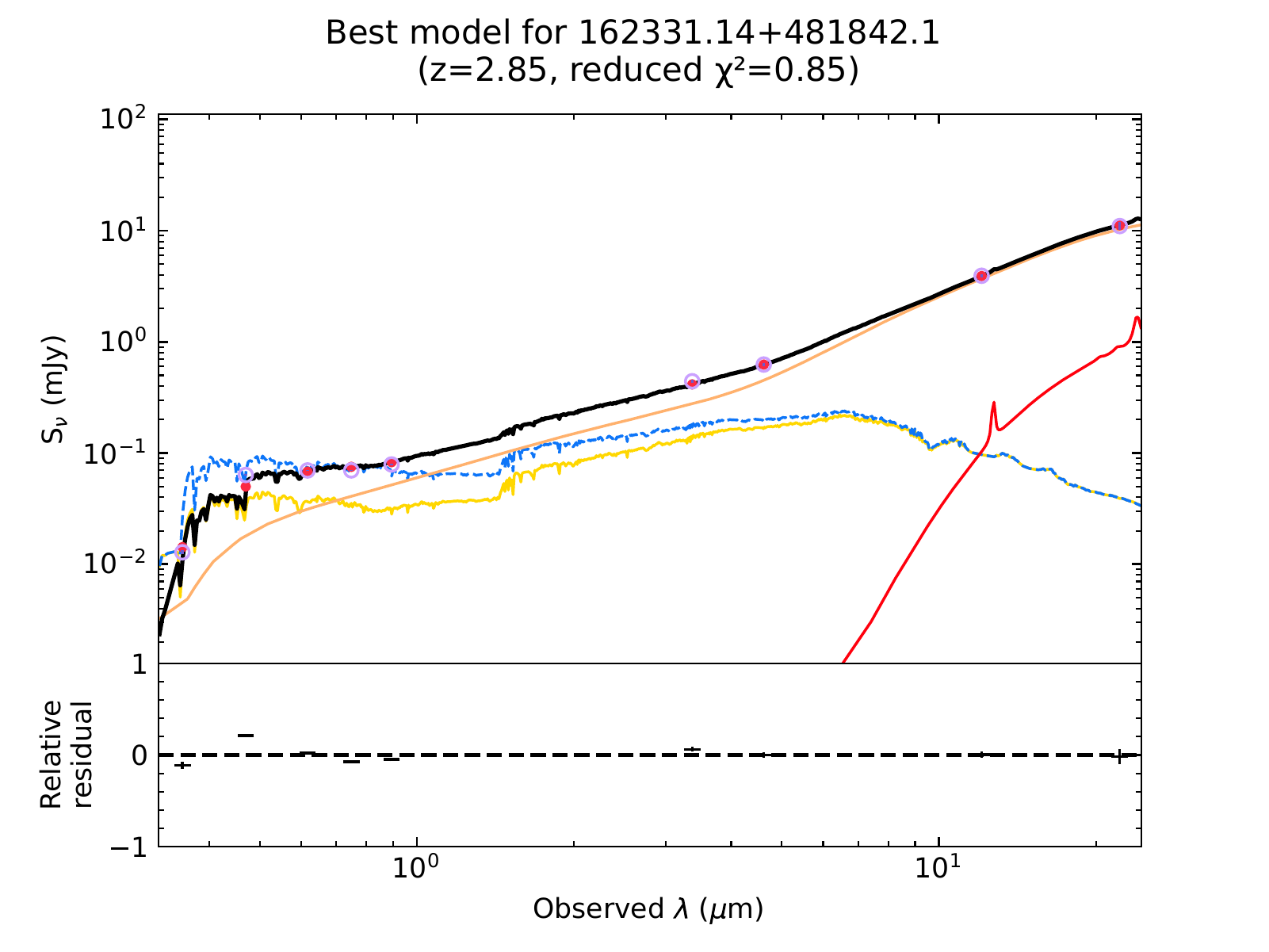}
\includegraphics[width=0.92\columnwidth]{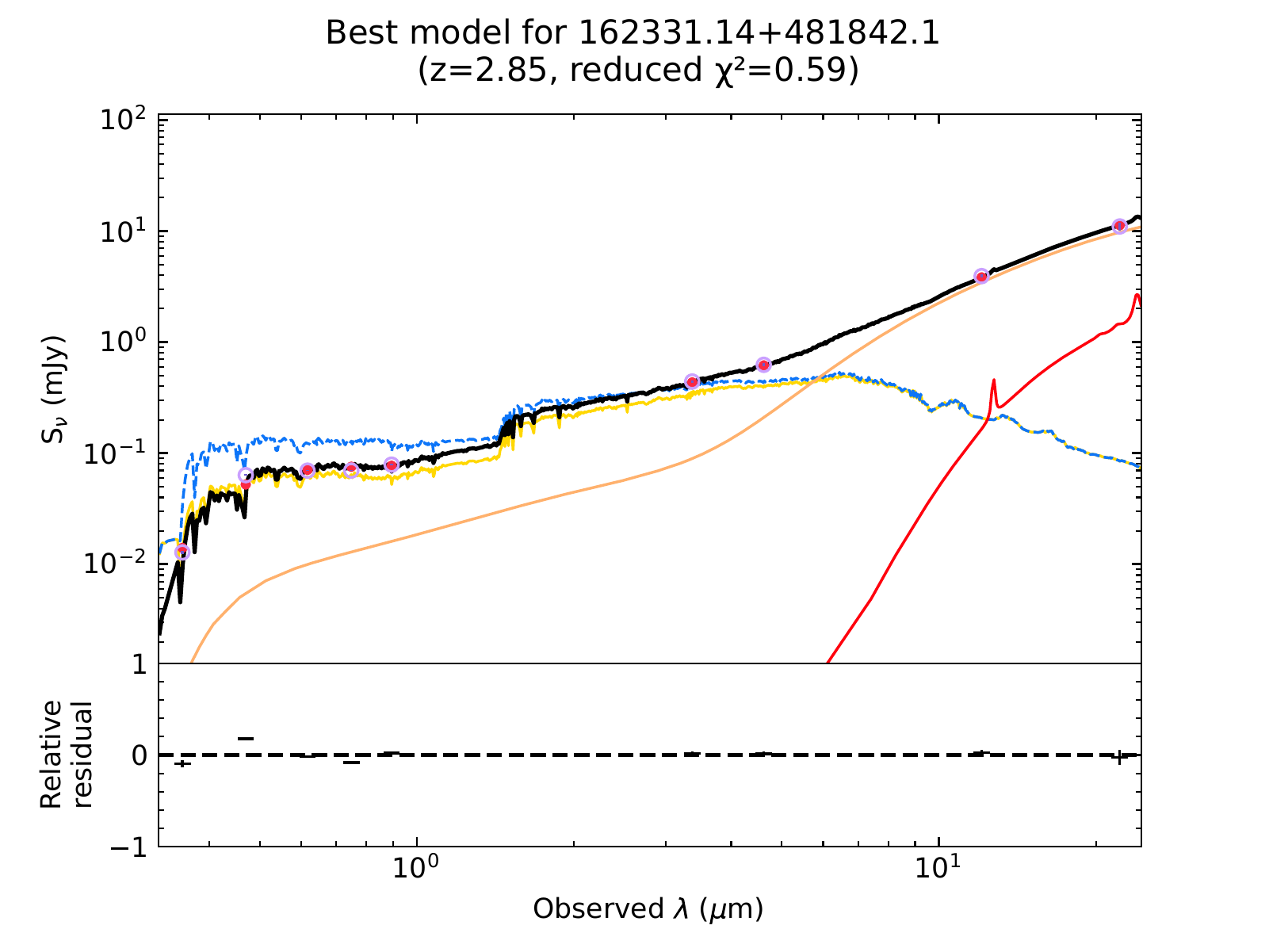}
\caption{Continued: bins 8 to 11.}
    \label{figA3}
\end{figure*}

\begin{figure*}
\ContinuedFloat
\includegraphics[width=0.92\columnwidth]{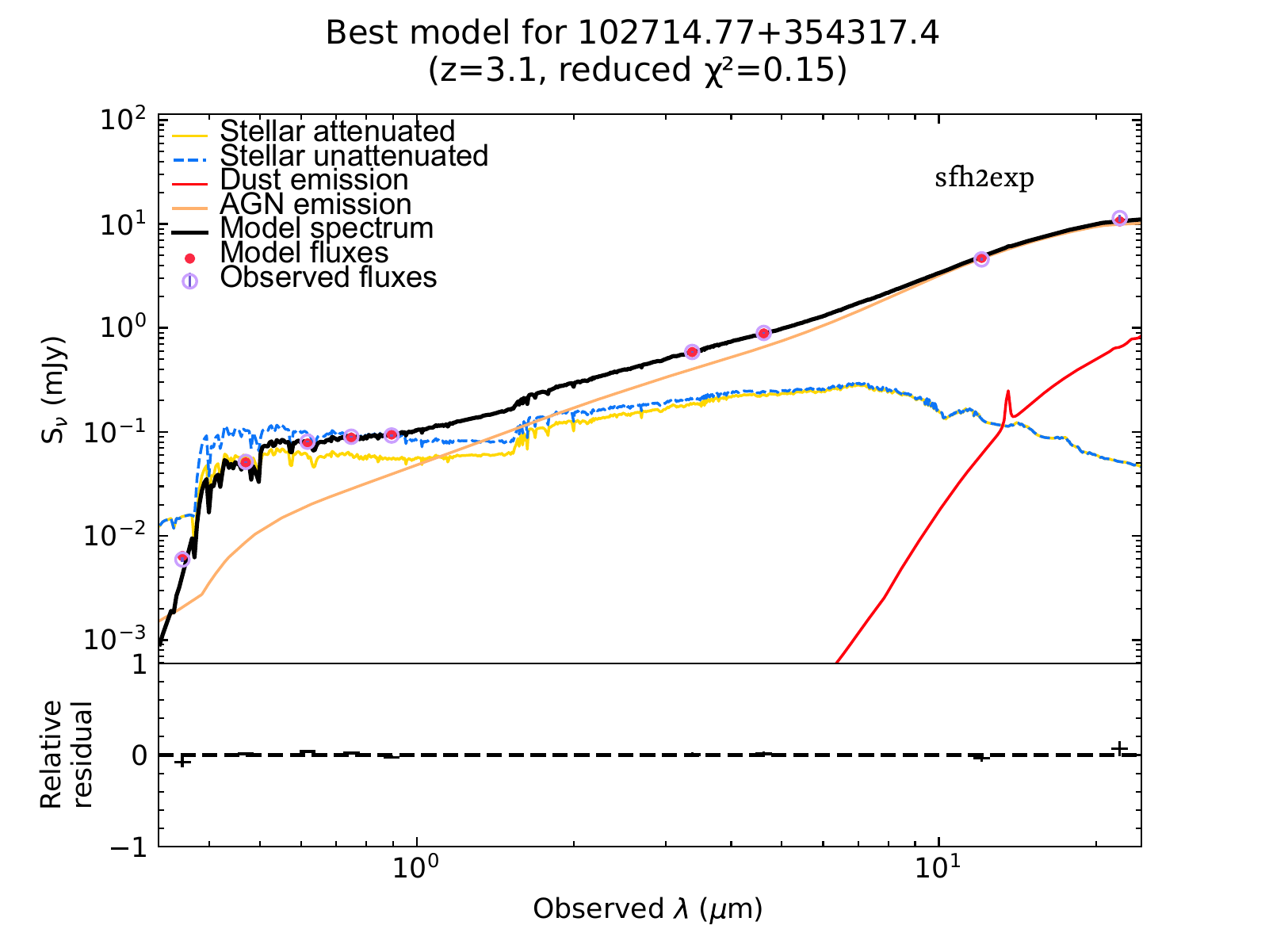}
\includegraphics[width=0.92\columnwidth]{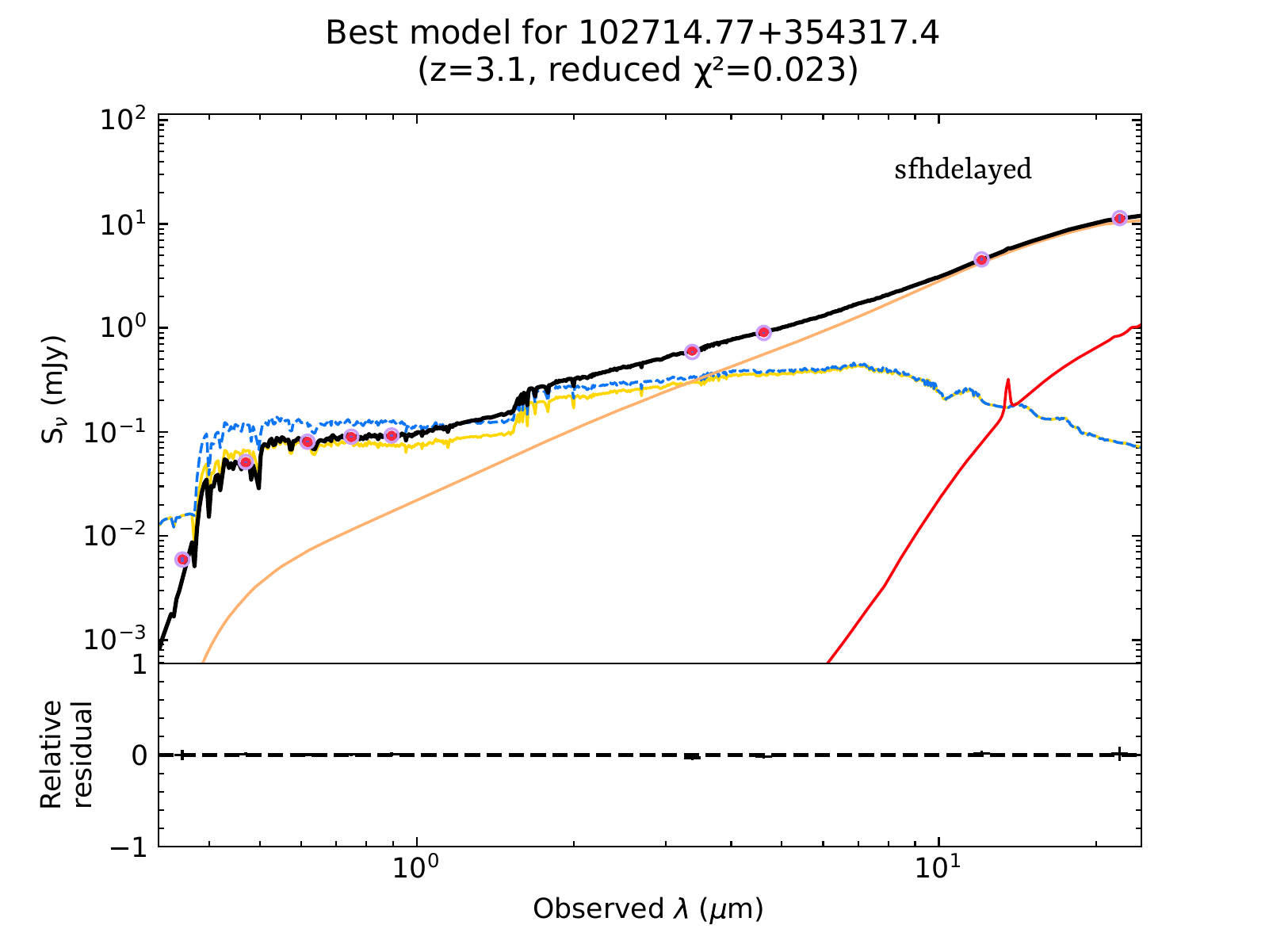}
\includegraphics[width=0.92\columnwidth]{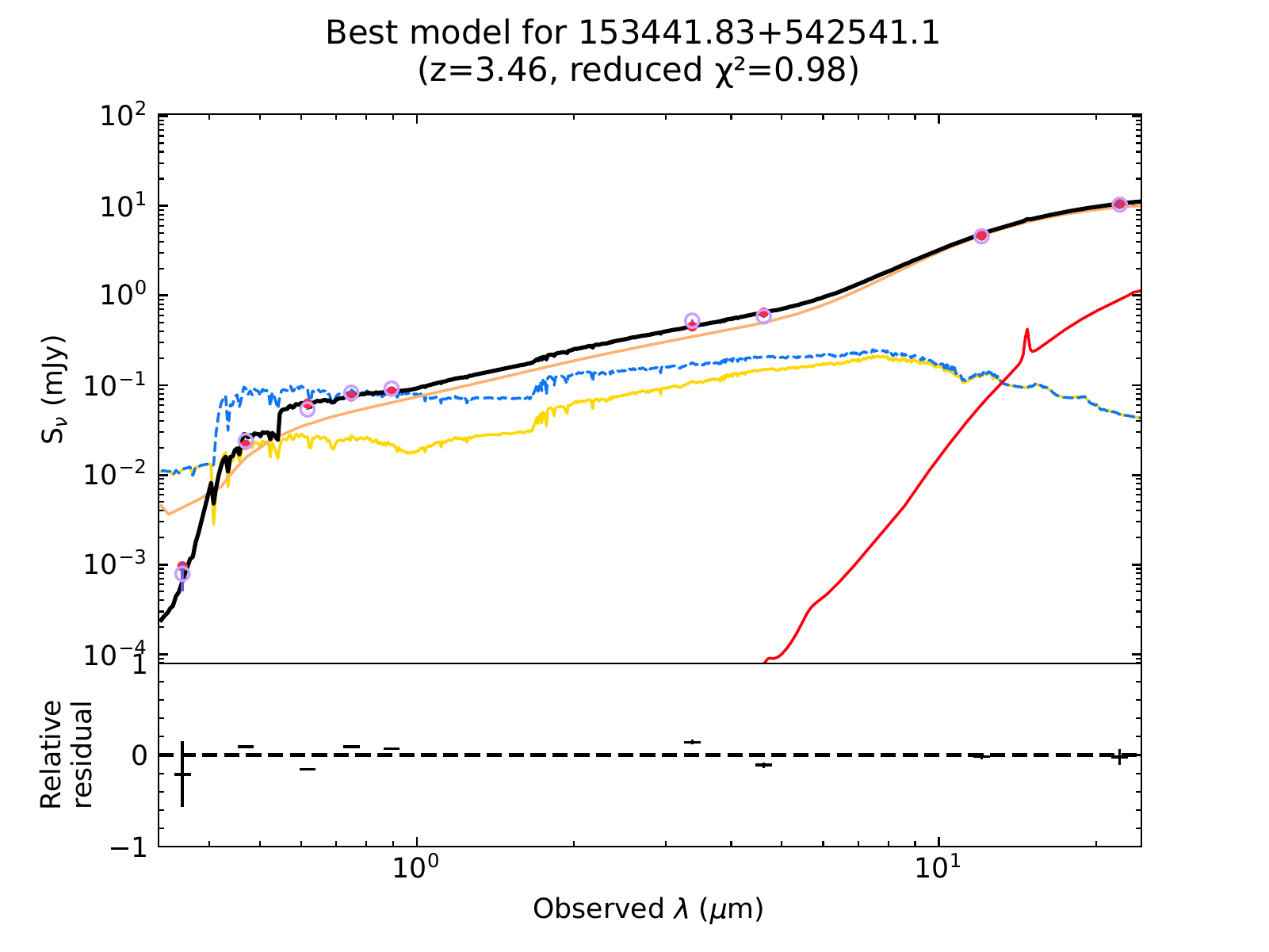}
\includegraphics[width=0.92\columnwidth]{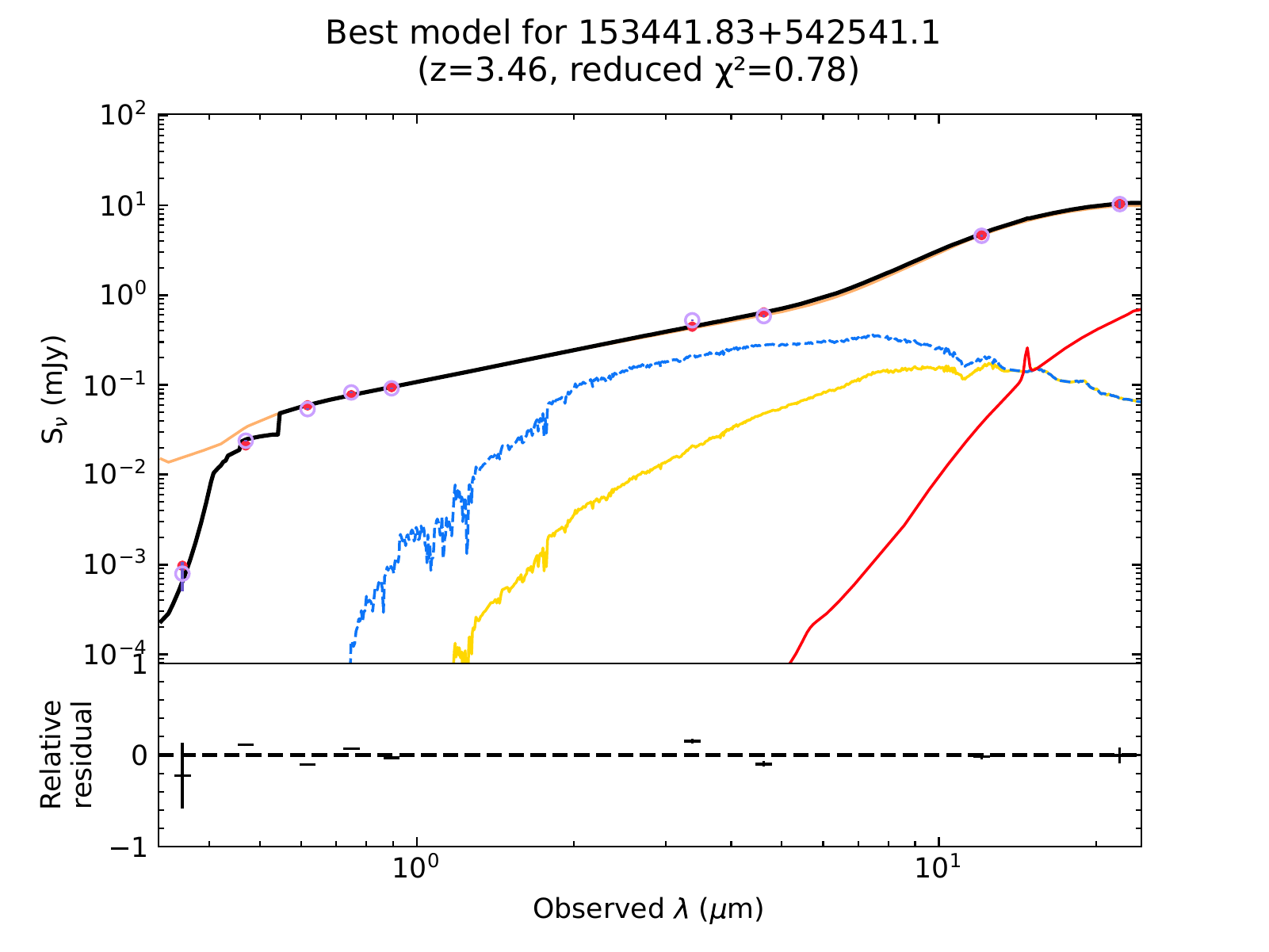}
\caption{Continued: bins 12 and 13.}
    \label{figA4}
\end{figure*}

In this section, we compare in parallel in Figure~\ref{figA1} the best solutions in each bin, as obtained using the two SFH functions \texttt{sfh2exp} (left panels) and \texttt{sfhdelayed} (right panels). In Tables~\ref{RES_sfhd} and ~\ref{RES_sfhe}, we summarize the results for both functions by giving the median values for important parameters in each bin, calculated considering only those models with $\chi^2 < 1$ (61\% of our sample for both solutions). Comparing col.~3 in the two tables, we conclude that both SFH functions yield equally reasonable solutions. In general, the AGN component dominates the SEDs (col.~10) and the SFR is high, increasing at high redshift (col.~13). The opening angle is huge (col.~8) and extinction relatively large (col.~11). These SEDs are consistent with obscured starbursts with dominant AGNs \citep{2018Marshall,2021Xie}.

Examining in detail col.~3 in Table~\ref{RES_sfhd} for the \texttt{sfhdelayed} function, the first three bins at low redshifts (\#0, \#1 and \#2) have slightly higher $\chi^2$, the fits getting better after that, up to bin \#6 ($z = 1.62$), the $\chi^2$ then fluctuating but still staying relatively low until the last three bins (\#11, \#12, \#13) where the numbers of QSOs is falling drastically. Disregarding the last three bins, we assert that the \texttt{sfhdelayed} function yields excellent solutions over a range of redshift $0.8<z\leq 2.8$. This conclusion is also supported by the fraction of models with $\chi^2 >3$ as indicated in col.~4, which does not pass 10\% within this redshift range. 

Doing the same analysis for the \texttt{sfh2exp} function in Table~\ref{RES_sfhe}, we also see higher $\chi^2$ values and higher fraction of models with $\chi^2 > 3$ in the first and last three bins, but comparably lower values, as observed for the \texttt{sfhdelayed} function, within the range $0.8 < z \leq 2.8$. This suggests that our solutions are degenerate, since we cannot differentiate between the two SFH functions. The reason for this degeneracy is easy to understand: the common traits of the SEDs at any redshift are the dominance of the AGN component (with some exceptions at high redshifts, for examples in bins \#9 to \#12) and the high SFR (col.~13). The high SFR is consistent in both models with relatively small e-folding (col.~14): not passing 1,000 Myrs in the \texttt{sfhdelayed} models and 750 Myr in the \texttt{sfh2exp}, suggesting that, independently of their SFH function (morphology), the host galaxies of HQWISE QSOs generally built the bulk of their stellar populations rapidly. 

\begin{table*}
\caption{\small Best fits for \texttt{sfhdelayed} function; the values reported are based on the medians of the models with $\chi^2 < 1$ (61\% of the sample).}
\label{RES_sfhd}
\begin{tabular}{ccccccccccccccc}
\hline
\hline
 (1) &(2) & (3)  & (4)      & (5)             & (6)   & (7)   & (8)   & (9)      & (10)      & (11)     & (12)     & (13)                       & (14)          &(15)\\
 bin & z  & CT   & $\chi^2$ & f($\chi^2 > 3$) & W1-W2 & W2-W3 & O.A.  & $\delta$ & f$_{AGN}$ & E$_{BV}$ & E$_{BV}$ & Log(SFR)                & $\tau_{main}$ & T$_{dust}$ \\
     &    & Gyrs &          & \%              &       &       & (deg) &          &           & Lines    & Factor   & (M$_\odot\ {\rm yr}^{-1}$) & (Myr)         & (K) \\
\hline\noalign{\smallskip}
0   & 0.18 &11.2 & 0.64 & 19 & 0.96 & 2.96 & 60  & -0.7  & 0.9 & 0.70 & 0.75 & -1.22 & 1000 & 100 \\
1   & 0.38 & 9.3 & 0.57 & 26 & 1.02 & 3.07 & 60  & -1.0  & 0.9 & 0.70 & 0.75 & -0.82 & 1000 & 100 \\
2   & 0.62 & 7.3 & 0.72 & 27 & 1.04 & 2.94 & 60  & -1.0  & 0.9 & 0.70 & 0.75 &  0.17 & 1000 & 100 \\
3   & 0.89 & 6.3 & 0.42 & 10 & 1.17 & 3.43 & 140 & -0.7  & 0.9 & 0.70 & 0.75 &  0.85 &  500 & 100 \\
4   & 1.11 & 5.3 & 0.33 &  8 & 1.31 & 3.00 & 100 & -0.7  & 0.9 & 0.70 & 0.75 &  1.89 &  500 & 100 \\
5   & 1.38 & 4.5 & 0.29 &  7 & 1.38 & 3.08 & 140 & -0.7  & 0.9 & 0.70 & 0.75 &  1.69 &  500 & 100 \\
6   & 1.62 & 3.9 & 0.18 &  6 & 1.37 & 3.25 & 140 & -0.5  & 0.9 & 0.70 & 0.75 &  2.08 & 1000 & 100 \\
7   & 1.86 & 3.4 & 0.24 &  5 & 1.30 & 3.38 & 140 & -0.7  & 0.9 & 0.50 & 0.50 &  2.52 & 1000 & 500 \\
8   & 2.12 & 3.0 & 0.28 &  6 & 1.23 & 3.53 & 100 & -0.5  & 0.9 & 0.50 & 0.50 &  2.76 & 1000 & 500 \\
9   & 2.38 & 2.7 & 0.14 & 10 & 1.18 & 3.61 & 140 & -0.6  & 0.9 & 0.50 & 0.50 &  3.02 & 1000 & 500 \\
10  & 2.63 & 2.4 & 0.16 &  9 & 1.14 & 3.65 & 140 & -0.5  & 0.9 & 0.50 & 0.50 &  3.33 & 1000 & 500 \\
11  & 2.85 & 2.2 & 0.42 & 11 & 1.08 & 3.58 & 140 & -0.7  & 0.9 & 0.10 & 0.25 &  3.39 &  500 & 500 \\
12  & 3.09 & 2.0 & 0.32 & 15 & 0.98 & 3.66 & 140 & -0.7  & 0.9 & 0.10 & 0.50 &  3.53 &  500 & 100 \\
13  & 3.41 & 1.7 & 0.44 &  8 & 1.03 & 3.73 & 140 & -0.5  & 0.9 & 0.10 & 0.50 &  3.54 &  500 & 100 \\
\hline
\end{tabular}
\end{table*}

\begin{table*}
\caption{\small Best fits for \texttt{sfh2exp} function; the values reported are based on the medians of the models with $\chi^2 < 1$ (61\% of the sample).}
\label{RES_sfhe}
\begin{tabular}{ccccccccccccccc}
\hline
\hline
 (1) &(2) & (3)  & (4) & (5)  & (6) & (7) & (8) & (9) & (10) & (11) & (12) & (13) & (14) & (15)\\
 bin & z  & CT   & $\chi^2$ & f($\chi^2 > 3$) & W1-W2 & W2-W3 & O.A.& $\delta$ & f$_{AGN}$ & E$_{BV}$ & E$_{BV}$ & Log(SFR) & $\tau_{main}$ & T$_{dust}$\\
     &    & Gyrs &  & \%  &  &    & (deg) &  &  & Lines & Factor & (M$_\odot\ {\rm yr}^{-1}$) & (Myr) & (K) \\
\hline\noalign{\smallskip}
0   & 0.18 &11.2 & 0.56 & 27 & 0.98 & 3.01 &  80 &  -0.7  &0.9 & 0.50 & 0.50 & 0.22 & 750 & 100 \\
1   & 0.40 & 9.3 & 0.62 & 32 & 1.02 & 3.05 &  60 &  -1.0  &0.9 & 0.70 & 0.50 & 0.91 &  50 & 100 \\
2   & 0.66 & 7.3 & 0.70 & 28 & 1.07 & 2.94 &  60 &  -1.0  &0.9 & 0.70 & 0.75 & 1.48 &  50 & 100 \\
3   & 0.88 & 6.3 & 0.41 & 14 & 1.22 & 2.97 & 140 &  -1.0  &0.9 & 0.70 & 0.75 & 1.72 &  50 & 100 \\
4   & 1.11 & 5.3 & 0.33 & 12 & 1.34 & 2.99 & 140 &  -0.7  &0.9 & 0.70 & 0.75 & 1.86 &  50 & 100 \\
5   & 1.38 & 4.5 & 0.30 &  9 & 1.39 & 3.09 & 140 &  -0.7  &0.9 & 0.70 & 0.75 & 2.15 & 100 & 100 \\
6   & 1.62 & 3.9 & 0.17 &  8 & 1.37 & 3.25 & 140 &  -0.5  &0.9 & 0.70 & 0.50 & 2.35 & 500 & 100 \\
7   & 1.86 & 3.4 & 0.17 &  6 & 1.27 & 3.41 & 140 &  -0.5  &0.9 & 0.50 & 0.50 & 2.70 & 500 & 100 \\
8   & 2.12 & 3.0 & 0.23 &  7 & 1.20 & 3.50 & 140 &  -0.3  &0.9 & 0.50 & 0.50 & 2.84 & 500 & 500 \\
9   & 2.38 & 2.7 & 0.10 & 16 & 1.17 & 3.64 & 100 &  -0.5  &0.9 & 0.50 & 0.50 & 2.98 & 500 & 500 \\
10  & 2.63 & 2.4 & 0.12 &  9 & 1.12 & 3.66 & 140 &  -0.5  &0.9 & 0.10 & 0.50 & 3.19 &  50 & 500 \\
11  & 2.85 & 2.2 & 0.39 & 30 & 1.08 & 3.60 & 140 &  -0.5  &0.9 & 0.50 & 0.25 & 3.26 & 500 & 500 \\
12  & 3.10 & 2.0 & 0.75 & 31 & 1.04 & 2.82 & 100 &   0.0  &0.9 & 0.10 & 0.50 & 3.59 & 500 & 100 \\
13  & 3.41 & 1.7 & 0.58 &  8 & 1.07 & 3.71 & 140 &   0.0  &0.9 & 0.10 & 0.75 & 3.23 & 500 & 100 \\
\hline
\end{tabular}
\end{table*}

In our sample the dominance of the AGN and the high SFRs explain the degeneracy. However, examining in detail the best fits in Figure~\ref{figA1} some important nuances must be added. More specifically, we observe significant differences in the first five bins at low redshifts in terms of the stellar populations (unattenuated curves in blue in the figures), suggesting they are systematically older using the \texttt{sfhdelayed} function than the \texttt{sfh2exp} function. This characteristic is consistent with slightly larger e-folding and much lower SFRs in the \texttt{sfhdelayed} SEDs than in the \texttt{sfh2exp} SEDs. On the other hand, starting at bin \#6 ($z \sim 1.62$) the stellar components in both models become similar. This happens as the cosmic time (CT in col.~3) becomes comparable to the e-folding time (CT falling from 3.9 to 1.7 Gyrs). 

Because the SEDs of the stellar populations in the \texttt{sfhdelayed} models at low redshifts are typical of early-type galaxies, the above difference could be interpreted as two solutions for two different morphological types for the host galaxies, Elliptical vs. Spiral. Considering the evolution in CT and assuming a standard scenario for the formation of galaxies \citep{2014Madau}, finding high SFRs in galaxies at low redshifts seem more natural in spiral than elliptical galaxies \citep[e.g.,][]{2020Bellstedt}. This means that the consequences of the two solutions are significantly different: in the \texttt{sfh2exp} models, the SFRs would be characterized as normal or mildly starburst-like for late type spiral galaxies, while in the \texttt{sfhdelayed} models the high SFRs would be definitely starburst-like for the morphology (consistent with blue Elliptical galaxies or merging galaxies).

\begin{figure*}
\includegraphics[width=\columnwidth]{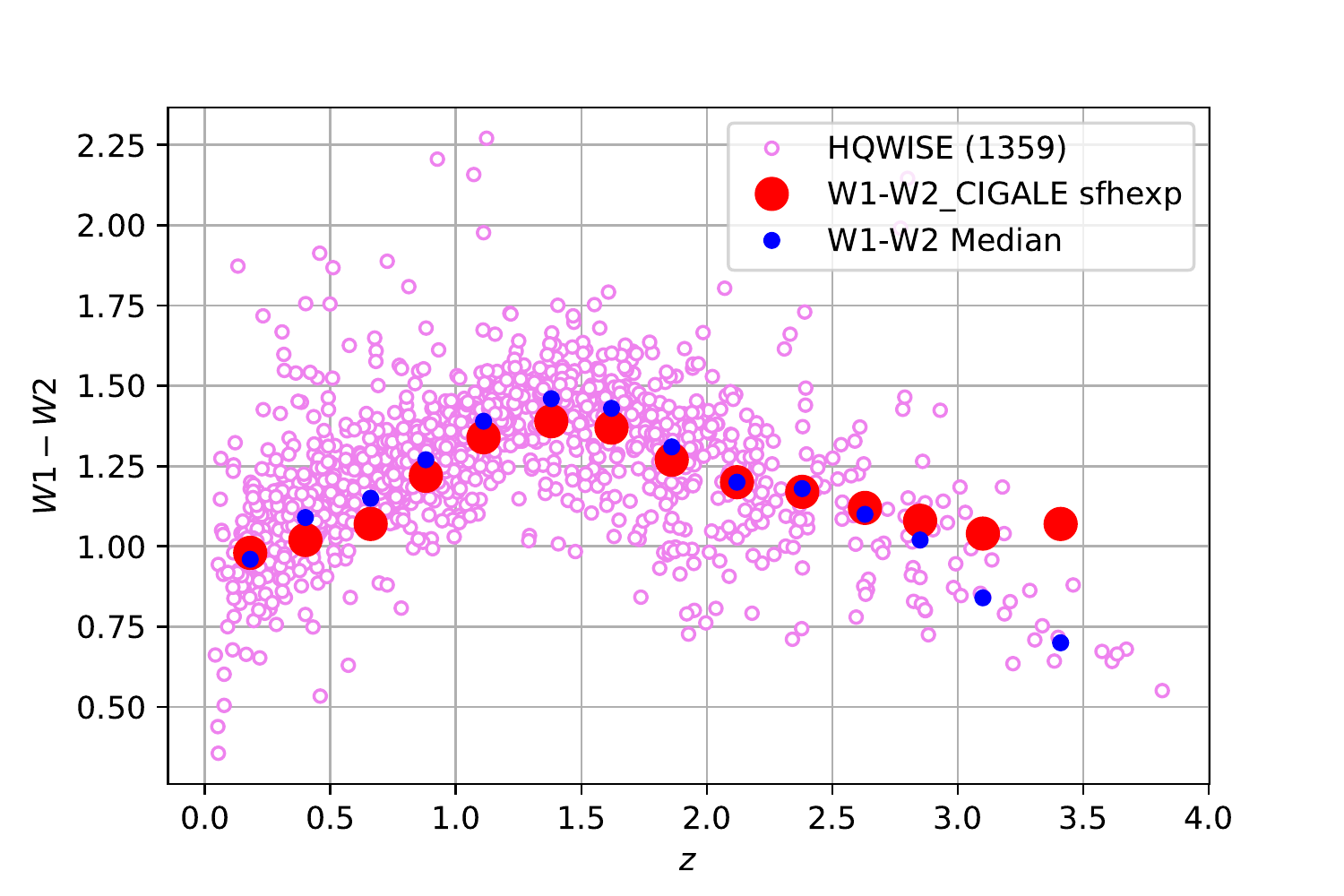}
\includegraphics[width=\columnwidth]{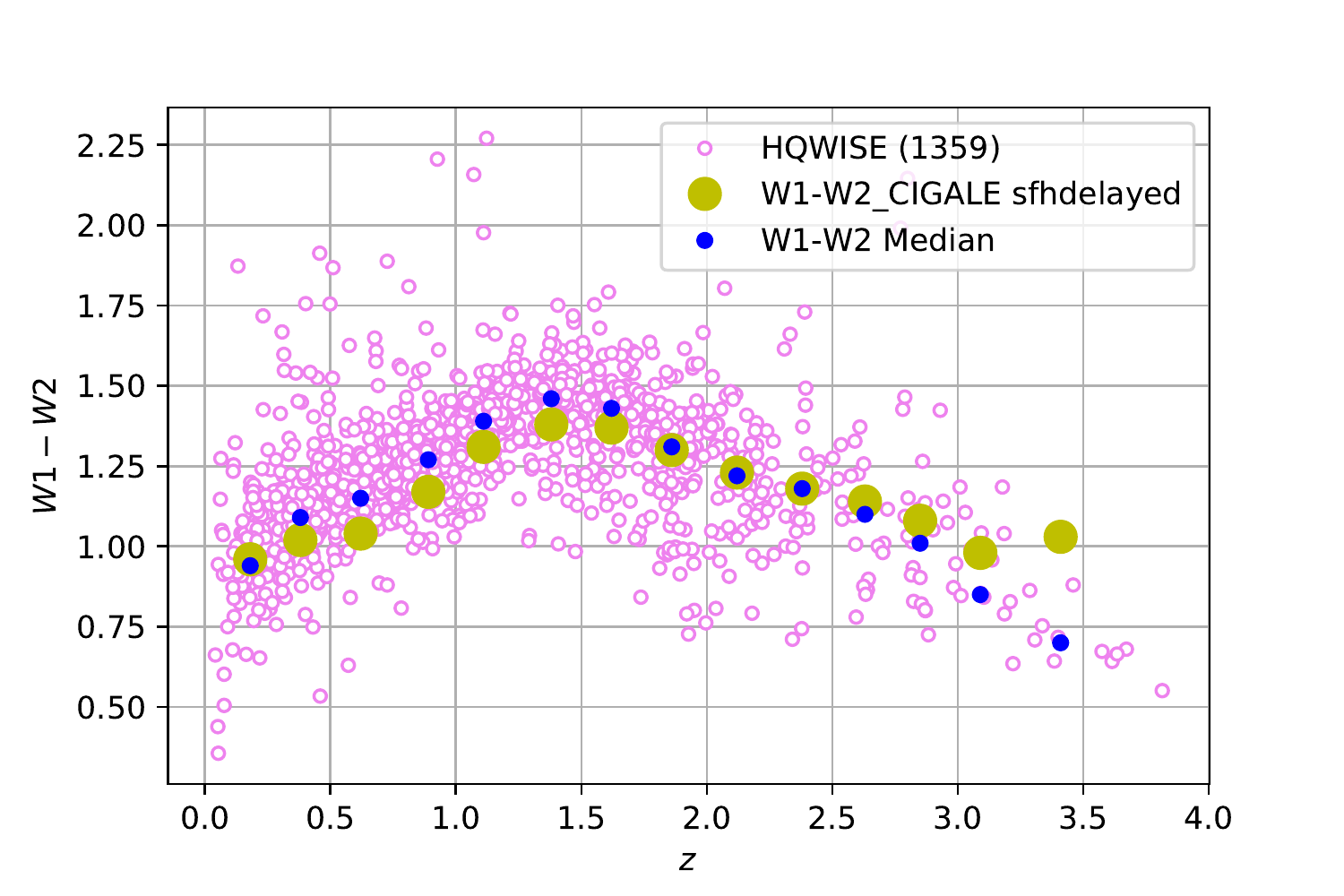}
\includegraphics[width=\columnwidth]{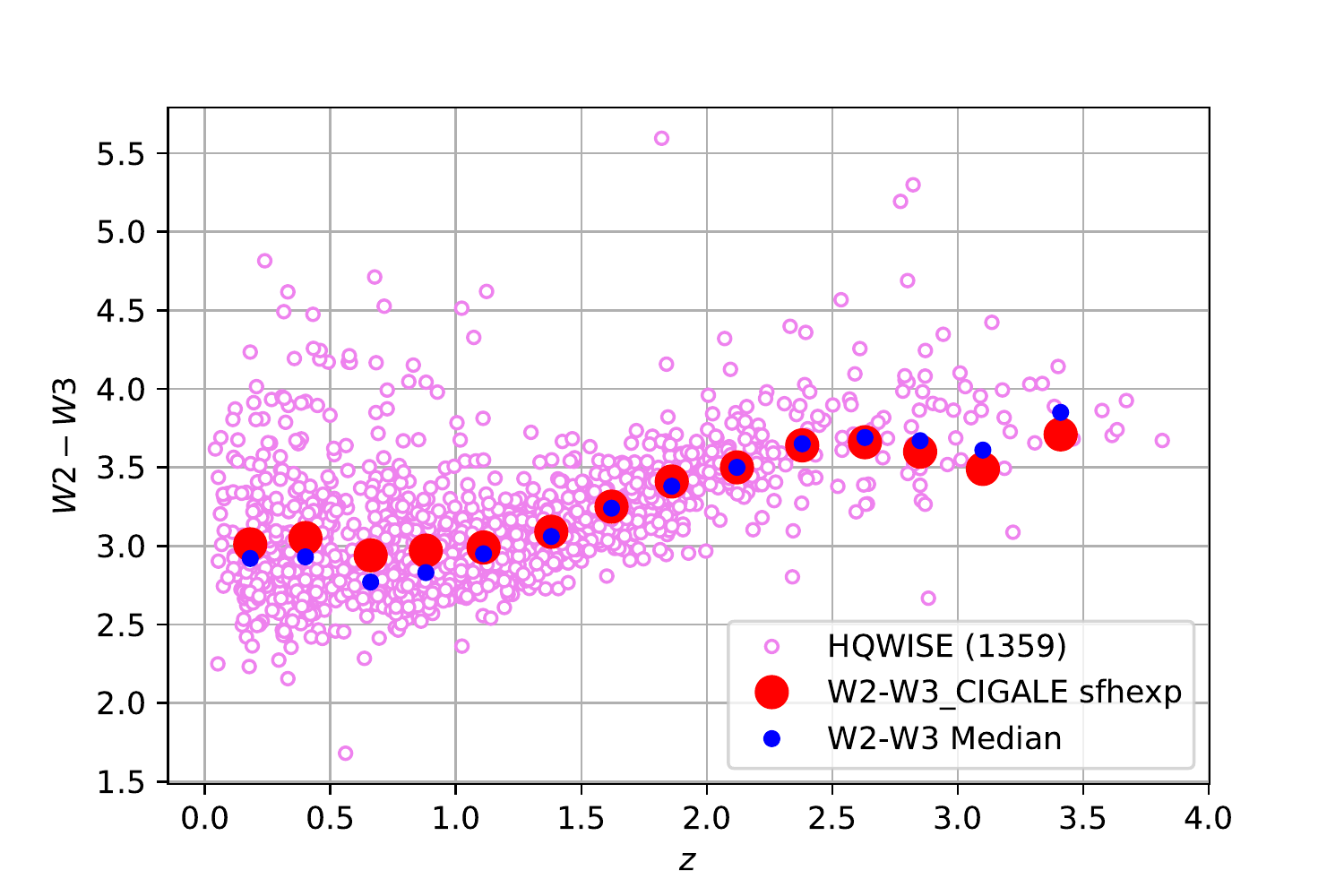}
\includegraphics[width=\columnwidth]{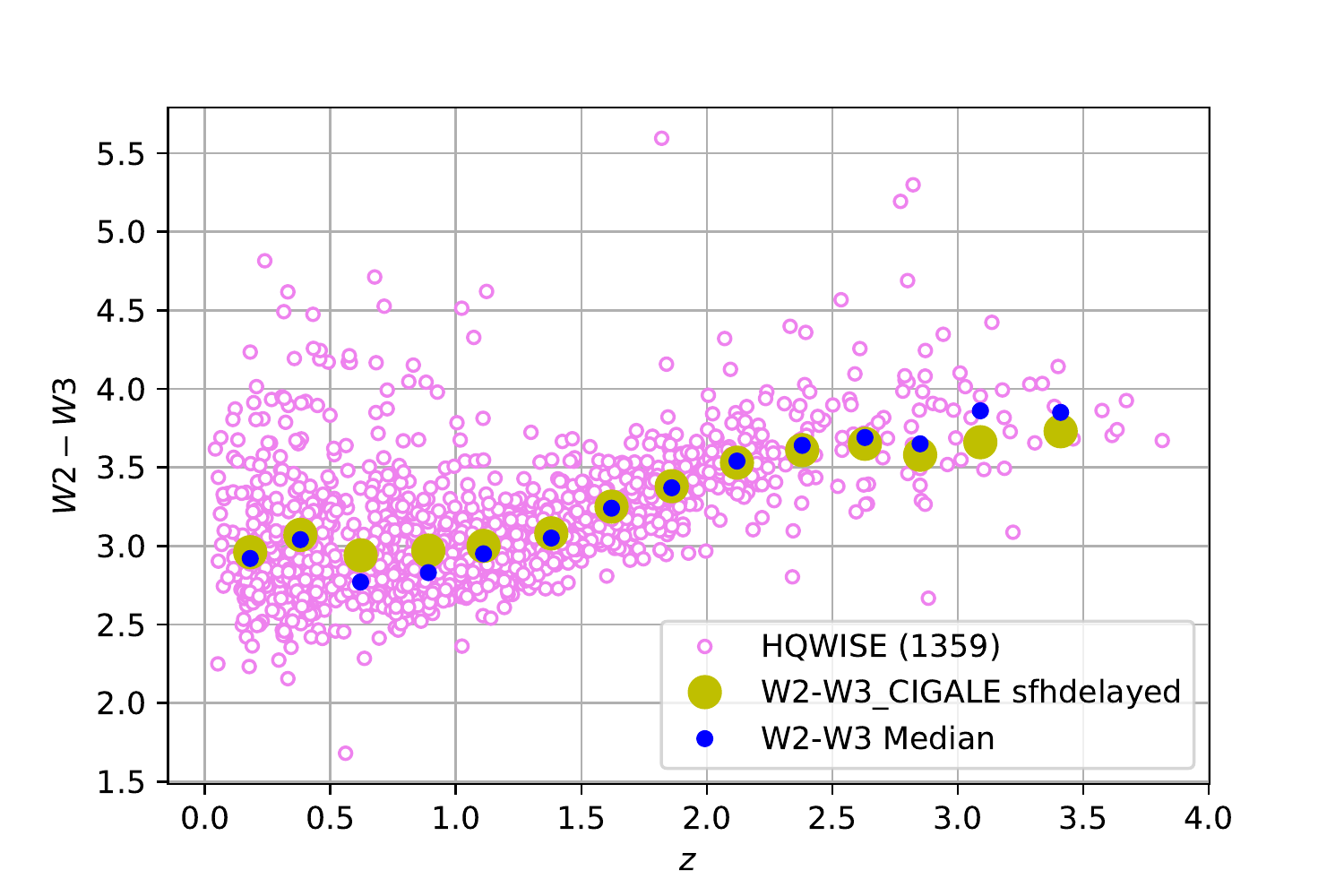}
    \caption{Reproducing the WISE colours with \texttt{X-CIGALE}: on the left, for the \texttt{sfh2exp} function and on the right for the \texttt{sfhdelayed} function. In all the graphics, the blue dots are the WISE medians in each bin and the red dots are the medians for the models with $\chi^2 < 1$.}
    \label{WISEcolours}
\end{figure*}

Continuing our analysis of the results, another important feature that is remarkable in Figure~\ref{figA1} is the similarity of the re-emission dust components in both models. This suggests that the WISE colours produced by the two SFH functions must also be similar. To check how well \texttt{X-CIGALE} reproduces the variations of these colours with the redshift, we trace in Figure~\ref{WISEcolours} the medians of the models with $\chi^2 < 1$ in each bin, comparing them with the medians measured by WISE. For both models, the fits are relatively good, except maybe in the last two bins where the number of QSOs are very low (only 13 and 12 QSOs respectively). In particular, both models reproduce relatively well the inflection to the blue of the W1-W2 colour at $z \sim 1.4$. Coincidentally, this happens in Figure~\ref{figA2} passing from bin \#5 to bin \#6 (medians $z = 1.38$ and $z = 1.62$, respectively), where the stellar contributions in the two models become comparable: the SED becoming gradually bluer (flatter). Consequently, the inflection of W1-W2 colours in both models could be interpreted as the signature of an increase of SFR in the host of QSOs at high redshifts. Based on the literature \citep{1994Elvis,2009Richards,2008Labita}, this might be recognized as a common trait of quasar evolution.  

\begin{figure}
\includegraphics[width=\columnwidth]{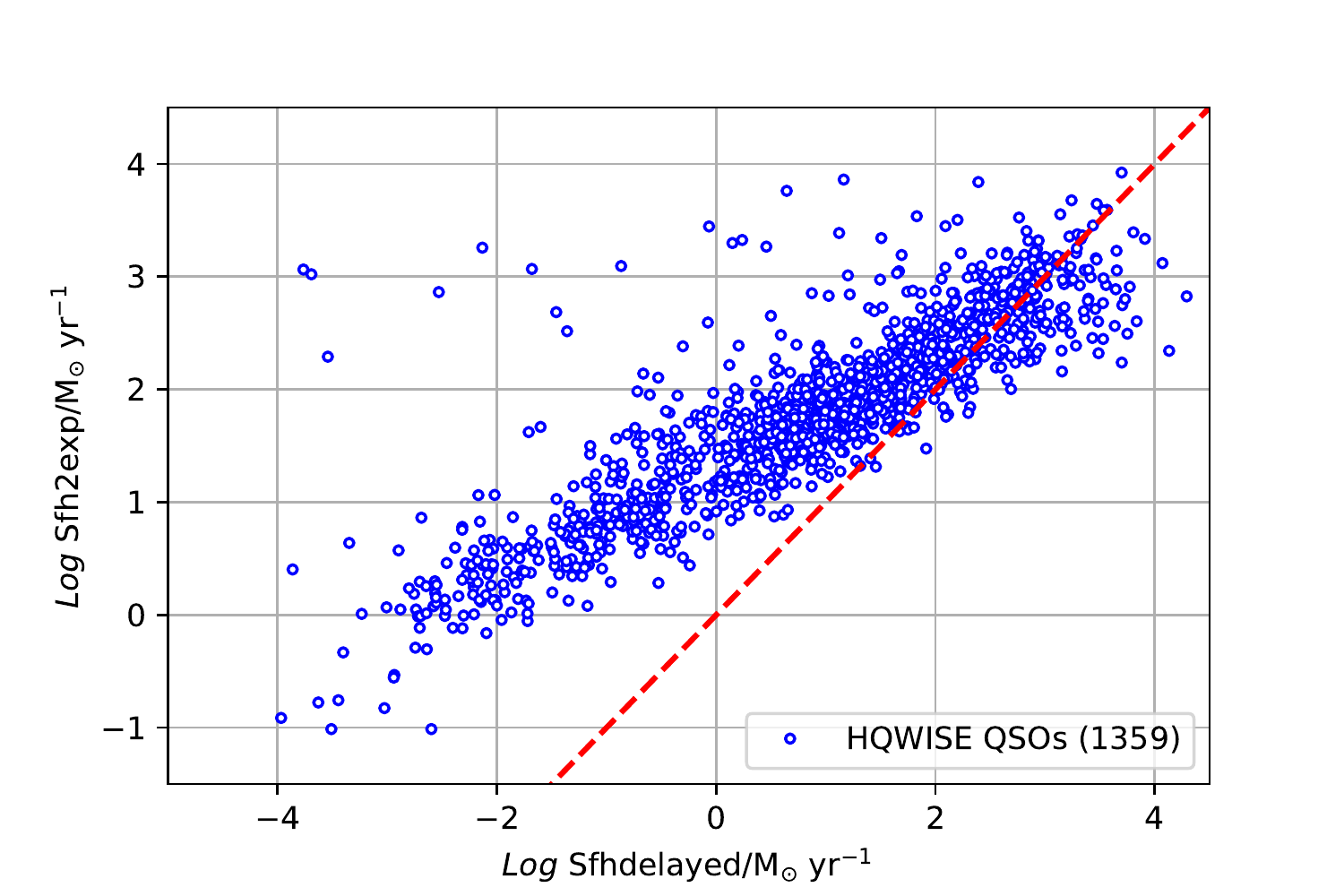}
    \caption{Comparing the \texttt{sfh2exp} SFRs with the \texttt{sfhdelayed} SFRs. Note that SFR values above 0 dex in the \texttt{sfhdelayed} models only appear in bin 2 ($z \sim 0.62$).}
    \label{CompSFR}
\end{figure}

Although both SFH functions predict high SFRs increasing with the redshifts, considering the possible difference in morphology, determining which SFR evolution is more realistic is not obvious. Comparing the two SFR solutions in Figure~\ref{CompSFR}, the \texttt{sfh2exp} SFRs are generally higher than the \texttt{sfhdelayed} SFRs. The differences being quite significant at low redshifts, varying from 100 to a few 10s below and up to $z = 1$ (first four bins), becoming slightly lower by only a few after bin \#4 (median $z \sim 1.11$), the differences decreasing in the ultimate 5 bins. 

 \begin{figure*}
 \includegraphics[width=\columnwidth]{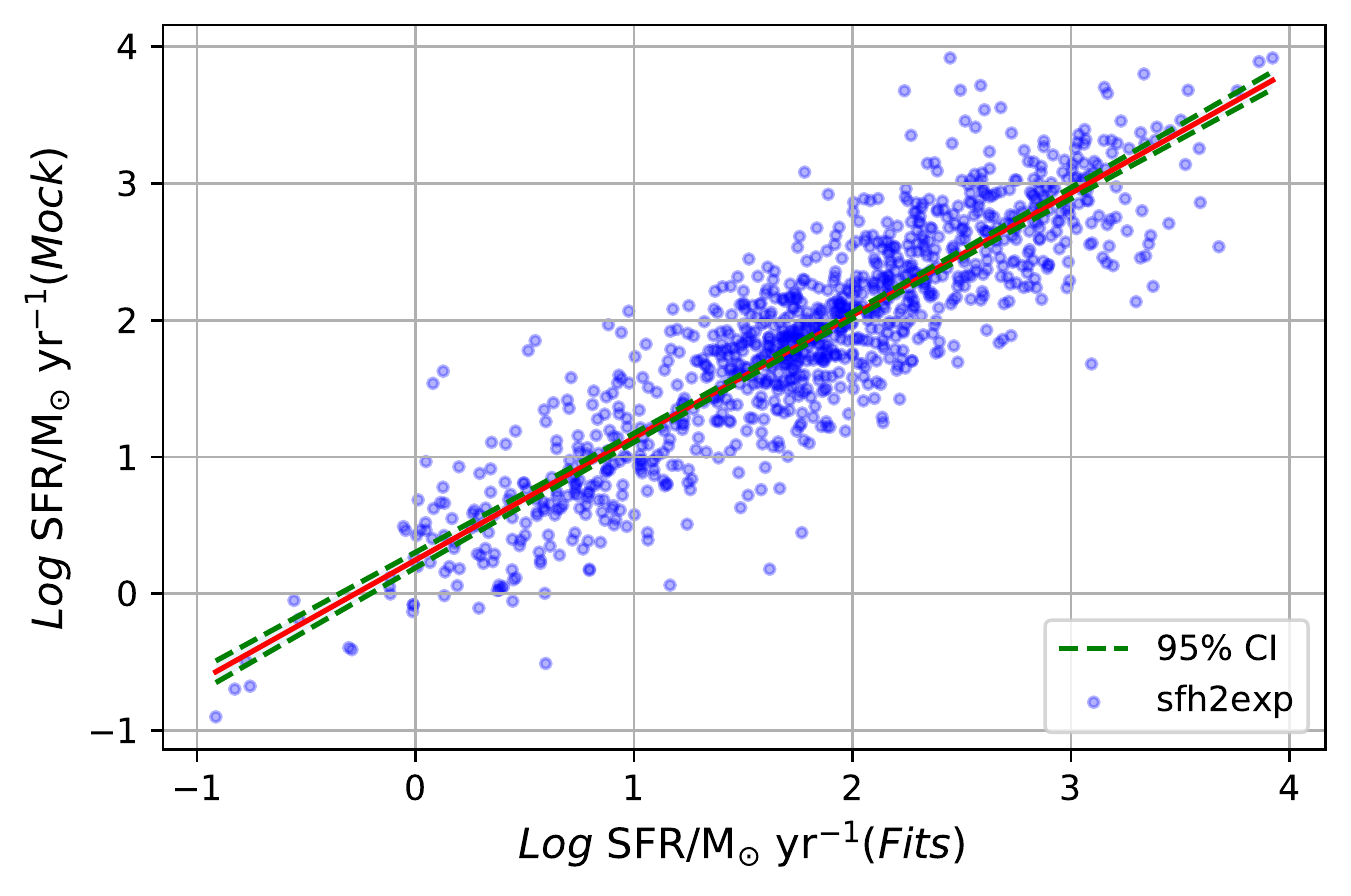}
 \includegraphics[width=\columnwidth]{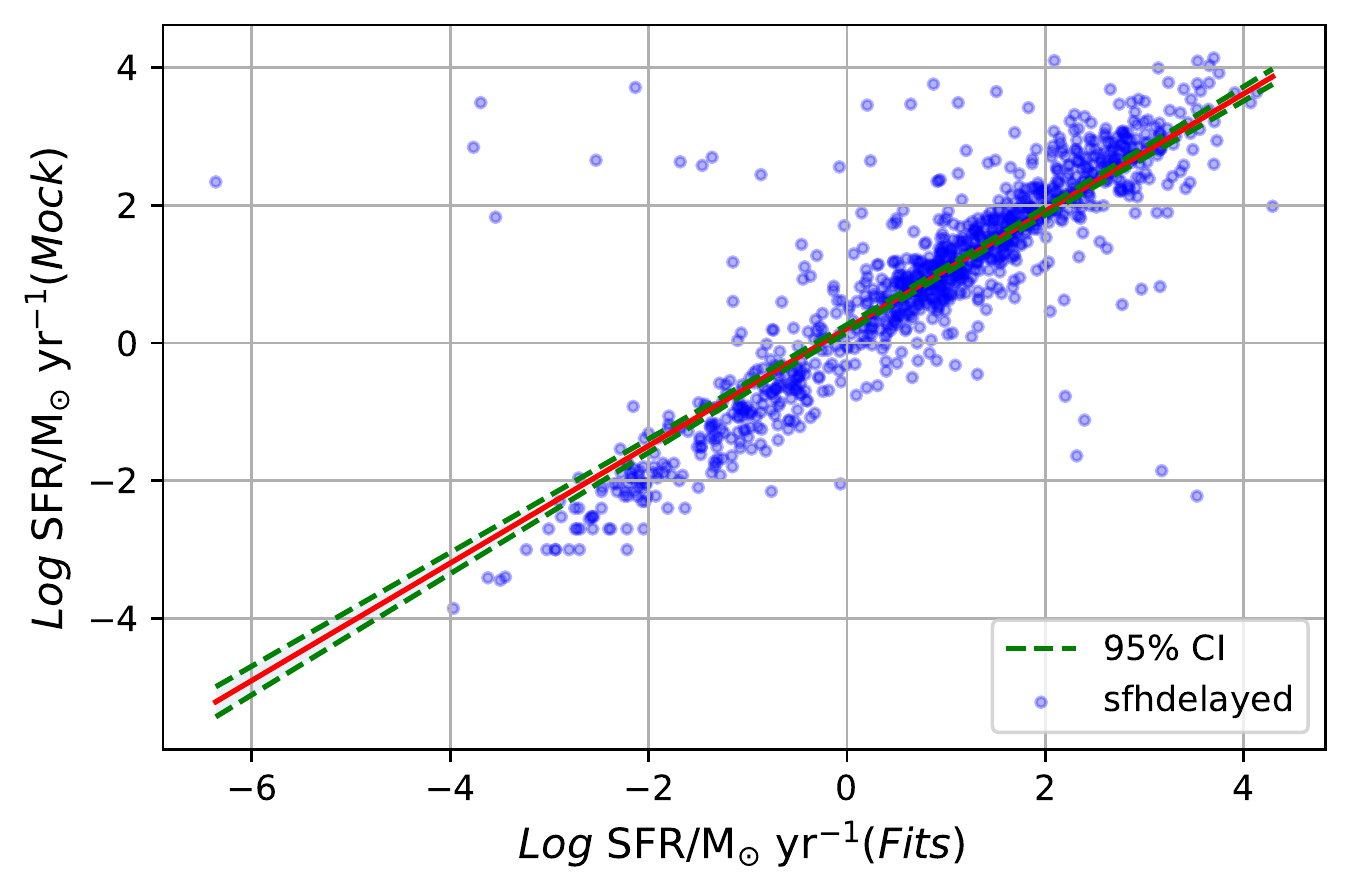}
 \includegraphics[width=\columnwidth]{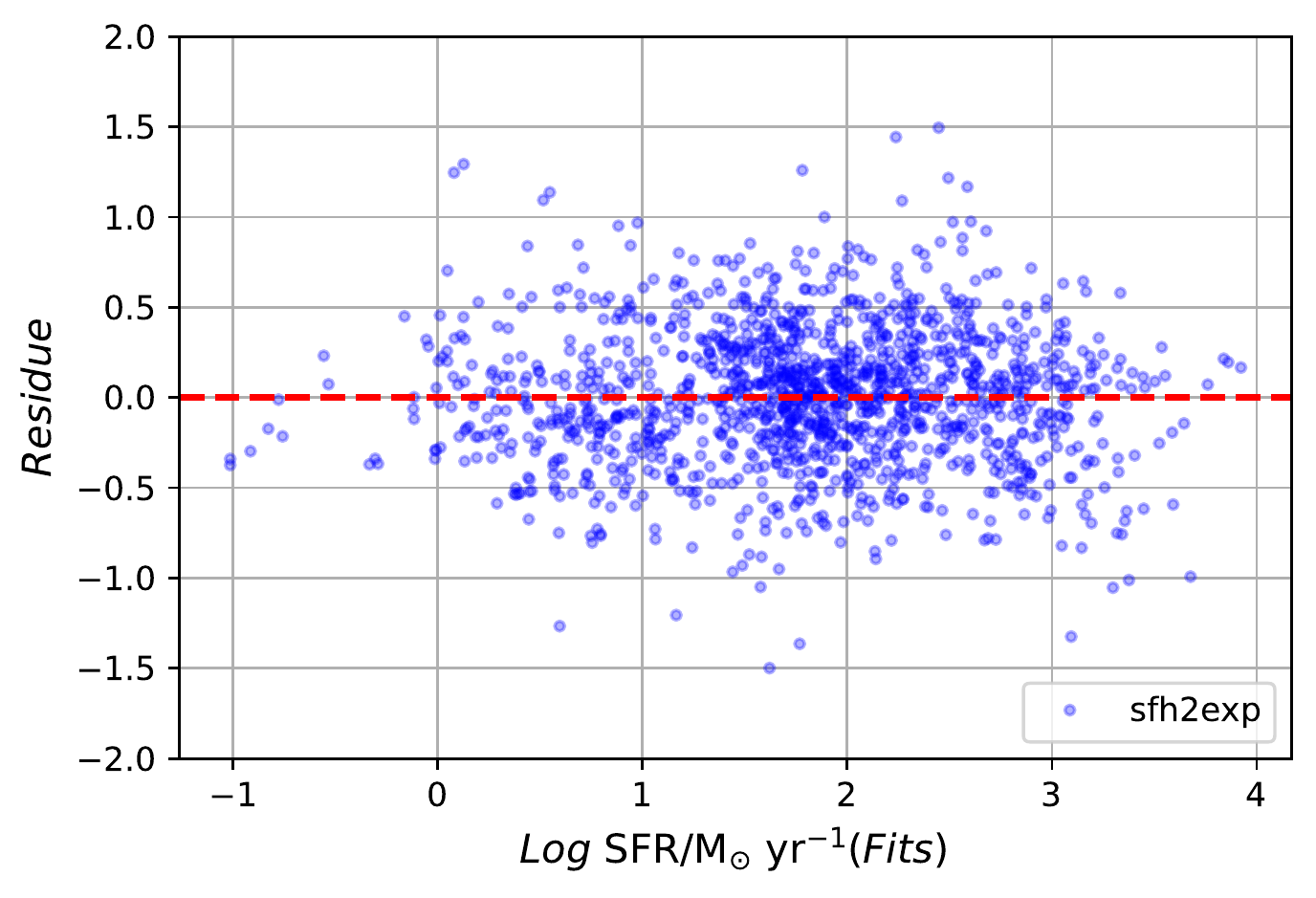}
 \includegraphics[width=\columnwidth]{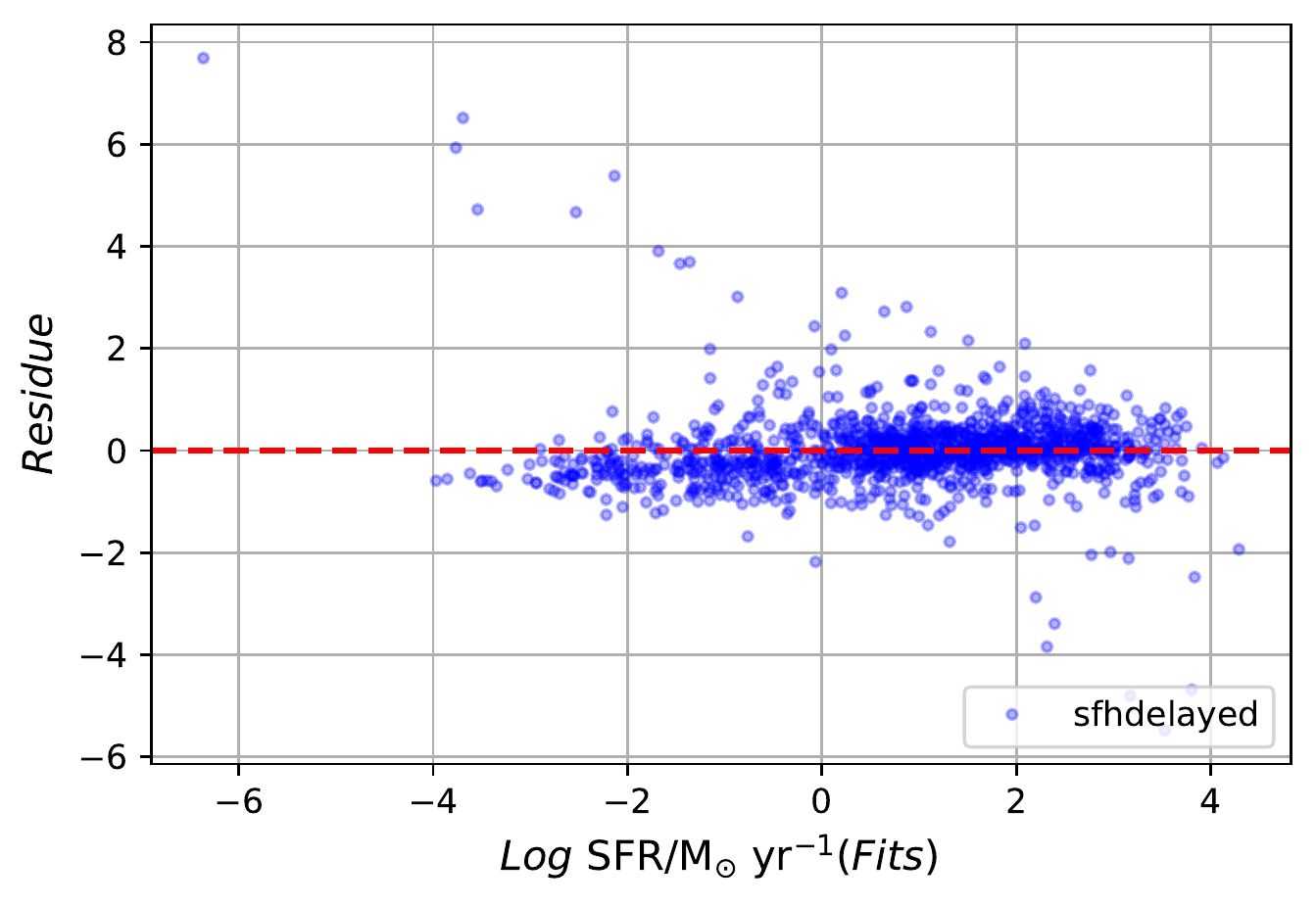}
    \caption{Comparing the SFRs fitted by our models with values from the mock analysis: for a) \texttt{sfh2exp}, b) \texttt{sfhdelayed}, and the two residual, for c) \texttt{sfh2exp}, and d) \texttt{sfhdelayed}. The line in red correspond to two simple linear regressions with in green the 95\% CI. The two simple linear regressions are $Y = 0.90 X + 0.22$ for \texttt{sfh2exp} and $Y = 0.87 X + 0.19$, with $R^2 = 0.80$ and $R^2 = 0.77$, respectively. The 95\% CI correspond to the intervals [0.88-0.93] and [0.84-0.90] respectively.}
    \label{WISEcolours}
\end{figure*}

To establish the reliability of each solution we used the option in \texttt{X-CIGALE} to generate two mock catalogues based on the best fit models for the two SFH functions \citep[see examples in][]{2021Mountrichas}. In Figure~\ref{WISEcolours}, we compare the mock SFRs with the fitted SFRs for the two SFH functions. The principle of this test is simple: using the fitted values as guesses, running \texttt{X-CIGALE} should converge to these values rapidly, showing that the solutions are robust. This is easily confirmed for the \texttt{sfh2exp}, the linear regression having a coefficient of correlation of R $=0.89$. Although the \texttt{sfhdelayed} also yields relatively good fits, with coefficient of correlation R $=0.88$, the residuals suggest the solutions at low redshifts where Log(SFR) $< 0$, are somewhat problematic, since the fitted SFRs are systematically below the mock SFRs and many QSOs show very large discrepant values at any redshift. This behavior is easy to explain: as the CT increases compared to the e-folding of the \texttt{sfhdelayed} function, the SFR cannot be kept as high as what the \texttt{sfh2exp} and mock values suggest. This also explains why the e-folding of the \texttt{sfhdelayed} function tend to be systematically higher than the e-folding of the \texttt{sfh2exp}. The difference between the two SFH functions has an important consequence on our interpretation. Although at high redshifts the results are degenerate, both requiring small e-folding and high SFR, the starburst like nature of QSOs hosts at low redshift \citep[compare with Figure~9 in][]{2012Kennicutt}, implies that, independent of the morphology of the galaxy host, their SEDs are best reproduced using an \texttt{sfh2exp} function \citep[e.g.,][]{2001Coziol}. Consequently, one would expect many of the nearby HQWISE QSO hosts to show a perturbed morphology, consistent with interactions or mergers \citep[e.g.,][]{1997Bahcall}. 

Based on the fact that the QSO hosts are starburst-like, and considering the solutions are degenerate at high redshifts, we decided to adopt the \texttt{sfh2exp} SFR in order to examine what could have been their evolution at different redshifts. 


\subsection{Possible presence of ULIRGs in QSOs at low redshifts}
\label{sub:ULIRG}

Adopting the \texttt{sfh2exp} SFRs, there is still one difficulty remaining, which is understanding why in Table~\ref{RES_sfhe} so many models (17\%) yield $\chi^2 > 3$. Since this happens more frequently at low redshifts, one possible explanation, consistent with spiral host galaxies, is that we see the QSOs edge-on, relative to the torus of dust (typical of Type~2 AGN). Alternatively, an early phase of evolution during which the BH is enshrouded by a veil of dust, as expected for ULIRGs, might also explain the observations. The fact that we find many HQWISE QSOs in Figure~\ref{fig1} with W2-W3 colours comparable to QSOs at high redshifts, where the SFRs were confirmed by our models to be high, would support such interpretation. According to the ULIRGs hypothesis, however, one would also expect these QSOs to have both higher than normal IR fluxes and SFRs \citep[e.g.,][]{2009Veilleux}. 

To test the above two possibilities, we have isolated all the QSOs with $\chi^2 > 3$ in Table~\ref{RES_sfhe} and run \texttt{X-CIGALE}, changing the $\psi$ angle using the smallest values possible. The medians in each redshift bin for these new models are reported in Table~\ref{RESsfh2exp_spy}. Comparing col.~5 with col.~4, we do see a significant improvement in the fits but only until bin \#6, after which the improvement becomes marginal, the new $\chi^2$, in fact, being worst in the last three bins. 

In general, lowering $\psi$ produces redder WISE colours and slightly higher SFRs but no systematic changes of the other parameters (most specifically, those related to extinction). Although these changes could be seen as favoring the ULIRG hypothesis, the fact that we see a significant improvement of the fits only at low redshifts could rather be interpreted as consistent with the geometry hypothesis, since AGN in spiral host galaxies are expected to be more numerous at low than high redshifts. In quantitative terms, the fraction of spiral-like hosts or edge-on (EO) QSOs up to $z \sim 1.6$ would amount to $\sim 19$\%, which is roughly consistent with what is usually observed in morphological studies of QSO hosts in nearby samples \citep[e.g.,][]{2010Letawe}.  

\begin{figure}
\includegraphics[width=\columnwidth]{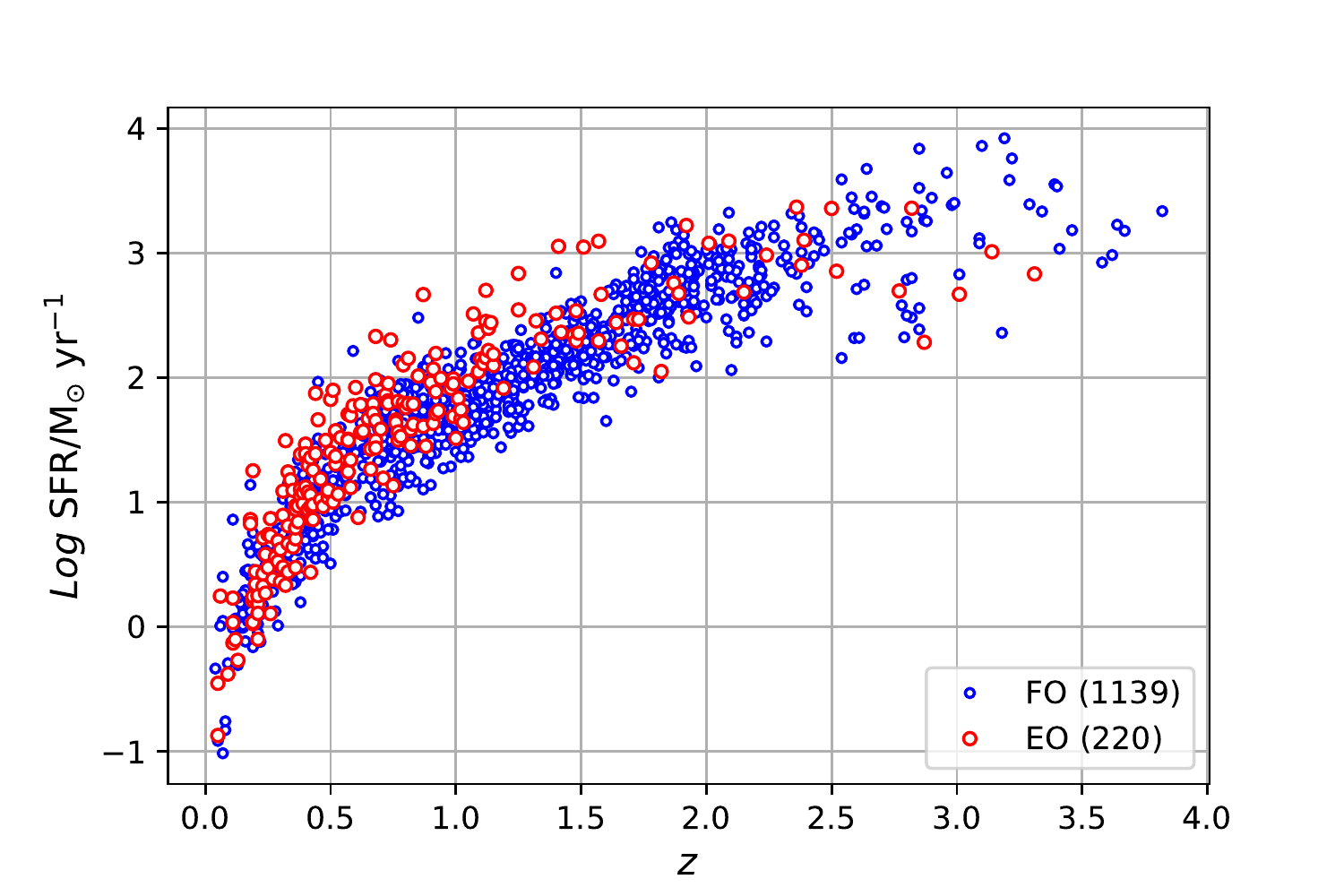}
   \caption{Variation of SFR with the redshift: blue, face-on (FO), red, edge-on (EO) QSOS.}
    \label{SFRvsz2}
\end{figure}

Adopting the \texttt{sfh2exp} function and correcting for EO cases we trace the variation of SFRs as a function of the redshift in Figure~\ref{SFRvsz2}. The models predict a significant increase of SFR by a factor 100 from $z = 0.5$ to $z \sim 2.5$, with no convincing evidence for a plateau or turn-over. 

\begin{table*}
\caption{\small Best fitted results for \texttt{sfh2exp} with $\chi^2 > 3$ varying the psy angle}
\label{RESsfh2exp_spy}
\begin{tabular}{cccccccccccccccc}
\hline
\hline
 (1) & (2) & (3) & (4)       & (5)      & (6)    & (7)   & (8)   & (9)   & (10)     & (11)      & (12)     & (13)     & (14)                      & (15)              & (16) \\
 bin & z   & N   & $\chi^2$  & $\chi^2$ & $\psi$ & W1-W2 & W2-W3 & O.A.  & $\delta$ & f$_{AGN}$ & E$_{BV}$ & E$_{BV}$ & Log(SFR)                  & $\tau_{\rm main}$ & T$_{dust}$  \\
     &     &     &  previous &    new   & (deg)  &       &       & (deg) &          &           & Lines    & Factor   & (M$_\odot\ {\rm yr}^{-1}$)& (Myr)             & (K) \\
\hline\noalign{\smallskip}
0   & 0.19 & 28 &  8.0 &  1.1 &  20 & 1.10 & 3.29 &  60 & -0.7 & 0.9 & 0.70 & 0.50 & 0.24 &  100 & 100 \\
1   & 0.39 & 60 &  6.4 &  1.4 &  20 & 1.11 & 3.19 &  60 & -0.7 & 0.9 & 0.50 & 0.75 & 0.99 & 1000 & 100 \\
2   & 0.66 & 38 &  5.5 &  2.1 &  20 & 1.24 & 3.22 &  60 & -1.0 & 0.9 & 0.70 & 0.75 & 1.58 & 1000 & 100 \\
3   & 0.86 & 32 &  6.7 &  3.4 &  30 & 1.25 & 3.25 & 100 & -0.5 & 0.9 & 0.70 & 0.75 & 1.75 &  100 & 100 \\
4   & 1.11 & 21 &  6.2 &  5.4 &  30 & 1.27 & 3.38 &  60 &  0.0 & 0.9 & 0.70 & 0.75 & 2.67 &  500 & 100 \\
5   & 1.40 & 11 &  7.3 &  6.0 &  30 & 1.16 & 3.23 &  60 &  0.0 & 0.9 & 0.70 & 0.75 & 2.36 & 1000 & 100 \\
6   & 1.60 &  9 &  7.8 &  5.8 &  30 & 1.34 & 3.31 &  60 &  0.0 & 0.9 & 0.70 & 0.75 & 2.50 &  500 & 100 \\
7   & 1.88 &  7 & 15.3 & 13.3 &  30 & 1.23 & 3.43 &  60 &  0.0 & 0.5 & 0.70 & 0.75 & 2.72 &  300 & 100 \\
8   & 2.12 &  5 & 18.8 & 18.5 &  25 & 1.26 & 3.61 &  60 &  0.0 & 0.5 & 0.70 & 0.75 & 3.03 &  100 & 300 \\
9   & 2.38 &  5 & 16.5 & 18.0 &  30 & 1.21 & 3.52 &  60 &  0.0 & 0.5 & 0.70 & 0.75 & 3.10 &  550 & 500 \\
10  & 2.51 &  2 & 22.5 & 17.6 &  25 & 1.47 & 3.69 & 100 &  0.0 & 0.5 & 0.70 & 0.75 & 3.11 &  550 & 750 \\
11  & 2.85 &  8 &  5.2 &  7.7 &  10 & 1.27 & 5.26 & 140 & -0.5 & 0.9 & 0.70 & 0.50 & 2.70 &  550 & 500 \\
12  & 3.08 &  4 &  5.8 &  6.9 &  10 & 1.31 & 4.30 &  80 &  0.0 & 0.9 & 0.30 & 0.50 & 2.84 &  300 & 750 \\
13  & 3.31 &  1 &  6.5 &  8.2 &  10 & 1.29 & 3.74 &  60 &  0.0 & 0.9 & 0.70 & 0.50 & 2.83 &  100 & 100 \\
\hline
\end{tabular}
\end{table*}

\section{Discussion: connecting the formation of the galaxy host to the formation of its SMBH}
\label{Disc}

The two most important results using \texttt{X-CIGALE} for our sample of QSOs are: 1- independently of the SFH function (morphology), the e-folding is small, of the order 1 Gyrs or less, and 2- the SFRs are relatively high and increasing with the redshift. These results imply that the galaxy hosts of HQWISE QSOs formed their star rapidly in a phase similar to starburst. What remains to be determined, however, is how the formation of the stellar populations are related to the formation of the SMBHs.  

Considering the rapid formation of both the SMBH and galaxy host stellar populations, and assuming the M$_{BH}-\sigma$ relation is established early during the formation of the galaxy, it seems natural to assume the relation established for AGN at low redshift extend at high redshifts. What should we expect then for the mass of the galaxy host, or most specifically the mass of its bulge, M$_{BH}-$M$_{bulge}$? According to Table~\ref{tab:bins} the SMBH in our sample of QSOs are quite massive, and consequently their galaxy host should also be massive. As a first approximation, we used the relation M$_{BH}$-M$_{bulge}$ recently determined by \citet{2022Ramsden} for a sample of SMBH with masses ranging from $10^6$ to $10^{10}$ M$_\odot$: 

\begin{equation}
\label{SFRrel}
\left( \frac{{\rm M}_{BH}}{10^9\ {\rm M}_\odot}\right) = (0.22) \left( \frac{{\rm M}_{bulge}}{10^{11}\ {\rm M}_\odot}\right)^{1.24}
\end{equation}

\begin{figure}
    \includegraphics[width=\columnwidth]{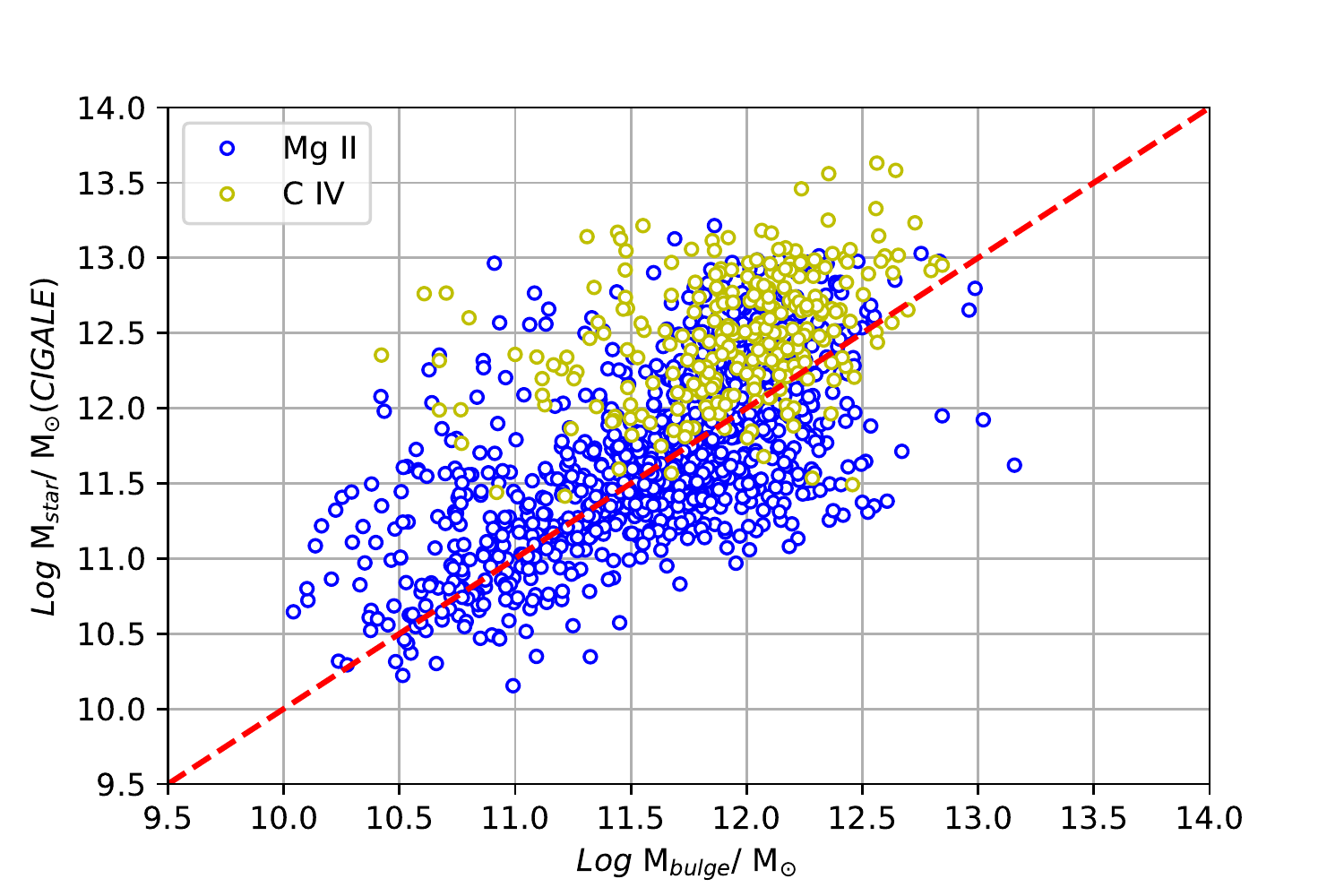}  
\caption{Comparing \texttt{X-CIGALE} stellar masses (young + old populations) with bulge masses obtained using Equation~2 as estimated by \citet{2022Ramsden}.}
    \label{McigaleVSMbulge}
\end{figure}

This yields masses ranging between $10^{10}$ and $\sim 10^{12.5}$ M$_\odot$ (very few above this value). Using the masses of stars (young + old populations) produced by \texttt{X-CIGALE} we can compare them with the masses obtained using Equation~2. This is done in Figure~\ref{McigaleVSMbulge}. Based on the one-to-one relation, we conclude that \texttt{X-CIGALE} masses are comparable to those produced by the relation of \citet{2022Ramsden}, which suggests that our SED solutions using \texttt{X-CIGALE} are consistent with the idea of a universal M$_{BH}-\sigma$ (at least up to $z \sim 4$). In general, the \texttt{X-CIGALE} masses seem slightly over-estimated above 11.5 dex. For this reason, we choose to use the masses determined by Equation~2 for the rest of our discussion.

\begin{figure}
    \includegraphics[width=\columnwidth]{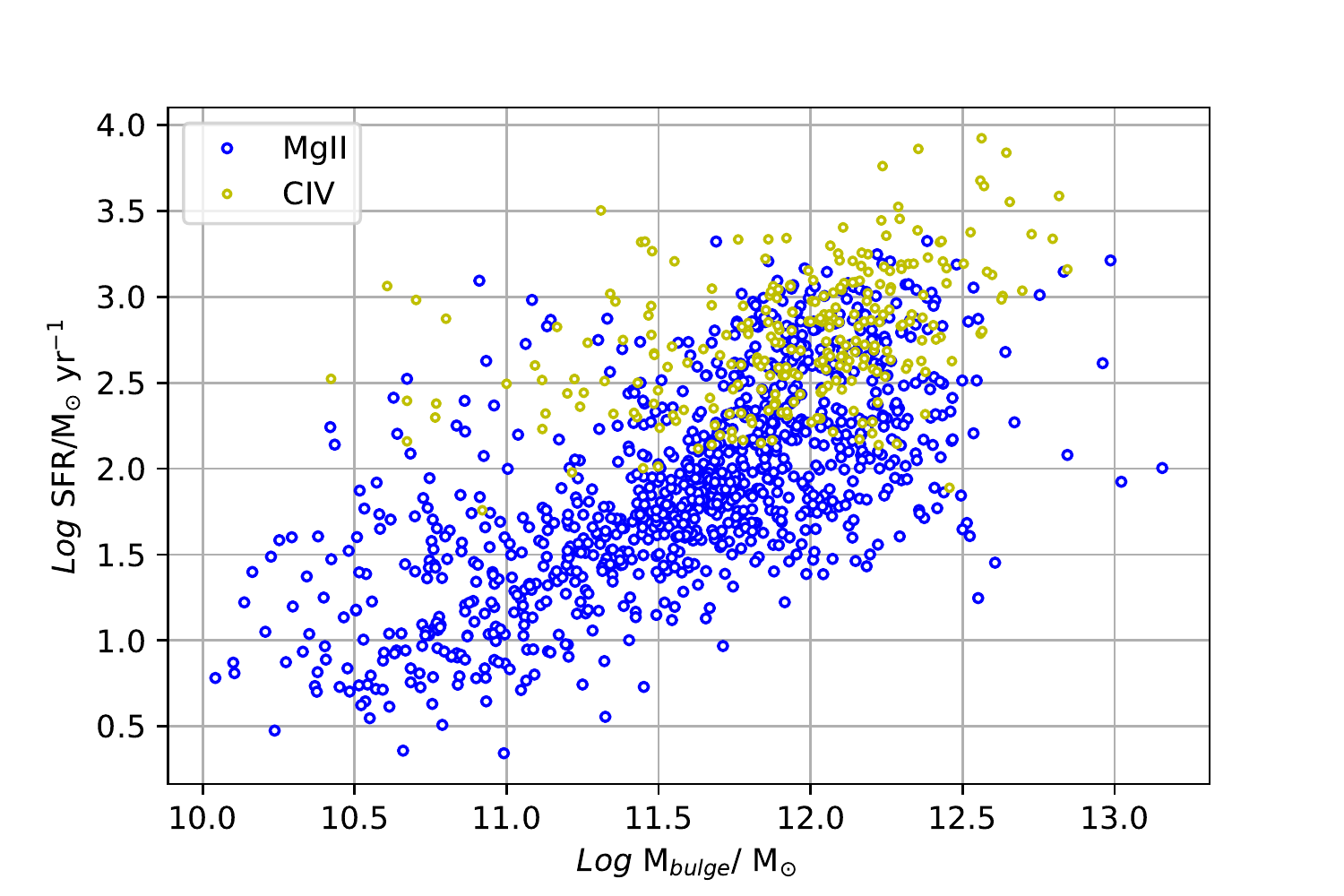}
\caption{Variation of SFR as function of bulge mass.}
    \label{SFRvsBulge}
\end{figure} 

\begin{figure}
    \includegraphics[width=\columnwidth]{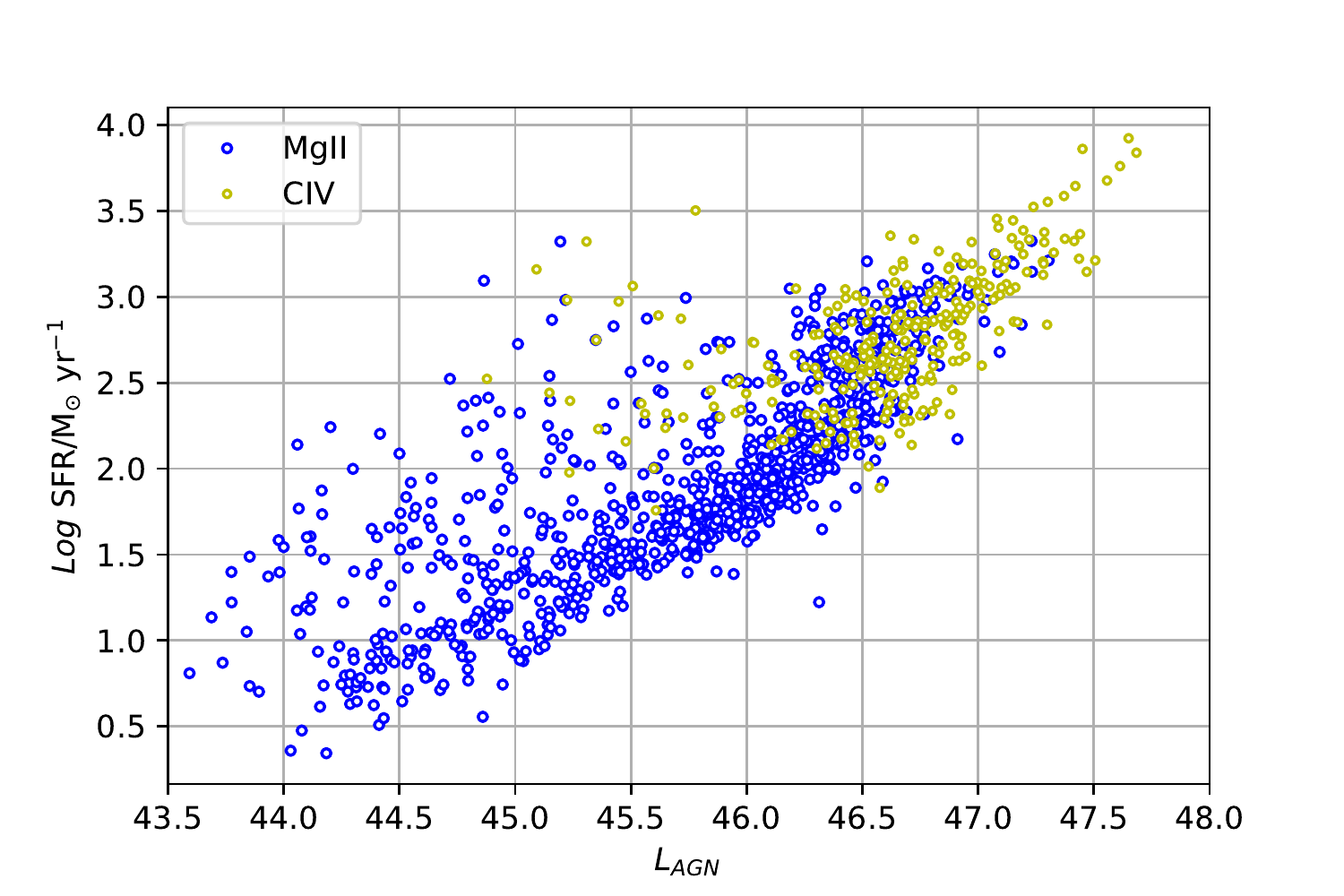}
    \includegraphics[width=\columnwidth]{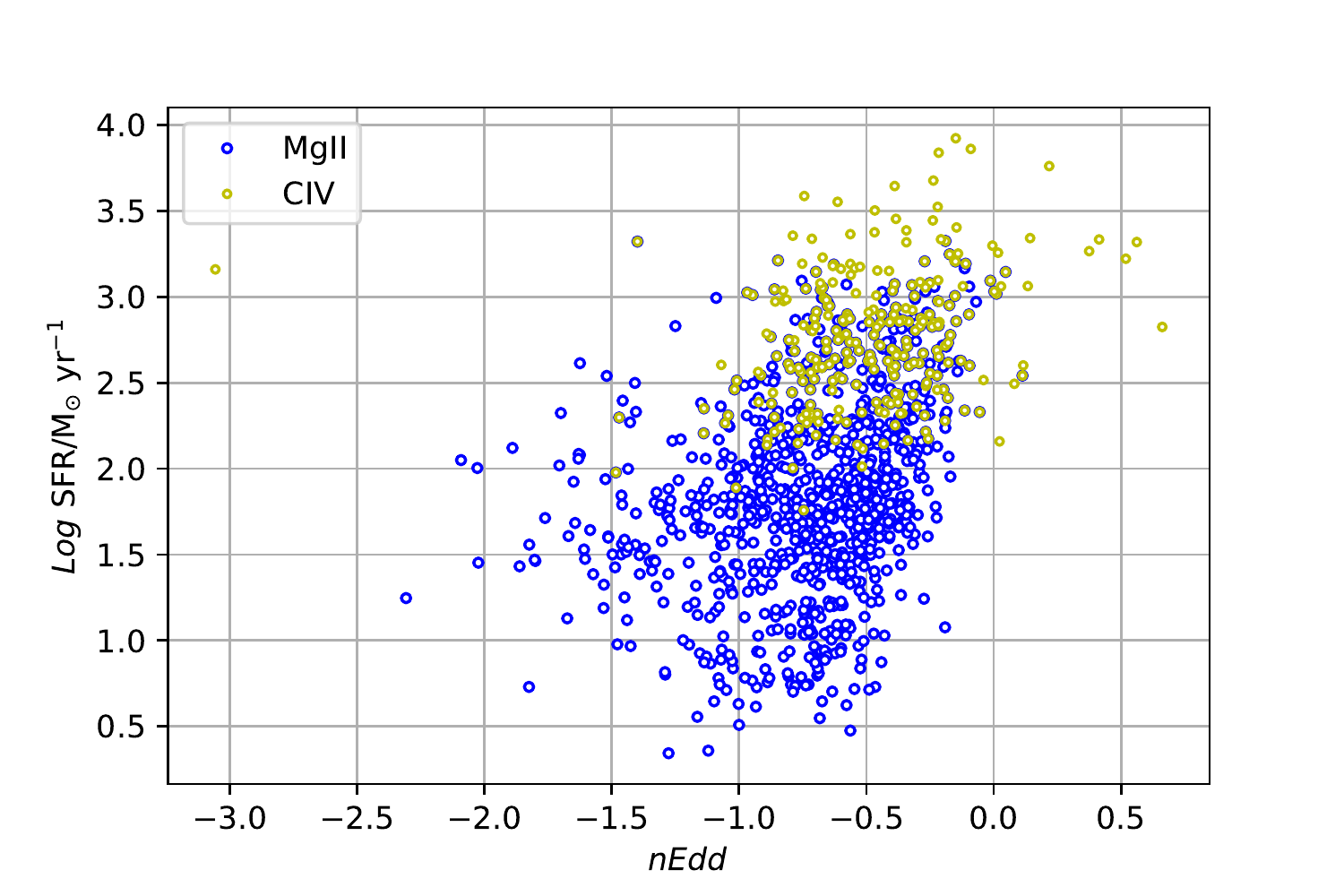}       
\caption{Variation of SFR as function of AGN luminosity (upper panel) and Eddington ratio (lower panel).}
    \label{SFRvsAGN}
\end{figure} 

Tracing in Figure~\ref{SFRvsBulge} the variation of SFR with the mass of the bulge, there is a clear trend for the SFR to increase with the mass, a factor 100 increase in mass corresponding to an increase by 1,000 in SFR. 
Because the BH mass in our sample of QSOs increases with the redshift (Figure~\ref{fig2}), this suggests that more massive BHs form in more massive galaxies at high redshifts, a phenomenon known as downsizing in the literature and which is accepted as a trait of hierarchical structure formation \citep[cf.][]{2006Neistein}.

However, the dispersion in SFR is large, especially after $\sim 11.7$\ dex in bulge mass. Note also that this dispersion does not depend on the emission line used, since we already observe it for \ion{Mg}{ii}. What explains this dispersion is not clear. This could either imply a variation in SFR related to different phases of evolution, or different SFHs due to different host morphologies (due to differences in environment densities). At 11.7~dex in bulge mass the BH mass is $10^{9.25}\ {\rm M}_\odot$, which in Figure~\ref{fig2} is a typical value in bin \#4 ($z \sim 1.1$), that is, close to where we observe the W1-W2 inflection and a possible change in SFH at high redshift, from late to early; due to the small e-folding and high SFR, the two solutions become indistinguishable at high redshifts.

In Figure~\ref{SFRvsAGN} we examine how the SFR correlates with two parameters related to the AGN activity: the luminosity (upper panel) and Eddington ratio (lower panel). In general, and clearer than for the bulge mass, the SFR increases with the luminosity. There also seems to be a slight inflection in SFR above L$_{AGN} = 10^{46}$ erg s$^{-1}$, the SFR increasing more rapidly at high luminosity. In Table~\ref{fig2}, L$_{AGN} = 10^{46}$ erg s$^{-1}$ is once again typical in bin \#4. At high redshifts, consequently, the higher the AGN activity the higher the stellar formation activity. Since higher AGN luminosity implies higher radiation pressure, this result is a strong argument against the AGN quenching hypothesis: contrary to expectation, AGN feedback related to radiation pressure does not quench SF in the AGN host galaxies. 

\begin{figure}
	\includegraphics[width=\columnwidth]{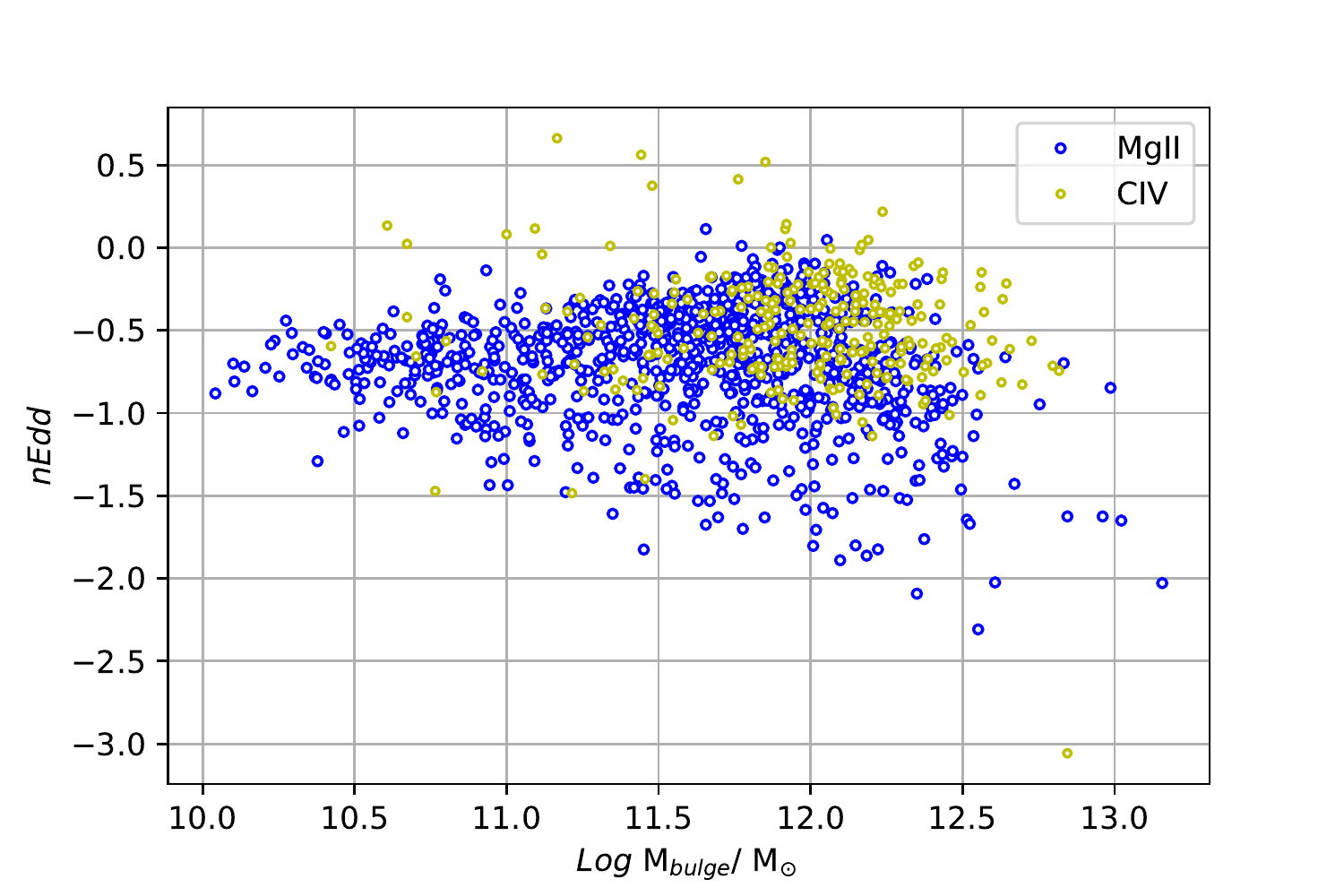}
    \caption{Variation of Eddington ratio with bulge mass.}
    \label{Nedd_vs_Mbulge}
\end{figure}

\begin{figure*}
	\includegraphics[width=0.95\textwidth]{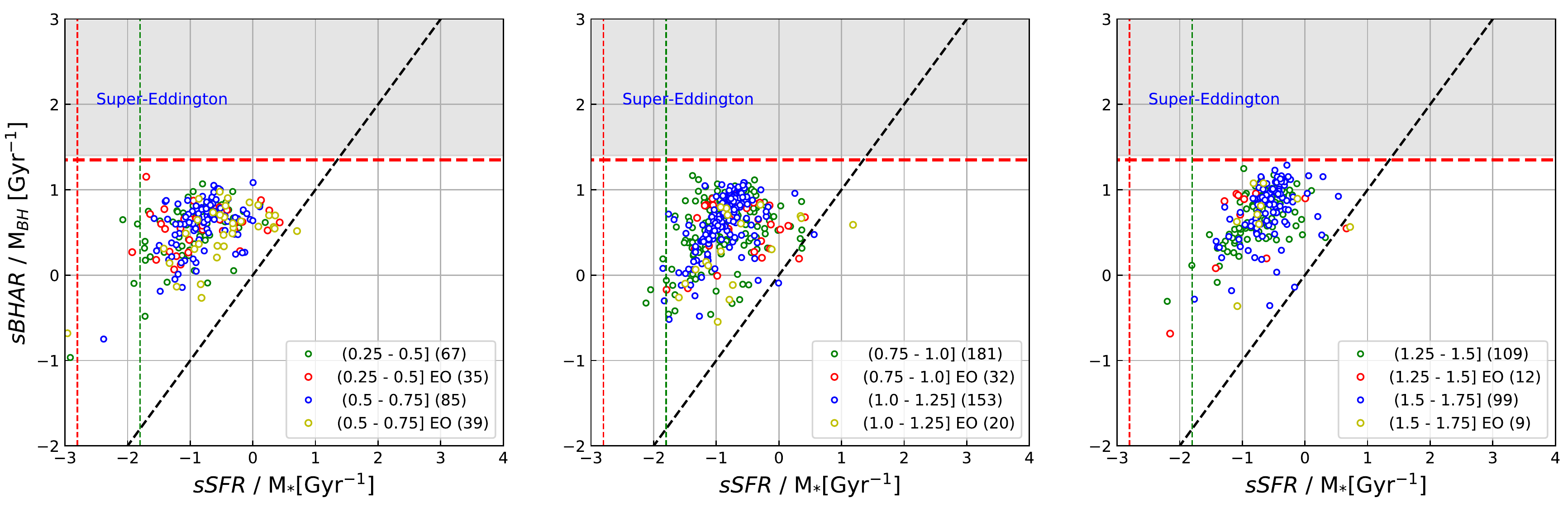}
	\includegraphics[width=0.95\textwidth]{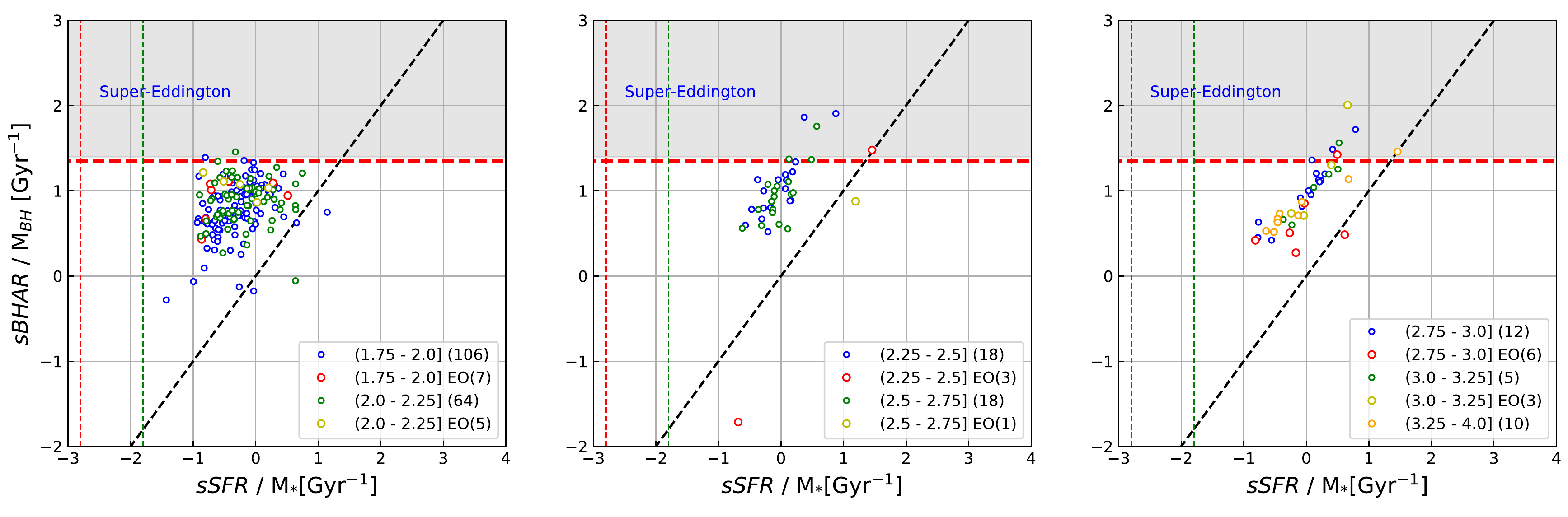}
    \caption{Variation of specific BH accretion rates (sBHAR) and specific star formation rates (sSFR) at different redshifts. The definition of Green valley and quenched region as defined by \citet{2017Bait} are also included: Green valley, $-1.8 >\ {\rm Log}(sSFR) > -2.8$, Quenched region, $ {\rm Log}(sSFR) \leq -2.8$.}
    \label{sDD}
\end{figure*}

Examining the lower panel, there is also a weak trend for the SFR to increase with the Eddington ratio. This is consistent with the relation between SFR and L$_{AGN}$, since higher accretion rates produce higher AGN luminosities. The reason why the relation is weak is also easy to comprehend, since, according to the definition of ${nEdd}$, any increase in BHAR is balanced by an increase in BH mass. Considering this factor, we can conclude that more massive BH form in more massive galaxies and that this implies higher BHAR and SFR. This is another argument against AGN quenching SF in their hosts. Another important trend we can deduced from Figure~\ref{SFRvsAGN} is that QSOs at high redshifts, those in yellow (identified as \ion{C}{IV}), have clearly higher ${nEdd}$, implying that BHAR increases with the redshift.  

From all the above relations, we conclude that the more massive the galaxies the more massive its BH, which translate into higher BHAR, and thus luminosity, and also higher SFR. This suggests that the mass of the galaxy host in QSOs is the primary parameter determining the AGN characteristics. 

Following this line of thought, we trace the Eddington ratio as a function of the mass of the bulges in Figure~\ref{Nedd_vs_Mbulge}. Only a few QSOs have super-Eddington ratios ($nEdd \ge 0$ dex). Those are Narrow line QSOs candidates candidates \citep[NLQSOs;][]{2001Krongold,2021Rakshit}, where lower than normal BH masses (not necessarily supper-Eddington accretion rates) explain their high $nEdd$. In general, therefore, the majority of the HQWISE QSOs have $nEdd$ below the Eddington limit, suggesting that the BH growth in these QSOs is strictly limited by radiation pressure. How do we reconcile this result with the masses of the galaxy hosts? Obviously, there is no associated limit in mass: the upper envelop of the $nEdd$-M$_{bulge}$ distribution peaks at a bulge mass of $10^{12}$ M$_\odot$, then decreases afterward as the mass continues to rise. Coincidentally, in galaxy formation simulations, $10^{12}$ M$_\odot$ is recognized as a limiting mass above which gas falling into an halo of dark matter would not have enough time to cool down and form stars \citep[][]{2018Wechsler,2021Das}. 

Although in Figure~\ref{SFRvsBulge} we do not see a clear cut in SFR above a mass of $10^{12}$ M$_\odot$, except for a few NLQSOs candidates, the SFR does not seem to grow much higher either. Assuming the galaxy host of QSOs form their stellar populations rapidly, over 1 Gyrs or less according to their e-folding, and considering the increase of SFR with the mass of the galaxies, then the higher the galaxy mass the faster must have been its formation \citep[][]{2021Zhang,2022Varma}. Such rapid formation would deplete the reservoir of gas rapidly and this would limit the mass of the bulge. A depletion rate of gas growing with the mass might also explain the decrease of BHAR above $10^{12}$ M$_\odot$ seen in Figure~\ref{SFRvsBulge}. Adopting \citet{2022Ramsden} relation, for example, the fraction of the mass of the BH to the mass of the bulge increases with the mass from 0.19\% to 0.5\%, suggesting faster BH growth, and thus more rapid depletion of gas to accrete as the mass of the bulge increases. This suggests that physical constraints on the formation process of galaxies could have limited the growth of the bulges and possibly also the growth in mass of their SMBHs. 

According to the AGN quenching hypothesis, the regulation of BH mass by AGN winds and outflows driven by radiation is assumed to have limited the mass of their host galaxies by quenching their SF. This process was proposed as the main explanation for the  M$_{BH}$-$\sigma$ relation. However, considering the rapid formation of the host galaxies the relation could be inverted: physical constraints on the formation process of galaxies limit the growth of the bulges and of their SMBHs. One way to distinguish which path the QSO hosts in our sample followed consists in comparing the variation of specific BH accretion rate (sBHAR) with specific star formation rate (sSFR) in different redshift bins. This is done in Figure~\ref{sDD}, starting with the lower bins (upper left panel) and ending with those in the highest bins (lower right panel). In this figure, the diagonal line represents the one-to-one relation, implying equal growth rates for the BH and its galaxy host. In our sample, very few QSOs are found on the diagonal, the majority being on the left side, suggesting BHs in QSOs grow more rapidly in mass than their host galaxies. Also indicated in Figure~\ref{sDD} are two different limits in sSFR identifying the Green valley and quenched region according to \citet{2017Bait}. Only a few QSOs in our sample cross the Green valley limit (only two cross the quenching limit), and they are all at low redshifts (bin \#6 or lower). The general trends as the redshift increases is for the sSFR to get farther away from the quenched region and for sBHAR to increase. There is consequently no evidence of AGN quenching of SF in the HQWISE QSOs.   
Comparing Figure~\ref{sDD} with Figure~11 in \citet{2021McDonald}, although the sBHAR of the HQWISE QSOs agree with those estimated by \citet{2016Dong}, 207 quasars detected in all three band of Herschel-SPIRE, our sSFR are much lower. Since the SFRs estimated by \texttt{X-CIGALE} are comparable to those estimated by these authors (see their Figure 6), the difference could only be in the masses of the bulges. This is confirmed in Figure~\ref{McigaleVSMbulge}, where the masses of our host galaxies are higher by a factor ten \citep[compared with Figure 8 in][]{2016Dong}. This is due to a difference of sample, not of model. Comparing now with the RGs (radio loud AGNs) observed by \citet{2014Drouart}, 70 galaxies at redshifts $1 < z < 5.2$, detected using the PACS and SPIRE instruments of Herschel Space Observatory, the main difference is in their sBHAR consistent with Super-Eddington QSOs, which are rare in the HQWISE QSOs. Since their BH masses are only slightly smaller than what we used, it is their higher BHARs that explain the difference. In a way, our solution makes more sense for radio quiet AGNs, since how a SMBH can accrete at higher rates than the Eddington limit is not physically explained. However, conditions could be more extreme in radio loud AGNs, which is what \citet{2014Drouart} considered. 

Also comparing with Figure~11 of \citet{2021McDonald}, the HQWISE QSOs at high redshifts have comparable sSFR as the Sy2s studied by \citet{2015Xu} at low redshift. However, the HQWISE QSOs have also higher sSFR than the mixed sample of local Seyfert studied by \citet[][]{2012Diamond-Stanic}. Once again, the main difference with these two samples seem to be in the masses of the galaxy hosts. Despite these differences, however, what is generally observed for AGNs is a sequence in sBHAR-sSFR parallel and to the left of the one-to-one relation suggesting that the AGN phenomenon is characteristics of a phase in galaxy evolution where the BHs grows in mass more rapidly than their galaxy hosts.

Considering the whole family of AGNs, we believe that the sequences in evolution left to the diagonal is physically significant. Depending how we estimate the stellar mass, the difference with a one-to-one relation growth could be smaller or larger, however, not enough to justify the AGN quenching hypothesis. Moreover, in the HQWISE QSOs the evolution of the sBHAR-sSFR sequence with redshift in Figure~\ref{sDD} is clear: after reaching the Eddington limit in bin \#6 the trend is for QSOs to move to the left (lower sSFR) and down (lower sBHAR), which is consistent with a decrease of sSFRs and sBHAR as the galaxy and BH mass decrease and CT increases, which is consistent with downsizing. Considering the fast formation of the bulge, the present state of QSOs to the left of the one-to-one relation could imply that the sSFR must have decreased more rapidly than the sBHAR. Since we see no evidence of AGN quenching in our sample, a rapid formation of the bulge, implying higher SFRs a few Gyrs before we observe the QSOs (as indicated by the e-folding), would decrease the reservoir of gas rapidly, reducing the sSFR below the one-to-one relation with sBHAR. A difference in evolution time scales might also be easy to justify physically, since the BH growth in a galaxy not only depends on the reservoir of gas a galaxy attracts through its mass but also on funneling this gas towards the center of the galaxies and how fast this gas is accreted by the BH, two processes which could easily be limited/regulated by radiation pressure through AGN feedback. 

In general, therefore, our results suggest that quenching of star formation through high astration rates (high star formation efficiencies) is the main process limiting the mass of galaxies during their formation. Consistent with downsizing, massive galaxies forming before less massive ones, we would thus naturally expect to see the efficiency in star formation to increase at high redshifts \citep{2022Inayoshi,2022Yang}, also possibly explaining massive galaxies forming at earlier epoch than previously expected according to the present cosmological paradigm \citep{2022Robertson,2022Curtis-Lake,2022Finkelstein}. 

 

\section{Conclusions}
\label{Con}

The modelling program \texttt{X-CIGALE} \citep{2019Boquien,2022Yang} integrates different physical modules, representing the quintessence of our understanding about galaxy formation and evolution processes, to reconstruct from various data available in different passbands, from the FUV to the FIR, the SED of galaxies at different redshifts, allowing us to better understand their specific assembly histories. By applying \texttt{X-CIGALE} on a carefully selected SDSS-WISE sample of 1,359 QSOs, within a range $0 < z \leq 4$, we were able to show that their host galaxies, 1- have high SFRs, consistent with starburst galaxies, that rise with the redshift and the mass of the galaxy hosts, and 2- based on their e-folding these hosts must have formed the bulk (69\%) of their stars rapidly, within 50 to 1000 Myrs of their formation. 

Some particularities of the SEDs also worth mentioning: 1- polar dust emission is present in all the QSOs at any redshift, 2- in 19\% of the QSOs at $z < 1.6$, small angle of view (20-30 degrees) are favored, consistent with a dust torus seen edge on, suggesting that 19\% of the QSO galaxy hosts at low redshifts have a spiral morphology, or are consistent with hidden AGNs in low redshift ULIRGs. 

Comparing the SFRs in the QSO hosts with the characteristics of their SMBHs, we were able to deduce that, 1- the SFR increases with the mass of the bulge and redshift, in a way that is consistent with downsizing, more massive galaxies, hosting more massive BHs, forming at higher redshifts than less massive galaxies, hosting less massive BHs, and 2- the SFR increases with the BHAR and AGN luminosity, suggesting there is no evidence of AGNs quenching SF in their hosts. 

Finally, comparing the sSFRs with sBHARs at different reshifts, we found that, 1- the sSFR is lower than the sBHAR, which implies that QSOs are a special phase in the evolution of galaxies during which the BH grow more rapidly in mass than its host galaxy, and 2- the sSFR and sBHAR increase in the past, the latter being limited by luminosity pressure at the Eddington limit. 

A simple interpretation in terms of variation of SFR consistent with the hierarchical model of galaxy formation and downsizing suggests that galaxy hosts of QSOs form their stars rapidly, exhausting their reservoir of gas over shorter time scales than the growth in mass of their SMBH, explaining why their sSFR are observed to be lower than their sBHAR. A higher efficiency of star formation in the past, rising with the mass of the galaxies, makes quenching by astration the dominant process by which the mass of galaxies is limited. 


\section*{Acknowledgements}

K. A. Cutiva-Alvarez acknowledges CONACyT for its support through grant CVU 940597. She would also like to thank Dr. Karla A. Alamo-Mart\'{\i}nez for helping her with the installation of \textsl{X-CIGALE}, her friends and colleagues, at Guanajuato, Dr. Aitor C. Robleto-Or\'us for the open discussions about her work and helps in learning Pyhton, and at Guadalajara, Juan Pablo Gutierrez for his support and guidance in programming with python. Finally, she would like to acknowledge her compatriot and friend Dr. Andres Felipe Ramos, for clarifying some of her doubts in using \textsl{X-CIGALE}, as well as Dr. Guang Yang and Dr. Jianwei Lyu for their openness in discussing some aspects of her study with them, most specifically, emphasizing the importance of polar dust components in AGNs. For his part, J. P. Torres-Papaqui acknowledges DAIP-UGTO (Mexico) for grant support 0077/2021. 

This research has made use of the VizieR catalogue access tool, CDS, Strasbourg, France (DOI : 10.26093/cds/vizier). The original description of the VizieR service was published in 2000, A\&AS 143, 23. Funding for SDSS-III has been provided by the Alfred P. Sloan Foundation, the Participating Institutions, the National Science Foundation, and the U.S. Department of Energy Office of Science. The SDSS-III web site is http://www.sdss3.org/.

\section*{Data Availability} The data used for this research are all available through the VizieR service (DOI: 10.26093/cds/vizier). 



\bibliographystyle{mnras}

\begin{thebibliography}{999}

\bibitem[Ramos Almeida et al.(2023)]{2023Almeida} Ramos Almeida, C., Esparza-Arredondo, D., Gonz{\'a}lez-Mart{\'\i}n, O., et al.\ 2023, Astronomy and Astrophysics, 669, L5. doi:10.1051/0004-6361/202245409
\bibitem[\protect\citeauthoryear{Bahcall et al.}{1997}]{1997Bahcall} Bahcall J.~N., Kirhakos S., Saxe D.~H., Schneider D.~P., 1997, ApJ, 479, 642. doi:10.1086/303926
\bibitem[\protect\citeauthoryear{Bait, Barway, \& Wadadekar}{2017}]{2017Bait} Bait O., Barway S., Wadadekar Y., 2017, MNRAS, 471, 2687. doi:10.1093/mnras/stx1688
\bibitem[\protect\citeauthoryear{Ba{\~n}ados et al.}{2018}]{2018Banados} Ba{\~n}ados E., Venemans B.~P., Mazzucchelli C., Farina E.~P., Walter F., Wang F., Decarli R., et al., 2018, Natur, 553, 473. doi:10.1038/nature25180
\bibitem[\protect\citeauthoryear{Barth et al.}{2003}]{2003Barth} Barth A.~J., Martini P., Nelson C.~H., Ho L.~C., 2003, ApJL, 594, L95. doi:10.1086/378735
\bibitem[\protect\citeauthoryear{Bellstedt et al.}{2020}]{2020Bellstedt} Bellstedt S., Robotham A.~S.~G., Driver S.~P., Thorne J.~E., Davies L.~J.~M., Lagos C. del P., Stevens A.~R.~H., et al., 2020, MNRAS, 498, 5581. doi:10.1093/mnras/staa2620
\bibitem[\protect\citeauthoryear{Bischetti et al.}{2021}]{2021Bischetti} Bischetti M., Feruglio C., Piconcelli E., Duras F., P{\'e}rez-Torres M., Herrero R., Venturi G., et al., 2021, A\&A, 645, A33. doi:10.1051/0004-6361/202039057
\bibitem[\protect\citeauthoryear{Boquien et al.}{2019}]{2019Boquien} Boquien M., Burgarella D., Roehlly Y., Buat V., Ciesla L., Corre D., Inoue A.~K., et al., 2019, A\&A, 622, A103. doi:10.1051/0004-6361/201834156
\bibitem[\protect\citeauthoryear{Boyle \& Terlevich}{1998}]{1998Boyle} Boyle B.~J., Terlevich R.~J., 1998, MNRAS, 293, L49. doi:10.1046/j.1365-8711.1998.01264.x
\bibitem[\protect\citeauthoryear{Bruzual \& Charlot}{2003}]{2003Bruzual} Bruzual G., Charlot S., 2003, MNRAS, 344, 1000. doi:10.1046/j.1365-8711.2003.06897.x

\bibitem[\protect\citeauthoryear{Calzetti et al.}{2000}]{2000Calzetti} Calzetti D., Armus L., Bohlin R.~C., Kinney A.~L., Koornneef J., Storchi-Bergmann T., 2000, ApJ, 533, 682. doi:10.1086/308692
\bibitem[\protect\citeauthoryear{Cavaliere \& Szalay}{1986}]{1986Cavaliere} Cavaliere A., Szalay A.~S., 1986, ApJ, 311, 589. doi:10.1086/164798
\bibitem[\protect\citeauthoryear{Cesarsky et al.}{1996}]{1996Cesarsky} Cesarsky D., Lequeux J., Abergel A., Perault M., Palazzi E., Madden S., Tran D., 1996, A\&A, 315, L305
\bibitem[\protect\citeauthoryear{Charlot \& Bruzual A}{1991}]{1991Charlot} Charlot S., Bruzual A G., 1991, ApJ, 367, 126. doi:10.1086/169608
\bibitem[\protect\citeauthoryear{Coziol, Doyon, \& Demers}{2001}]{2001Coziol} Coziol R., Doyon R., Demers S., 2001, MNRAS, 325, 1081. doi:10.1046/j.1365-8711.2001.04512.x
\bibitem[\protect\citeauthoryear{Coziol et al.}{2011}]{2011Coziol} Coziol R., Torres-Papaqui J.~P., Plauchu-Frayn I., Islas-Islas J.~M., Ortega-Minakata R.~A., Neri-Larios D.~M., Andernach H., 2011, RMxAA, 47, 361
\bibitem[\protect\citeauthoryear{Coziol et al.}{2014}]{2014Coziol} Coziol R., Torres-Papaqui J.~P., Plauchu-Frayn I., Andernach H., Neri-Larios D.~M., Ortega-Minakata R.~A., Islas-Islas J.~M., 2014, RMxAA, 50, 255
\bibitem[\protect\citeauthoryear{Coziol, Torres-Papaqui, \& Andernach}{2015}]{2015Coziol} Coziol R., Torres-Papaqui J.~P., Andernach H., 2015a, AJ, 149, 192. doi:10.1088/0004-6256/149/6/192
\bibitem[\protect\citeauthoryear{Coziol et al.}{2017}]{2017Coziol} Coziol R., Andernach H., Torres-Papaqui J.~P., Ortega-Minakata R.~A., Moreno del Rio F., 2017, MNRAS, 466, 921. doi:10.1093/mnras/stw3164
\bibitem[\protect\citeauthoryear{Croom et al.}{2009}]{2009Croom} Croom S.~M., Richards G.~T., Shanks T., Boyle B.~J., Strauss M.~A., Myers A.~D., Nichol R.~C., et al., 2009, MNRAS, 399, 1755. doi:10.1111/j.1365-2966.2009.15398.x
\bibitem[\protect\citeauthoryear{Curtis-Lake et al.}{2022}]{2022Curtis-Lake} Curtis-Lake E., Carniani S., Cameron A., Charlot S., Jakobsen P., Maiolino R., Bunker A., et al., 2022, arXiv, arXiv:2212.04568
\bibitem[\protect\citeauthoryear{Cutri et al.}{2021}]{2014Cutri} Cutri R.~M., Wright E.~L., Conrow T., Fowler J.~W., Eisenhardt P.~R.~M., Grillmair C., Kirkpatrick J.~D., et al., 2021, yCat, II/328

\bibitem[\protect\citeauthoryear{Daddi et al.}{2005}]{2005Daddi} Daddi E., Dickinson M., Chary R., Pope A., Morrison G., Alexander D.~M., Bauer F.~E., et al., 2005, ApJL, 631, L13. doi:10.1086/496918
\bibitem[\protect\citeauthoryear{Das, Pandey, \& Sarkar}{2021}]{2021Das} Das A., Pandey B., Sarkar S., 2021, JCAP, 2021, 045. doi:10.1088/1475-7516/2021/06/045
\bibitem[\protect\citeauthoryear{De Rosa et al.}{2014}]{2014DeRosa} De Rosa G., Venemans B.~P., Decarli R., Gennaro M., Simcoe R.~A., Dietrich M., Peterson B.~M., et al., 2014, ApJ, 790, 145. doi:10.1088/0004-637X/790/2/145
 \bibitem[\protect\citeauthoryear{Devecchi \& Volonteri}{2009}]{2009Devecchi} Devecchi B., Volonteri M., 2009, ApJ, 694, 302. doi:10.1088/0004-637X/694/1/302
\bibitem[\protect\citeauthoryear{Diamond-Stanic \& Rieke}{2012}]{2012Diamond-Stanic} Diamond-Stanic A.~M., Rieke G.~H., 2012, ApJ, 746, 168. doi:10.1088/0004-637X/746/2/168
\bibitem[\protect\citeauthoryear{Dietrich et al.}{2003a}]{2003DietrichA} Dietrich M., Hamann F., Shields J.~C., Constantin A., Heidt J., J{\"a}ger K., Vestergaard M., et al., 2003, ApJ, 589, 722. doi:10.1086/374662
\bibitem[\protect\citeauthoryear{Dietrich et al.}{2003b}]{2003DietrichB} Dietrich M., Hamann F., Appenzeller I., Vestergaard M., 2003, ApJ, 596, 817. doi:10.1086/378045
\bibitem[\protect\citeauthoryear{Dong \& Wu}{2016}]{2016Dong} Dong X.~Y., Wu X.-B., 2016, ApJ, 824, 70. doi:10.3847/0004-637X/824/2/70
\bibitem[\protect\citeauthoryear{Draine \& Li}{2007}]{2007Draine} Draine B.~T., Li A., 2007, ApJ, 657, 810. doi:10.1086/511055
\bibitem[\protect\citeauthoryear{Draine et al.}{2014}]{2014Draine} Draine B.~T., Aniano G., Krause O., Groves B., Sandstrom K., Braun R., Leroy A., et al., 2014, ApJ, 780, 172. doi:10.1088/0004-637X/780/2/172
\bibitem[\protect\citeauthoryear{Drouart et al.}{2014}]{2014Drouart} Drouart G., De Breuck C., Vernet J., Seymour N., Lehnert M., Barthel P., Bauer F.~E., et al., 2014, A\&A, 566, A53. doi:10.1051/0004-6361/201323310

\bibitem[\protect\citeauthoryear{Elvis et al.}{1994}]{1994Elvis} Elvis M., Wilkes B.~J., McDowell J.~C., Green R.~F., Bechtold J., Willner S.~P., Oey M.~S., et al., 1994, ApJS, 95, 1. doi:10.1086/192093
\bibitem[\protect\citeauthoryear{Engelbracht et al.}{2010}]{2010Engelbracht} Engelbracht C.~W., Hunt L.~K., Skibba R.~A., Hinz J.~L., Calzetti D., Gordon K.~D., Roussel H., et al., 2010, A\&A, 518, L56. doi:10.1051/0004-6361/201014677

\bibitem[\protect\citeauthoryear{Fan et al.}{2003}]{2003Fan} Fan X., Strauss M.~A., Schneider D.~P., Becker R.~H., White R.~L., Haiman Z., Gregg M., et al., 2003, AJ, 125, 1649. doi:10.1086/368246
\bibitem[\protect\citeauthoryear{Fan}{2006}]{2006Fan} Fan X., 2006, NewAR, 50, 665. doi:10.1016/j.newar.2006.06.077
\bibitem[\protect\citeauthoryear{Farrah et al.}{2002}]{2002Farrah} Farrah D., Verma A., Oliver S., Rowan-Robinson M., McMahon R., 2002, MNRAS, 329, 605. doi:10.1046/j.1365-8711.2002.04991.x
\bibitem[\protect\citeauthoryear{Ferrarese \& Merritt}{2000}]{2000Ferrarese} Ferrarese L., Merritt D., 2000, ApJL, 539, L9. doi:10.1086/312838
\bibitem[\protect\citeauthoryear{Finkelstein et al.}{2022}]{2022Finkelstein} Finkelstein S.~L., Bagley M.~B., Haro P.~A., Dickinson M., Ferguson H.~C., Kartaltepe J.~S., Papovich C., et al., 2022, ApJL, 940, L55. doi:10.3847/2041-8213/ac966e
\bibitem[\protect\citeauthoryear{Fritz, Franceschini, \& Hatziminaoglou}{2006}]{2006Fritz} Fritz J., Franceschini A., Hatziminaoglou E., 2006, MNRAS, 366, 767. doi:10.1111/j.1365-2966.2006.09866.x

\bibitem[\protect\citeauthoryear{Gold}{1967}]{1967Gold} Gold T., 1967, ApJ, 147, 832. doi:10.1086/149066
\bibitem[\protect\citeauthoryear{Graham et al.}{2011}]{2011Graham} Graham A.~W., Onken C.~A., Athanassoula E., Combes F., 2011, MNRAS, 412, 2211. doi:10.1111/j.1365-2966.2010.18045.x
\bibitem[\protect\citeauthoryear{G{\"u}ltekin et al.}{2009}]{2009Gultekin} G{\"u}ltekin K., Richstone D.~O., Gebhardt K., Lauer T.~R., Tremaine S., Aller M.~C., Bender R., et al., 2009, ApJ, 698, 198. doi:10.1088/0004-637X/698/1/198

\bibitem[\protect\citeauthoryear{Hamann \& Ferland}{1993}]{1993Hamann} Hamann F., Ferland G., 1993, ApJ, 418, 11. doi:10.1086/173366
\bibitem[\protect\citeauthoryear{Harrison et al.}{2018}]{2018Harrison} Harrison C.~M., Costa T., Tadhunter C.~N., Fl{\"u}tsch A., Kakkad D., Perna M., Vietri G., 2018, NatAs, 2, 198. doi:10.1038/s41550-018-0403-6
\bibitem[\protect\citeauthoryear{H{\"a}ring \& Rix}{2004}]{2004Haring} H{\"a}ring N., Rix H.-W., 2004, ApJL, 604, L89. doi:10.1086/383567
\bibitem[\protect\citeauthoryear{Heger \& Woosley}{2002}]{2002Heger} Heger A., Woosley S.~E., 2002, ApJ, 567, 532. doi:10.1086/338487
\bibitem[\protect\citeauthoryear{Hogan et al.}{2022}]{2022Hogan} Hogan L., Rigopoulou D., Garc{\'\i}a-Burillo S., Alonso-Herrero A., Barrufet L., Combes F., Garc{\'\i}a-Bernete I., et al., 2022, MNRAS, 512, 2371. doi:10.1093/mnras/stac520
\bibitem[\protect\citeauthoryear{Hopkins et al.}{2010}]{2010Hopkins} Hopkins P.~F., Younger J.~D., Hayward C.~C., Narayanan D., Hernquist L., 2010, MNRAS, 402, 1693. doi:10.1111/j.1365-2966.2009.15990.x

\bibitem[\protect\citeauthoryear{Inayoshi, Visbal, \& Haiman}{2020}]{2020Inayoshi} Inayoshi K., Visbal E., Haiman Z., 2020, ARA\&A, 58, 27. doi:10.1146/annurev-astro-120419-014455
\bibitem[\protect\citeauthoryear{Inayoshi et al.}{2022}]{2022Inayoshi} Inayoshi K., Harikane Y., Inoue A.~K., Li W., Ho L.~C., 2022, ApJL, 938, L10. doi:10.3847/2041-8213/ac9310
\bibitem[\protect\citeauthoryear{Isbell et al.}{2022}]{2022Isbell} Isbell J.~W., Meisenheimer K., Pott J.-U., Stalevski M., Tristram K.~R.~W., Sanchez-Bermudez J., Hofmann K.-H., et al., 2022, arXiv, arXiv:2205.01575

\bibitem[\protect\citeauthoryear{Jahnke \& Macci{\`o}}{2011}]{2011Jahnke} Jahnke K., Macci{\`o} A.~V., 2011, ApJ, 734, 92. doi:10.1088/0004-637X/734/2/92
\bibitem[\protect\citeauthoryear{Juarez et al.}{2009}]{2009Juarez} Juarez Y., Maiolino R., Mujica R., Pedani M., Marinoni S., Nagao T., Marconi A., et al., 2009, A\&A, 494, L25. doi:10.1051/0004-6361:200811415
\bibitem[\protect\citeauthoryear{Jiang et al.}{2007}]{2007Jiang} Jiang L., Fan X., Vestergaard M., Kurk J.~D., Walter F., Kelly B.~C., Strauss M.~A., 2007, AJ, 134, 1150. doi:10.1086/520811

\bibitem[\protect\citeauthoryear{Kennicutt \& Evans}{2012}]{2012Kennicutt} Kennicutt R.~C., Evans N.~J., 2012, ARA\&A, 50, 531. doi:10.1146/annurev-astro-081811-125610
\bibitem[\protect\citeauthoryear{Koz{\l}owski}{2017}]{2017Kozlowski} Koz{\l}owski S., 2017, ApJS, 228, 9. doi:10.3847/1538-4365/228/1/9
\bibitem[\protect\citeauthoryear{Krongold, Dultzin-Hacyan, \& Marziani}{2001}]{2001Krongold} Krongold Y., Dultzin-Hacyan D., Marziani P., 2001, qhte.conf, 273
\bibitem[\protect\citeauthoryear{Kurk et al.}{2007}]{2007Kurk} Kurk J.~D., Walter F., Fan X., Jiang L., Riechers D.~A., Rix H.-W., Pentericci L., et al., 2007, ApJ, 669, 32. doi:10.1086/521596

\bibitem[\protect\citeauthoryear{Labita, Treves, \& Falomo}{2008}]{2008Labita} Labita M., Treves A., Falomo R., 2008, MNRAS, 383, 1513. doi:10.1111/j.1365-2966.2007.12656.x
\bibitem[\protect\citeauthoryear{Lapi et al.}{2006}]{2006Lapi} Lapi A., Shankar F., Mao J., Granato G.~L., Silva L., De Zotti G., Danese L., 2006, ApJ, 650, 42. doi:10.1086/507122
\bibitem[\protect\citeauthoryear{Leitherer, Calzetti, \& Martins}{2002}]{2002Leitherer} Leitherer C., Calzetti D., Martins L.~P., 2002, ApJ, 574, 114. doi:10.1086/340902
\bibitem[\protect\citeauthoryear{Letawe, Letawe, \& Magain}{2010}]{2010Letawe} Letawe Y., Letawe G., Magain P., 2010, MNRAS, 403, 2088. doi:10.1111/j.1365-2966.2010.16244.x
\bibitem[\protect\citeauthoryear{Ling \& Yan}{2022}]{2022Ling} Ling C., Yan H., 2022, ApJ, 929, 40. doi:10.3847/1538-4357/ac57c1
\bibitem[\protect\citeauthoryear{Lupton, Gunn, \& Szalay}{1999}]{1999Lupton} Lupton R.~H., Gunn J.~E., Szalay A.~S., 1999, AJ, 118, 1406. doi:10.1086/301004
\bibitem[\protect\citeauthoryear{Lyu \& Rieke}{2022}]{2022Lyu} Lyu J., Rieke G., 2022, arXiv, arXiv:2205.14172

\bibitem[\protect\citeauthoryear{Madau, Pozzetti, \& Dickinson}{1998}]{1998Madau} Madau P., Pozzetti L., Dickinson M., 1998, ApJ, 498, 106. doi:10.1086/305523
\bibitem[\protect\citeauthoryear{Madau \& Dickinson}{2014}]{2014Madau} Madau P., Dickinson M., 2014, ARA\&A, 52, 415. doi:10.1146/annurev-astro-081811-125615
\bibitem[\protect\citeauthoryear{Magorrian et al.}{1998}]{1998Magorrian} Magorrian J., Tremaine S., Richstone D., Bender R., Bower G., Dressler A., Faber S.~M., et al., 1998, AJ, 115, 2285. doi:10.1086/300353
\bibitem[\protect\citeauthoryear{Marshall et al.}{2018}]{2018Marshall} Marshall J.~A., Elitzur M., Armus L., Diaz-Santos T., Charmandaris V., 2018, ApJ, 858, 59. doi:10.3847/1538-4357/aabcc0
\bibitem[\protect\citeauthoryear{McDonald et al.}{2021}]{2021McDonald} McDonald M., McNamara B.~R., Calzadilla M.~S., Chen C.-T., Gaspari M., Hickox R.~C., Kara E., et al., 2021, ApJ, 908, 85. doi:10.3847/1538-4357/abd47f
\bibitem[\protect\citeauthoryear{Matsuoka et al.}{2011}]{2011Matsuoka} Matsuoka K., Nagao T., Marconi A., Maiolino R., Taniguchi Y., 2011, A\&A, 527, A100. doi:10.1051/0004-6361/201015584
\bibitem[\protect\citeauthoryear{Mortlock et al.}{2011}]{2011Mortlock} Mortlock D.~J., Warren S.~J., Venemans B.~P., Patel M., Hewett P.~C., McMahon R.~G., Simpson C., et al., 2011, Natur, 474, 616. doi:10.1038/nature10159
\bibitem[\protect\citeauthoryear{Mountrichas et al.}{2021}]{2021Mountrichas} Mountrichas G., Buat V., Yang G., Boquien M., Burgarella D., Ciesla L., Malek K., et al., 2021, A\&A, 653, A74. doi:10.1051/0004-6361/202140630

\bibitem[\protect\citeauthoryear{Nagao, Maiolino, \& Marconi}{2006a}]{2006NagaoA} Nagao T., Maiolino R., Marconi A., 2006, A\&A, 447, 863. doi:10.1051/0004-6361:20054127
\bibitem[\protect\citeauthoryear{Nardini et al.}{2009}]{2009Nardini} Nardini E., Risaliti G., Salvati M., Sani E., Watabe Y., Marconi A., Maiolino R., 2009, MNRAS, 399, 1373. doi:10.1111/j.1365-2966.2009.15357.x
\bibitem[\protect\citeauthoryear{Neistein, van den Bosch, \& Dekel}{2006}]{2006Neistein} Neistein E., van den Bosch F.~C., Dekel A., 2006, MNRAS, 372, 933. doi:10.1111/j.1365-2966.2006.10918.x
\bibitem[\protect\citeauthoryear{Neri-Larios et al.}{2011}]{2011NeriLarios} Neri-Larios D.~M., Torres-Papaqui J.~P., Coziol R., Islas-Islas J.~M., Ortega-Minakata R.~A., 2011, RMxAC, 40, 80
\bibitem[\protect\citeauthoryear{Matsuoka et al.}{2018}]{2018Matsuoka} Matsuoka K., Nagao T., Marconi A., Maiolino R., Mannucci F., Cresci G., Terao K., et al., 2018, A\&A, 616, L4. doi:10.1051/0004-6361/201833418


\bibitem[\protect\citeauthoryear{Pacucci \& Loeb}{2022}]{2022Pacucci} Pacucci F., Loeb A., 2022, MNRAS, 509, 1885. doi:10.1093/mnras/stab3071
\bibitem[\protect\citeauthoryear{P{\^a}ris et al.}{2017}]{2017Paris} P{\^a}ris I., Petitjean P., Ross N.~P., Myers A.~D., Aubourg {\'E}., Streblyanska A., Bailey S., et al., 2017, A\&A, 597, A79. doi:10.1051/0004-6361/201527999
\bibitem[\protect\citeauthoryear{Peng}{2007}]{2007Peng} Peng C.~Y., 2007, ApJ, 671, 1098. doi:10.1086/522774
\bibitem[\protect\citeauthoryear{Perna et al.}{2021}]{2021Perna} Perna M., Arribas S., Pereira Santaella M., Colina L., Bellocchi E., Catal{\'a}n-Torrecilla C., Cazzoli S., et al., 2021, A\&A, 646, A101. doi:10.1051/0004-6361/202039702


\bibitem[\protect\citeauthoryear{Rakshit, Stalin, \& Kotilainen}{2020}]{2020Rakshit} Rakshit S., Stalin C.~S., Kotilainen J., 2020, ApJS, 249, 17. doi:10.3847/1538-4365/ab99c5
\bibitem[\protect\citeauthoryear{Rakshit et al.}{2021}]{2021Rakshit} Rakshit S., Stalin C.~S., Kotilainen J., Shin J., 2021, ApJS, 253, 28. doi:10.3847/1538-4365/abd9bb
\bibitem[\protect\citeauthoryear{Ramsden et al.}{2022}]{2022Ramsden} Ramsden P., Lanning D., Nicholl M., McGee S.~L., 2022, arXiv, arXiv:2201.02650
\bibitem[\protect\citeauthoryear{Rees}{1978}]{1978Rees} Rees M.~J., 1978, Obs, 98, 210
\bibitem[\protect\citeauthoryear{Reines \& Volonteri}{2015}]{2015Reines} Reines A.~E., Volonteri M., 2015, ApJ, 813, 82. doi:10.1088/0004-637X/813/2/82
\bibitem[\protect\citeauthoryear{Reinoso et al.}{2018}]{2018Reinoso} Reinoso B., Schleicher D.~R.~G., Fellhauer M., Klessen R.~S., Boekholt T.~C.~N., 2018, A\&A, 614, A14. doi:10.1051/0004-6361/201732224
\bibitem[\protect\citeauthoryear{Richards et al.}{2009}]{2009Richards} Richards G.~T., Deo R.~P., Lacy M., Myers A.~D., Nichol R.~C., Zakamska N.~L., Brunner R.~J., et al., 2009, AJ, 137, 3884. doi:10.1088/0004-6256/137/4/3884
\bibitem[\protect\citeauthoryear{Robertson et al.}{2022}]{2022Robertson} Robertson B.~E., Tacchella S., Johnson B.~D., Hainline K., Whitler L., Eisenstein D.~J., Endsley R., et al., 2022, arXiv, arXiv:2212.04480
\bibitem[\protect\citeauthoryear{Robotham et al.}{2020}]{2020Robotham} Robotham A.~S.~G., Bellstedt S., Lagos C. del P., Thorne J.~E., Davies L.~J., Driver S.~P., Bravo M., 2020, MNRAS, 495, 905. doi:10.1093/mnras/staa1116

\bibitem[\protect\citeauthoryear{Sandage}{1965}]{1965Sandage} Sandage A., 1965, ApJ, 141, 1560. doi:10.1086/148245
\bibitem[\protect\citeauthoryear{Sandage}{1969}]{1969Sandage} Sandage A., 1969, tsra.conf, 103
\bibitem[\protect\citeauthoryear{Sandage}{1986}]{1986Sandage} Sandage A., 1986, A\&A, 161, 89
\bibitem[\protect\citeauthoryear{Sanders et al.}{1988}]{1988Sanders} Sanders D.~B., Soifer B.~T., Elias J.~H., Madore B.~F., Matthews K., Neugebauer G., Scoville N.~Z., 1988, ApJ, 325, 74. doi:10.1086/165983
\bibitem[\protect\citeauthoryear{Sakurai et al.}{2017}]{2017Sakurai} Sakurai Y., Yoshida N., Fujii M.~S., Hirano S., 2017, MNRAS, 472, 1677. doi:10.1093/mnras/stx2044
\bibitem[\protect\citeauthoryear{Shen et al.}{2019}]{2019Shen} Shen Y., Wu J., Jiang L., Ba{\~n}ados E., Fan X., Ho L.~C., Riechers D.~A., et al., 2019, ApJ, 873, 35. doi:10.3847/1538-4357/ab03d9
\bibitem[\protect\citeauthoryear{Shimasaku \& Izumi}{2019}]{2019Shimasaku} Shimasaku K., Izumi T., 2019, ApJL, 872, L29. doi:10.3847/2041-8213/ab053f
\bibitem[\protect\citeauthoryear{Shin et al.}{2019}]{2019Shin} Shin J., Nagao T., Woo J.-H., Le H.~A.~N., 2019, ApJ, 874, 22. doi:10.3847/1538-4357/ab05da
\bibitem[\protect\citeauthoryear{Sijacki et al.}{2007}]{2007Sijacki} Sijacki D., Springel V., Di Matteo T., Hernquist L., 2007, MNRAS, 380, 877. doi:10.1111/j.1365-2966.2007.12153.x
\bibitem[\protect\citeauthoryear{Silk \& Norman}{1981}]{1981Silk} Silk J., Norman C., 1981, ApJ, 247, 59. doi:10.1086/159010
\bibitem[\protect\citeauthoryear{Silk \& Rees}{1998}]{1998Silk} Silk J., Rees M.~J., 1998, A\&A, 331, L1
\bibitem[\protect\citeauthoryear{{\'S}niegowska et al.}{2021}]{2021Sniegowska} {\'S}niegowska M., Marziani P., Czerny B., Panda S., Mart{\'\i}nez-Aldama M.~L., del Olmo A., D'Onofrio M., 2021, ApJ, 910, 115. doi:10.3847/1538-4357/abe1c8
\bibitem[\protect\citeauthoryear{Soltan}{1982}]{1982Soltan} Soltan A., 1982, MNRAS, 200, 115. doi:10.1093/mnras/200.1.115
\bibitem[\protect\citeauthoryear{Stalevski et al.}{2016}]{2016Stalevski} Stalevski M., Ricci C., Ueda Y., Lira P., Fritz J., Baes M., 2016, MNRAS, 458, 2288. doi:10.1093/mnras/stw444
\bibitem[\protect\citeauthoryear{Stalevski, Tristram, \& Asmus}{2019}]{2019Stalevski} Stalevski M., Tristram K.~R.~W., Asmus D., 2019, MNRAS, 484, 3334. doi:10.1093/mnras/stz220

\bibitem[\protect\citeauthoryear{Thomas}{1999}]{1999Thomas} Thomas D., 1999, MNRAS, 306, 655. doi:10.1046/j.1365-8711.1999.02552.x
\bibitem[\protect\citeauthoryear{Tinsley \& Larson}{1979}]{1979Tinsley} Tinsley B.~M., Larson R.~B., 1979, MNRAS, 186, 503. doi:10.1093/mnras/186.3.503

\bibitem[\protect\citeauthoryear{Toba et al.}{2021}]{2021Toba} Toba Y., Ueda Y., Gandhi P., Ricci C., Burgarella D., Buat V., Nagao T., et al., 2021, ApJ, 912, 91. doi:10.3847/1538-4357/abe94a

\bibitem[\protect\citeauthoryear{Torres-Papaqui et al.}{2013}]{2013Torres-Papaqui} Torres-Papaqui J.~P., Coziol R., Plauchu-Frayn I., Andernach H., Ortega-Minakata R.~A., 2013, RMxAA, 49, 311
\bibitem[\protect\citeauthoryear{Torres-Papaqui et al.}{2020}]{2020Torres-Papaqui} Torres-Papaqui J.~P., Coziol R., Romero-Cruz F.~J., Robleto-Or{\'u}s A.~C., Escobar-V{\'a}squez G., Morales-Vargas A., Trejo-Alonso J.~J., et al., 2020, AJ, 160, 176. doi:10.3847/1538-3881/abae5a

\bibitem[\protect\citeauthoryear{Tristram et al.}{2014}]{2014Tristram} Tristram K.~R.~W., Burtscher L., Jaffe W., Meisenheimer K., H{\"o}nig S.~F., Kishimoto M., Schartmann M., et al., 2014, A\&A, 563, A82. doi:10.1051/0004-6361/201322698

\bibitem[\protect\citeauthoryear{Varma et al.}{2022}]{2022Varma} Varma S., Huertas-Company M., Pillepich A., Nelson D., Rodriguez-Gomez V., Dekel A., Faber S.~M., et al., 2022, MNRAS, 509, 2654. doi:10.1093/mnras/stab3149
\bibitem[\protect\citeauthoryear{Vayner et al.}{2021}]{2021Vayner} Vayner A., Wright S.~A., Murray N., Armus L., Boehle A., Cosens M., Larkin J.~E., et al., 2021, ApJ, 910, 44. doi:10.3847/1538-4357/abddc1
\bibitem[\protect\citeauthoryear{Veilleux et al.}{2009}]{2009Veilleux} Veilleux S., Rupke D.~S.~N., Kim D.-C., Genzel R., Sturm E., Lutz D., Contursi A., et al., 2009, ApJS, 182, 628. doi:10.1088/0067-0049/182/2/628

\bibitem[\protect\citeauthoryear{Wang et al.}{2021}]{2021Wang} Wang F., Yang J., Fan X., Hennawi J.~F., Barth A.~J., Banados E., Bian F., et al., 2021, ApJL, 907, L1. doi:10.3847/2041-8213/abd8c6
\bibitem[\protect\citeauthoryear{Wang et al.}{2022}]{2022ChuWang} Wang S., Jiang L., Shen Y., Ho L.~C., Vestergaard M., Ba{\~n}ados E., Willott C.~J., et al., 2022, ApJ, 925, 121. doi:10.3847/1538-4357/ac3a69
\bibitem[\protect\citeauthoryear{Warner, Hamann, \& Dietrich}{2003}]{2003Warner} Warner C., Hamann F., Dietrich M., 2003, ApJ, 596, 72. doi:10.1086/377710
\bibitem[\protect\citeauthoryear{Wechsler \& Tinker}{2018}]{2018Wechsler} Wechsler R.~H., Tinker J.~L., 2018, ARA\&A, 56, 435. doi:10.1146/annurev-astro-081817-051756
\bibitem[\protect\citeauthoryear{Woods et al.}{2019}]{2019Woods} Woods T.~E., Agarwal B., Bromm V., Bunker A., Chen K.-J., Chon S., Ferrara A., et al., 2019, PASA, 36, e027. doi:10.1017/pasa.2019.14
\bibitem[\protect\citeauthoryear{Wu et al.}{2015}]{2015Wu} Wu X.-B., Wang F., Fan X., Yi W., Zuo W., Bian F., Jiang L., et al., 2015, Natur, 518, 512. doi:10.1038/nature14241
\bibitem[\protect\citeauthoryear{Wright et al.}{2010}]{2010Wright} Wright E.~L., Eisenhardt P.~R.~M., Mainzer A.~K., Ressler M.~E., Cutri R.~M., Jarrett T., Kirkpatrick J.~D., et al., 2010, AJ, 140, 1868. doi:10.1088/0004-6256/140/6/1868

\bibitem[\protect\citeauthoryear{Xie et al.}{2021}]{2021Xie} Xie Y., Ho L.~C., Zhuang M.-Y., Shangguan J., 2021, ApJ, 910, 124. doi:10.3847/1538-4357/abe404
\bibitem[\protect\citeauthoryear{Xu et al.}{2015}]{2015Xu} Xu L., Rieke G.~H., Egami E., Haines C.~P., Pereira M.~J., Smith G.~P., 2015, ApJ, 808, 159. doi:10.1088/0004-637X/808/2/159

\bibitem[\protect\citeauthoryear{Yajima \& Khochfar}{2016}]{2016Yajima} Yajima H., Khochfar S., 2016, MNRAS, 457, 2423. doi:10.1093/mnras/stw058
\bibitem[\protect\citeauthoryear{Yang et al.}{2020}]{2020Yang} Yang G., Boquien M., Buat V., Burgarella D., Ciesla L., Duras F., Stalevski M., et al., 2020, MNRAS, 491, 740. doi:10.1093/mnras/stz3001
\bibitem[\protect\citeauthoryear{Yang et al.}{2020}]{2020YangJ} Yang J., Wang F., Fan X., Hennawi J.~F., Davies F.~B., Yue M., Banados E., et al., 2020, ApJL, 897, L14. doi:10.3847/2041-8213/ab9c26
\bibitem[Yang et al.(2022)]{2022Yang} Yang, G., Boquien, M., Brandt, W.~N., et al.\ 2022, The Astrophysical Journal, 927, 192. doi:10.3847/1538-4357/ac4971

\bibitem[\protect\citeauthoryear{Zhang, Yang, \& Guo}{2021}]{2021Zhang} Zhang Y., Yang X., Guo H., 2021, MNRAS, 507, 5320. doi:10.1093/mnras/stab2487

\end{thebibliography}
\bsp	
\label{lastpage}
\end{document}